\documentclass[%
superscriptaddress,
twocolumn,
 amsmath,amssymb,
 aps,
]{revtex4-2}

\usepackage{graphicx}
\usepackage{dcolumn}
\usepackage{bm}
\usepackage{color}
\usepackage[dvipsnames]{xcolor}
\usepackage[colorlinks,citecolor=blue]{hyperref}

\newcommand{\gcc}{g$\,$cm$^{-3}$}
\definecolor{darkviolet}{rgb}{0.58, 0.0, 0.83}

\begin{document}


\title{First-Principles Equation of State Database for Warm Dense Matter Computation}

\author{Burkhard Militzer}
\email{militzer@berkeley.edu}
\affiliation{Department of Earth and Planetary Science, University of California, Berkeley, CA 94720, USA}
\affiliation{Department of Astronomy, University of California, Berkeley, CA 94720, USA}

\author{Felipe Gonz\'alez-Cataldo}
\affiliation{Department of Earth and Planetary Science, University of California, Berkeley, CA 94720, USA}

\author{Shuai Zhang}
\affiliation{Department of Earth and Planetary Science, University of California, Berkeley, CA 94720, USA}
\affiliation{Lawrence Livermore National Laboratory, Livermore, California 94550, USA}
\affiliation{Laboratory for Laser Energetics, University of Rochester, Rochester, NY 14623, USA}

\author{Kevin P. Driver}
\affiliation{Department of Earth and Planetary Science, University of California, Berkeley, CA 94720, USA}
\affiliation{Lawrence Livermore National Laboratory, Livermore, California 94550, USA}

\author{Fran\c{c}ois Soubiran}
\affiliation{Department of Earth and Planetary Science, University of California, Berkeley, CA 94720, USA}
\affiliation{CEA DAM-DIF, 91297 Arpajon, France}

\begin{abstract}
We put together a first-principles equation of state (FPEOS) database for matter at extreme conditions by combining results from path integral Monte Carlo and density functional molecular dynamics simulations of the elements H, He, B, C, N, O, Ne, Na, Mg, Al and Si as well as the compounds LiF, B$_4$C, BN, CH$_4$, CH$_2$, C$_2$H$_3$, CH, C$_2$H, MgO, and MgSiO$_3$. For all these materials, we provide the pressure and internal energy over a density-temperature range from $\sim$0.5 to 50 \gcc~and from $\sim$10$^4$ to 10$^{9}$ K, which are based on $\sim$5000 different first-principles simulations. We compute isobars, adiabats and shock Hugoniot curves in the regime of L and K shell ionization. Invoking the linear mixing approximation, we study the properties of mixtures at high density and temperature. We derive the Hugoniot curves for water and alumina as well as for carbon-oxygen, helium-neon, and CH-silicon mixtures. We predict the maximal shock compression ratios of H$_2$O, H$_2$O$_2$, Al$_2$O$_3$, CO, and CO$_2$ to be 4.61, 4.64, 4.64, 4.89, and 4.83, respectively. Finally we use the FPEOS database to determine the points of maximum shock compression for all available binary mixtures. We identify mixtures that reach higher shock compression ratios than their endmembers. We discuss trends common to all mixtures in pressure-temperature and particle-shock velocity spaces. In the supplementary material, we provide all FPEOS tables as well as computer codes for interpolation, Hugoniot calculations, and plots of various thermodynamic functions. 
\end{abstract}


\maketitle

\section{Introduction}

A rigorous and consistent theoretical description of warm dense matter (WDM) has been identified~\cite{FESAC2009,HEDLP2009,FCWDM} as a central goal to the development of key plasma technologies, such as inertial and magnetic confinement fusion~\cite{Seidl2009,Hammel2010,Lindl2014,Miyanishi2015,Betti2016,Gaffney2018}, shock physics~\cite{Ze66,Remington2006,Fortov2009}, and high energy astrophysics~\cite{SC95,ChabrierBaraffe2000}. WDM represents materials at solid-state densities and elevated temperatures of 10$^4$--10$^7$ K $\approx$ 1--10$^3$ eV. This regime is particularly challenging to characterize with analytical methods because there is no small parameter for perturbative methods to be applicable. The densities are too high and the interaction effects too strong for typical plasma theory models~\cite{Ebeling1976,ER85a,Ro86,Ro90,PotekhinChabrier2000,Rozsnyai2001}, such as Saha ionization models, or the Debye plasma model~\cite{Debye1923}, to be applicable. On the other hand, the temperatures are too high and the fraction of excited electrons too large for conventional condensed matter theory to apply. Chemical bonds are short-lived but cannot be neglected. One expects these systems to be partially ionized and some of the electrons to occupy excited and free states. Because of the high density, Pauli exclusion effects are relevant when the ionization equilibrium is established, which renders these systems partially degenerate~\cite{Surh01}. A fraction of the electrons occupy core states because density is orders of magnitude too low for them to form a rigid, neutralizing background. In this regard, a one-component plasma model would be a poor description of WDM. 

Despite these challenges, the development of a rigorous and consistent theoretical framework to describe WDM remains to be of high importance because it represents states of matter on the pathway to reaching fusion conditions. Predicting with high accuracy the equation of state (EOS) as well as transport and optical properties at extreme pressure and temperature conditions is the primary motivation for developing new methods. Significant progress towards this goal has been made with laboratory experiments~\cite{Nellis1991,Weir1996,Remington2006,Knudson2012,Gomez2020} and first-principles (FP) computer simulations~\cite{Graziani2014Book}. Since hydrodynamic simulations typically guide the design of dynamic compression experiments and they rely on accurate EOS tables to be predictive, computer simulations of any material and thermodynamic condition that can be probed with laboratory experiments are of high interest. FP computer simulations, that are based on the fundamental laws of quantum mechanics, enabled us to compute the EOS of materials over a wide range of conditions that also include planetary and stellar interiors. In giant planets~\cite{hubbard_planets}, not only hydrogen-helium mixtures~\cite{Vorberger2007,Militzer2013b,Militzer2013c,Wahl2017,CMS_EOS,Militzer2019b} but also rocky materials~\cite{Wilson2012,Wahl2013a,REOS3,Soubiran2016,Gonzalez-Cataldo2014,Gonzalez-Cataldo2016,Soubiran2017} are exposed to pressures of tens of megabars and temperature of $\sim$10$^4$ K. Accurate EOSs are needed to complete the spacecraft measurements of giant planets in our solar system to better characterize their interior structure and evolution~\cite{Saumon2004,Baraffe2014,Militzer2016b}. The discovery of thousands of exoplanets with ground-based observations and space missions~\cite{Guillot1999,Borucki2011,Deming2020} has considerably broadened the range of conditions and materials of interest~\cite{Madhusudhan2011,Wagner2012,Wilson2014}. 

Stellar interiors encompass a wide range of temperatures from  10$^4-$10$^8\,$K. The most detailed information came from observing the normal mode oscillations of our Sun~\cite{Christensen1996,Christensen2002,Schumacher2020}. Such astero-seismological observations now improve our understanding of distant stars~\cite{Aerts2019}. For the first time, the frequencies of a number of normal modes in a giant planet have been determined with high precision through the detection of spiral density waves in Saturn's ring by the Cassini spacecraft~\cite{HedmanNicholson2013}.


In this article, we build a FPEOS database for WDM computation by combining the results of two computer simulation methods, path integral Monte Carlo (PIMC) calculations and density functional theory molecular dynamics (DFT-MD) simulations. Alternative methods to perform these calculations include orbital-free density functional theory~\cite{Lambert2006,Karasiev2013,Sjostrom2014}, Thomas-Fermi molecular dynamics~\cite{Mazevet2019}, or average atom models~\cite{Sterne2007,Rozsnyai2014,Starrett2016Si,Pain2007} or their combination~\cite{Danel2012,Danel2014}. With their approximations, all these methods enable one to compute the properties of WDM independently for one set of temperature-density condition at a time. In this regard, they differ from conventional EOS models that start from a cold curve and then introduce nuclear and electronic excitations by constructing elaborate free energy models. Multi-material databases like the Quotidian EOS~\cite{QEOS} and many Sesame models~\cite{Sesame} rely on that approach. 

Here we instead rely exclusively on predictions from FP computer simulations in order to build an FPEOS database to characterize 11 elements and 10 compounds over a wide range of temperature and density conditions. We exclude nuclear reactions from consideration even though they occur at the highest temperatures that we study. We predict the shock Hugoniot curves and study a variety of binary mixtures by invoking the ideal mixing approximation at constant pressure, $P$, and temperature, $T$. In Ref.~\cite{Militzer2020}, this approximation has been shown to work remarkably well for WDM computations for temperatures above $2 \times 10^5\,$K and the shock compression ratio exceeding $\sim$ 3.2. With the goal of making WDM computations more reliable and efficient, we make available as supplemental material all EOS tables as well as the C++ computer codes for their interpolation. Python code is provided to generate graphs of shock Hugoniot curve, isentropes, isobars, and isotherms for compounds and user-defined mixtures~\cite{SupplementalMaterial}. 

\section{Methods}

\subsection{PIMC Simulations}

\begin{table*}[]
    \begin{center}
\begin{tabular}{l |c r r|r r | c | c } 
Material   & Number       & Minimum & Maximum &  Minimum       &  Maximum      &  Number of   & References\\
           & of isochores & density & density &  temperature   &  temperature  &  EOS points              \\
           &              & [\gcc]  & [\gcc]  &  [K]~     &  [K]~~~~~ &                          \\
\hline
Hydrogen   &  33  &  0.001  & 798.913   &  15625   &  6.400$\times 10^{7}$  &  401 & ~\cite{MC00,MC01,Hu2010,Hu2011,Mi01,Mi99}\\
Helium     &   9  &  0.387  &  10.457   &    500   &  2.048$\times 10^{9}$  &  228 & \cite{Mi06,Militzer2009} \\
Boron      &  16  &  0.247  &  49.303   &   2000   &  5.174$\times 10^{8}$  &  314 & \cite{Zhang2018} \\
Carbon     &   9  &  0.100  &  25.832   &   5000   &  1.035$\times 10^{9}$  &  162 & \cite{Driver2012,Benedict2014}\\
Nitrogen   &  17  &  1.500  &  13.946   &   1000   &  1.035$\times 10^{9}$  &  234 & \cite{DriverNitrogen2016}  \\
Oxygen     &   6  &  2.486  & 100.019   &  10000   &  1.035$\times 10^{9}$  &   76 & \cite{Driver2015b}\\
Neon       &   4  &  0.895  &  15.026   &   1000   &  1.035$\times 10^{9}$  &   67 & \cite{Driver2015}  \\
Sodium     &   9  &  1.933  &  11.600   &   1000   &  1.293$\times 10^{8}$  &  193 & \cite{ZhangSodium2017,Zhang2016b} \\
Magnesium  &  23  &  0.431  &  86.110   &  20000   &  5.174$\times 10^{8}$  &  371 & \cite{Gonzalez-Cataldo2020}\\
Aluminum   &  15  &  0.270  &  32.383   &  10000   &  2.156$\times 10^{8}$  &  240 & \cite{Driver2018}\\
Silicon    &   7  &  2.329  &  18.632   &  50000   &  1.293$\times 10^{8}$  &   85 & \cite{MilitzerDriver2015,Hu2016}\\
 \hline
LiF        &   8  &  2.082  &  15.701   &  10000   &  1.035$\times 10^{9}$  &   91 & ~\cite{Driver2017}\\
B$_4$C     &  16  &  0.251  &  50.174   &   2000   &  5.174$\times 10^{8}$  &  291 & \cite{Zhang_B4C_2020}\\
BN         &  16  &  0.226  &  45.161   &   2000   &  5.174$\times 10^{8}$  &  311 & \cite{ZhangBN2019}\\	
CH$_4$     &  16  &  0.072  &  14.376   &   6736   &  1.293$\times 10^{8}$  &  247 & \cite{ZhangCH2017,ZhangCH2018} \\
CH$_2$     &  16  &  0.088  &  17.598   &   6736   &  1.293$\times 10^{8}$  &  248 & \cite{ZhangCH2017,ZhangCH2018}  \\
C$_2$H$_3$ &  16  &  0.097  &  19.389   &   6736   &  1.293$\times 10^{8}$  &  247 & \cite{ZhangCH2017,ZhangCH2018} \\
CH         &  16  &  0.105  &  21.000   &   6736   &  1.293$\times 10^{8}$  &  248 & \cite{ZhangCH2017,ZhangCH2018} \\
C$_2$H     &  16  &  0.112  &  22.430   &   6736   &  1.293$\times 10^{8}$  &  245 & \cite{ZhangCH2017,ZhangCH2018} \\
MgO        &  19  &  0.357  &  71.397   &  20000   &  5.174$\times 10^{8}$  &  286 & \cite{Soubiran2019}\\
MgSiO$_3$  &  16  &  0.321  &  64.158   &   6736   &  5.174$\times 10^{8}$  &  284 & \cite{Gonzalez2020,GonzalezMilitzer2020}\\
 \hline
\end{tabular}
    \end{center}
    \caption{Parameters of the 21 EOS tables in this database. A total of 4869 first-principles calculations were combined. }
    \label{tab:1}
\end{table*}

The equations of state in Table~\ref{tab:1} were assembled by combining published results from PIMC simulations at high temperature and from Kohn-Sham density DFT-MD simulations at lower temperature. The PIMC simulation method is based on the early work on superfluid $^4$He that introduced the permutation sampling to the path integral computations~\cite{PC84,PC87,Ce95}. The algorithm was subsequently extended to fermionic systems by introducing the {\em restricted} paths approach~\cite{Ce91,Ce92,Ce96}. The first results of this simulation method were reported in the seminal works on liquid $^3$He~\cite{Ce92} and dense hydrogen~\cite{PC94,Ma96}. Simulations of one-component plasmas~\cite{JC96,MP04,MP05} and of hydrogen-helium mixtures~\cite{Mi05} followed. In Ref.~\cite{Driver2012}, it was demonstrated that the free-particle nodal approximation worked sufficiently well to study hot, dense carbon and water, which paved the way for performing the PIMC simulations of elements from hydrogen to neon (Z=10), as shown in Tab.~\ref{tab:1}. In Ref.~\cite{MilitzerDriver2015}, Hartree-Fock orbitals were used to efficiently incorporate localized electronic states into the nodal structure, which extended the applicability of fermionic PIMC simulations to heavier elements up to silicon (Z=14).

The PIMC method is based on the thermal density matrix of a quantum
system, $\hat\rho=e^{-\beta \hat{\cal H}}$, that is expressed as a
product of higher-temperature matrices, $e^{-\beta \hat{\cal H}}=(e^{-\tau \hat{\cal H}})^M$. The integer $M$ represents the number of steps along the path in imaginary time. $\tau=\beta/M$ is the corresponding time step. The path integral emerges when the operator $\hat\rho$
is evaluated in real space,
\begin{equation}
\left<\mathbf R|\hat\rho| \mathbf R'\right>=\frac{1}{N!}\sum_{\mathcal P}(-1)^{\mathcal P}\oint_{\mathbf R\to\mathcal P\mathbf R'}\mathbf{dR}_t\, e^{-S[\mathbf R_t]}.
\label{PI}
\end{equation}
The sum over $\mathcal P$ represents all permutations of identical fermions that in combination with the $(-1)^P$ factor project out the antisymmetric states.  For sufficiently small time steps, $\tau$, all many-body correlation effects vanish and the action, $S[\mathbf R_t]$, can be
computed by solving a series of two-particle problems~\cite{PC84,Na95,BM2016}. The advantage of this approach is that  all many-body quantum correlations are recovered through the integration over paths. The integration also enables one to compute quantum mechanical expectation values of thermodynamic observables, such as the kinetic and potential energies, pressure,
pair correlation functions and the momentum distribution~\cite{Ce95,Militzer2019}. Most practical many-body implementations of the path integral method rely on Monte Carlo sampling techniques
because the integral has $D \times N \times M$ dimensions in
addition to sum over permutations. ($D$ is the number of spatial dimensions; $N$ is the number of particles.) The method becomes increasingly efficient at high temperature because the length of the paths scales like $1/T$. In the limit of low temperature, where few electronic excitations are present, the PIMC method becomes computationally demanding and the Monte Carlo sampling can become inefficient. Still, the PIMC method avoids any exchange-correlation approximation and the calculation of single-particle eigenstates, which are embedded in all standard Kohn-Sham DFT calculations. 

The PIMC simulations were performed with the CUPID code~\cite{MilitzerThesis} using periodic boundary conditions. The necessary computer time foremost depends on the number of electrons and the number of path integral time steps. Earlier simulations of hydrogen were performed with 32 electrons~\cite{MC00} while later calculations~\cite{Hu2011} used between 64 and 256 electrons depending on density. The helium calculations in Ref.~\cite{Militzer2009} employed 64 and 114 electrons. For the simulations of boron~\cite{Zhang2018} and B$_4$C~\cite{Zhang_B4C_2020}, slightly larger cells with 30 nuclei and 150 (B) or 156 (B$_4$C) electrons were used. For the simulations of elemental nitrogen, oxygen, magnesium, silicon, we employed cells with 8 nuclei and 56, 64, 96, and 112 electrons, respectively. For BN and MgSiO$_3$, PIMC simulations with 144 electrons were performed. A detailed finite-size study is provided in the supplementary material of Ref.~\cite{Driver2015}. 

\subsection{DFT-MD Simulations}

All DFT-MD simulations were performed with the Vienna Ab initio Simulation Package (VASP)~\cite{Kresse1999}. We used the hardest projector augmented wave~\cite{Blochl1994} pseudopotentials that were available for that code. The Perdew-Burke-Ernzerhof (PBE)~\cite{PBE} functional or the local density approximation~\cite{Ceperley1980,Perdew81} were employed to incorporate exchange-correlation effects. 
We used cubic simulation cells with periodic boundary conditions and, to improve efficiency, we used a smaller number of atoms at the highest temperatures than we employ at lower temperature. As shown in our previous work~\cite{Driver2015b,ZhangCH2017,Driver2018,Soubiran2019}, this is not detrimental to the accuracy of the EOS data at high temperatures. 

The Mermin functional~\cite{Mermin1965} was used throughout to incorporate the effects of excited electronic states at elevated temperatures. The temperature condition where we switched from DFT-MD to PIMC depends on the material. For low-Z materials like helium, we already switched to PIMC at 10$^5$ K while for elements from Na through Si, we performed DFT-MD simulation for temperatures as high as 2$\times 10^6\,$K. The agreement between the EOSs derived with PIMC and DFT-MD methods is fairly good. The deviations in pressure were found to be 2\% or less while the internal energies typically deviated by $\sim$5 Ha/nucleus or less. This means that fundamental approximations, like the nodal structure in PIMC and the choice an exchange-correlation functional in DFT methods, do not prevent us from constructing consistent EOS tables for all 21 materials under consideration. This also suggests that the most fundamental electronic properties were accurately described with both method and that the numerical approximations were reasonably well controlled. Moreover, in Ref.~\cite{surprapa2018,ZhangBN2019,Zhang_B4C_2020} it was shown that alternate DFT methods such Fermi operator expansion and spectral quadrature as well as different pseudopotentials and exchange-correlation functionals gave consistent results. 

With the described approximations, one typically finds that the predictions from FP simulations agree with results from laboratory experiments at extreme temperatures and pressures because at the present time, these measurements have error bars that are larger those of high-pressure experiments at room temperature, which have enabled us to benchmark the accuracy of different FP methods~\cite{Shulenburger2013}. Still as experiments at high temperature and pressure become more precise in the future, one will need to revisit the fundamental approximations (Fermion nodes in PIMC and exchange-correlation description in DFT) and the controllable approximations (finite size effects in both methods, convergence with respect to simulation duration in DFT-MD and PIMC, pseudopotential approximation in DFT-MD) that are employed in state-of-the-art simulation methods today. 

\section{Results for Single Compounds}

\begin{figure}[h]
    \centering
    \includegraphics[width=1.0\columnwidth]{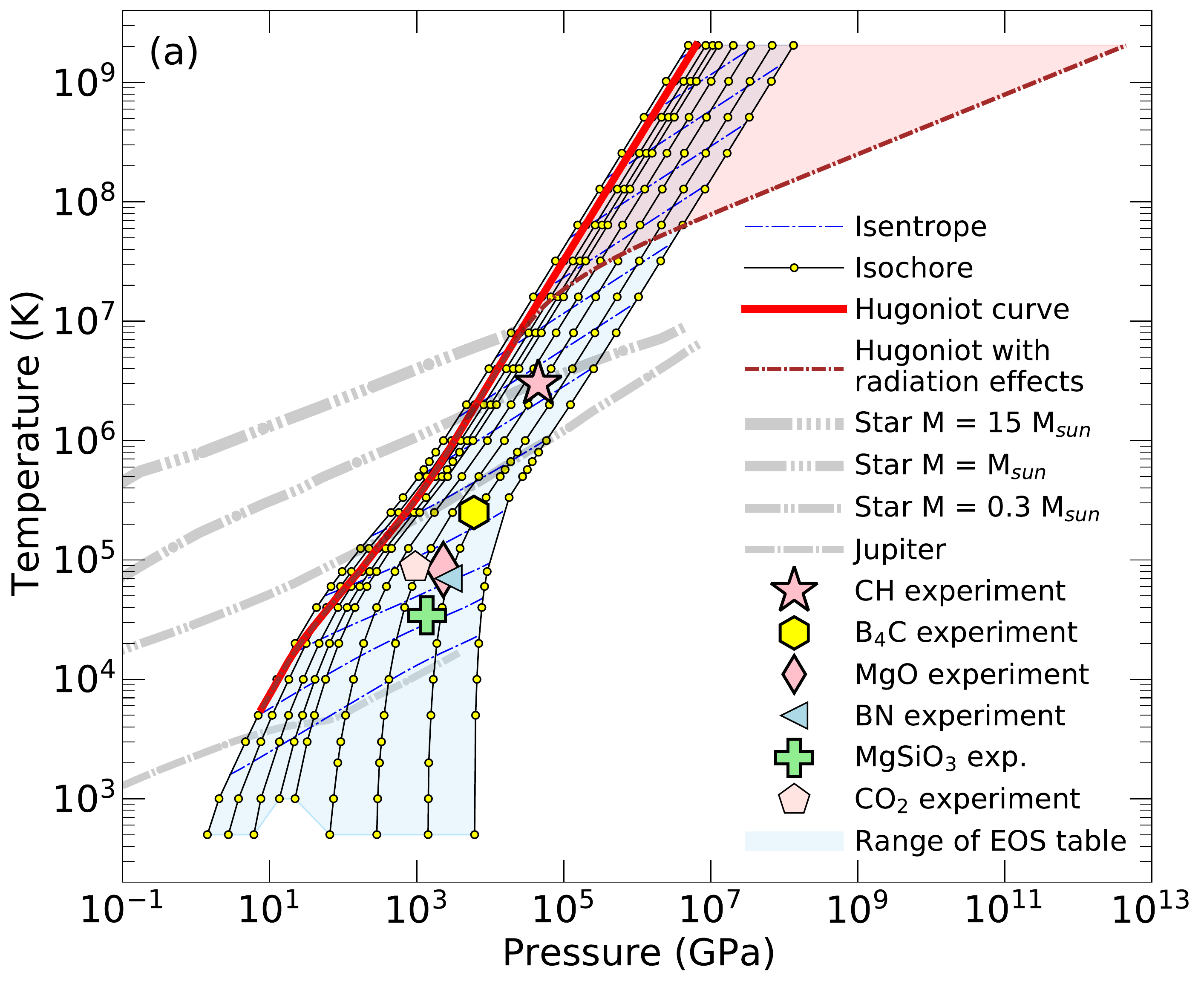}
    \includegraphics[width=1.0\columnwidth]{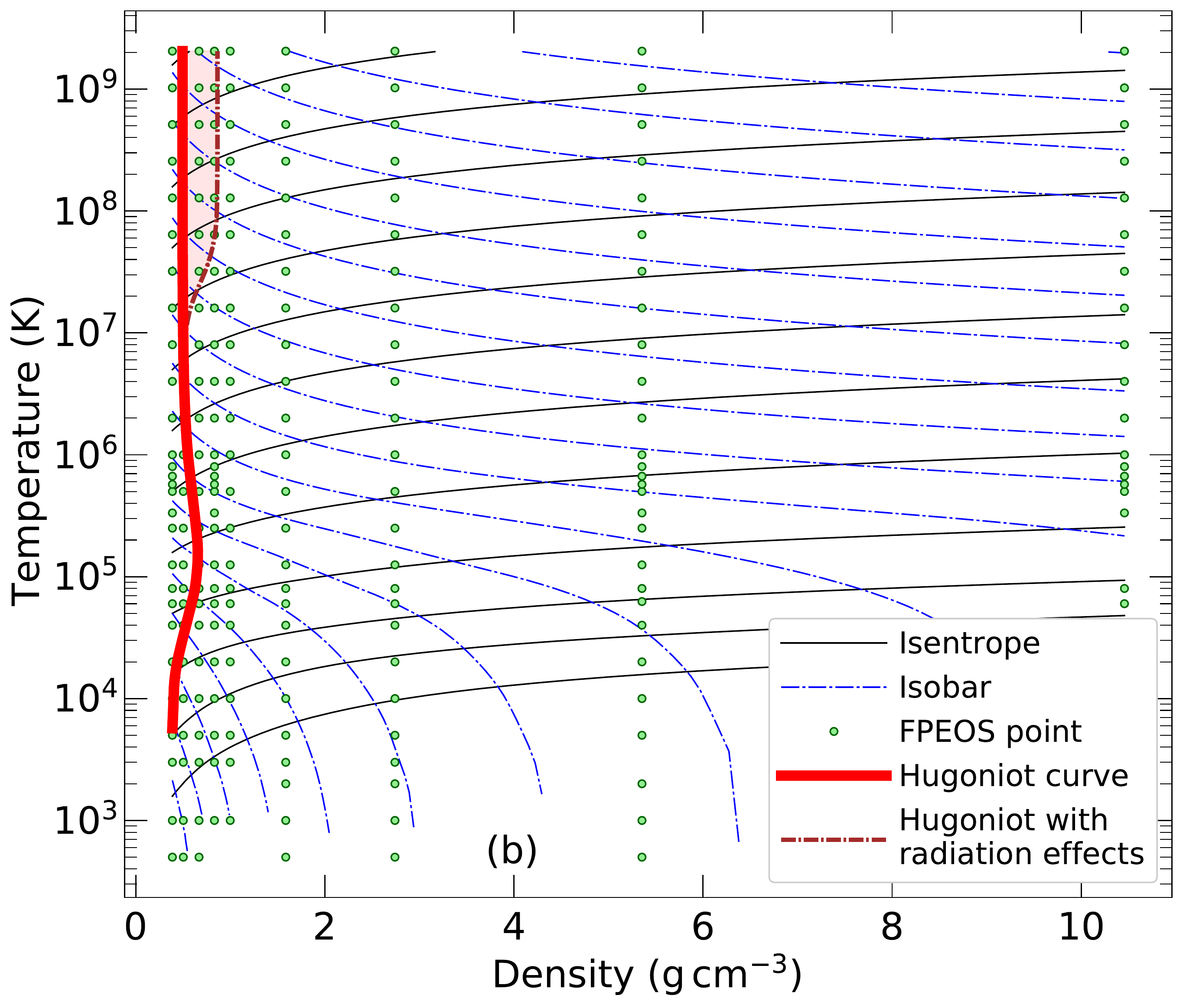}
    \caption{Density-temperature-pressure conditions of our first-principles simulations of helium~\cite{Militzer2009}. Isobars, isentropes, and shock Hugoniot curves are shown. In the upper panel, we include the interior conditions of Jupiter~\cite{Militzer2016b,Wahl2017a} and main-sequence stars of different masses~\cite{SC95} as well as the highest-pressure conditions reached in recent shock wave experiments on CH~\cite{Kritcher2020}, B$_4$C~\cite{Zhang_B4C_2020}, MgO~\cite{McCoy2019}, BN~\cite{ZhangBN2019}, and MgSiO$_3$~\cite{Millot2020}, and CO$_2$~\cite{Crandall2020}. The corresponding temperatures were not measured but derived from simulations.} 
    \label{fig:EOS_helium} 
\end{figure}

In this section, we outline the basic functions of our FPEOS database for single compounds. In the following two sections, we will discuss the properties of specific mixtures and then query the database to compute properties of all binary mixtures. In Table~\ref{tab:1}, we provide the density, $\rho$, and temperature ranges of the EOS tables of eleven elements and ten compounds in our database. We chose helium as an example to illustrate the calculations and plots that our database provides for all these 21 materials. In Fig.~\ref{fig:EOS_helium}, we directly plot the EOS points from the first-principles simulations in $T$-$P$ and $T$-$\rho$ spaces. We added isobars that we obtained via a 2D spline interpolation of $P(\rho,T)$ that we also employ to interpolate the internal energy, $E(\rho,T)$. As a guide for future ramp compression experiments, we also plotted a collection of isentropes that we derived from the relationship
$\left.\frac{\partial T}{\partial V}\right|_S=-T\left.\frac{\partial P}{\partial T}\right|_V/\left.\frac{\partial E}{\partial T}\right|_V$. 

\begin{figure}
    \centering
    \includegraphics[width=1.0\columnwidth]{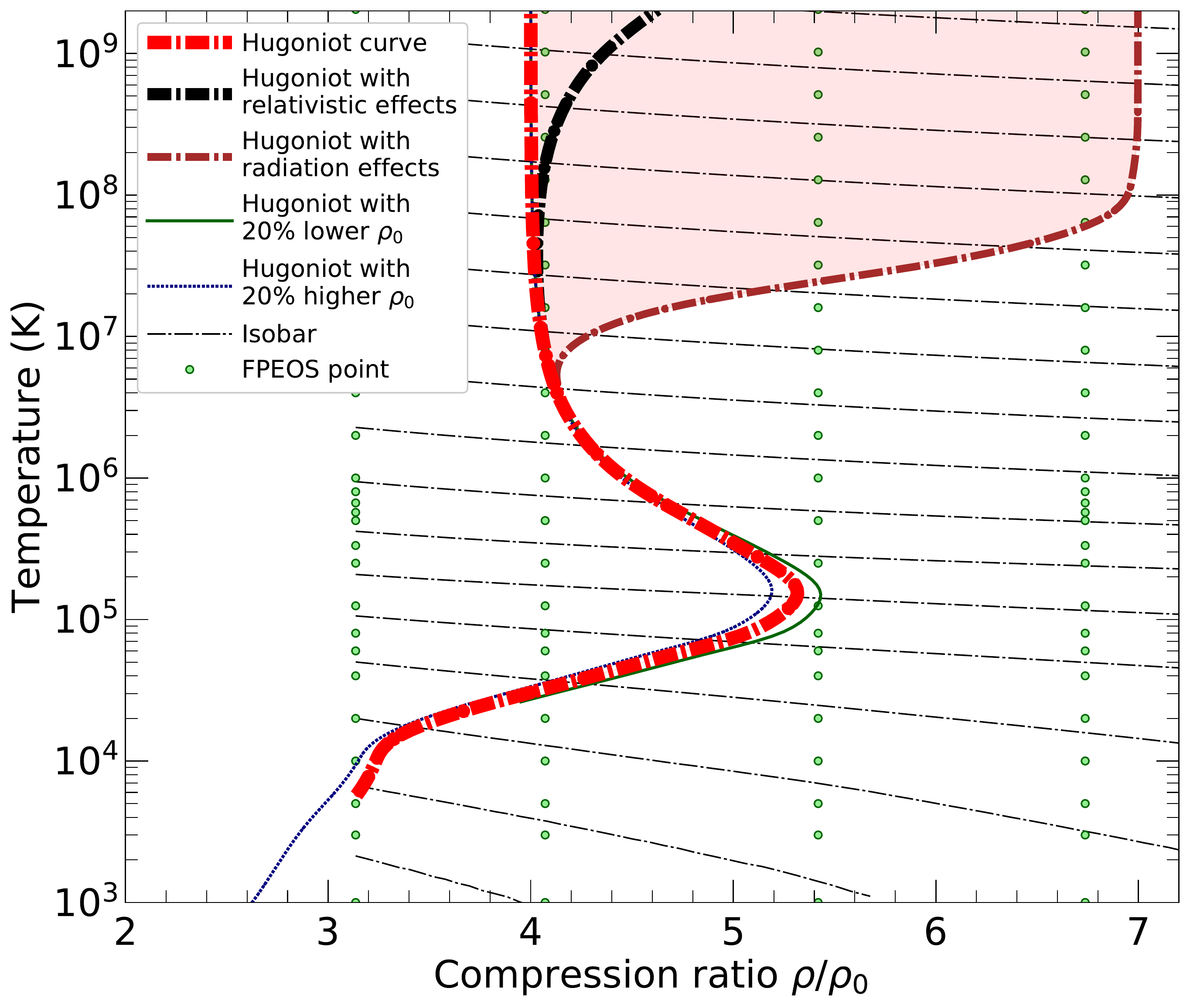}
    \caption{Shock Hugoniot curves for helium at different initial densities ($\rho_0$=0.1235 \gcc) are shown in compression-temperature space. Relativistic and radiation effects have also been introduced. 
    \label{fig:Hug_helium}
    }
\end{figure}

Then we added different shock Hugoniot curves that predict the states generated in dynamic compression experiments. By only measuring the shock
and particle velocities, they provide a direct way to determine the EOS of materials at extreme conditions. The sample material initially has the internal energy, pressure, and
volume, $\{E_0,P_0,V_0\}$. Under shock compression, the material
reaches a final state denoted by $\{E(\rho,T),P(\rho,T),V\}$.  The
conservation of mass, momentum, and energy across the shock front leads
to the Rankine-Hugoniot relation~\cite{Hugoniot1887,Hugoniot1889,Ze66},
\begin{equation}
\left[E(\rho,T)-E_0\right] + \frac{1}{2} \left[P(\rho,T)+P_0\right]\left[V-V_0\right] = 0.
\label{eq:hug}
\end{equation}
The volume, $V$, follows from the density, $\rho=m/V$. For helium, we set $\rho_0=0.1235$ \gcc. In Fig.~\ref{fig:Hug_helium}, we show the resulting shock Hugoniot curve that has a pronounced compression maximum of $\rho/\rho_0= 5.32$ at $T=151\,$000$\,$K and $P=370$ GPa. If internal degrees of freedom are excited at high $T$ and $P$, the typical compression ratio of an ideal gas ($\rho/\rho_0=4$) can be exceeded because these excitations increase the internal energy $E$, which is then compensated by a decrease in volume to satisfy Eq.~\eqref{eq:hug}. For conditions under consideration in this article, it is the excitation of K and L shell electrons that introduce one or two compression maxima into the Hugoniot curves that we compute here.

In Fig.~\ref{fig:Hug_helium}, we also show the effect of relativistic electrons that increases the shock compression for temperatures above 10$^8\,$K. Since relativistic effects are not included in our PIMC computations, we derived them for an ideal electron gas assuming complete ionization. We also show a Hugoniot curve with radiation effects. Assuming an ideal black body behavior, we very approximately derived the photon contribution to the EOS using $P_\text{rad}=(4\sigma/3c)T^4$ and $E_\text{rad}=3 P_\text{rad}V$, where $\sigma$ is the Stefan-Boltzmann constant and $c$ is the speed of light in vacuum. We find that radiative effects are important for temperatures above $5\times10^6$ K, which are well above the temperature necessary to completely ionize the K shell electrons of the helium nuclei. 

\begin{figure}
    \centering
    \includegraphics[width=1.0\columnwidth]{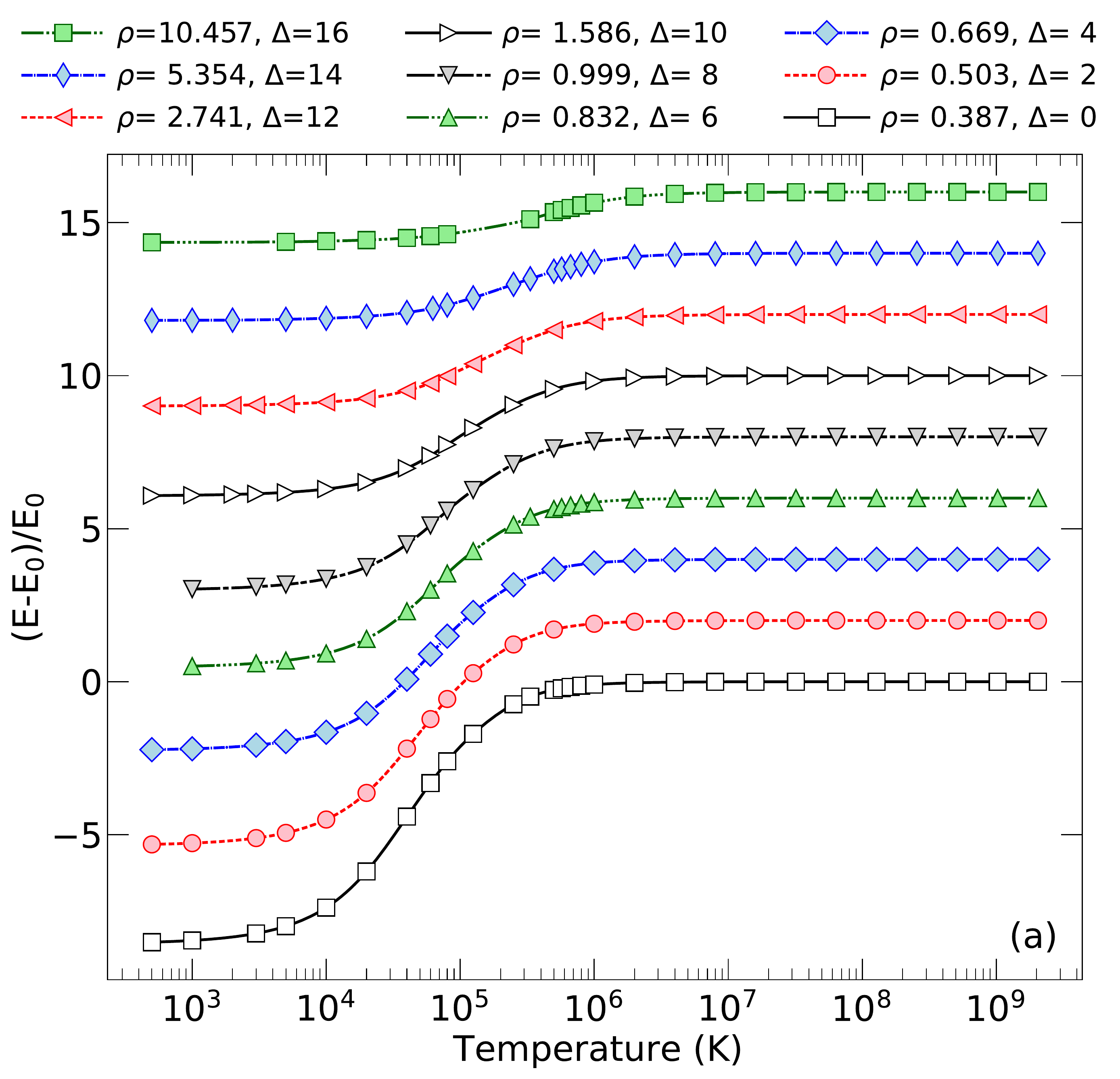}
    \includegraphics[width=1.0\columnwidth]{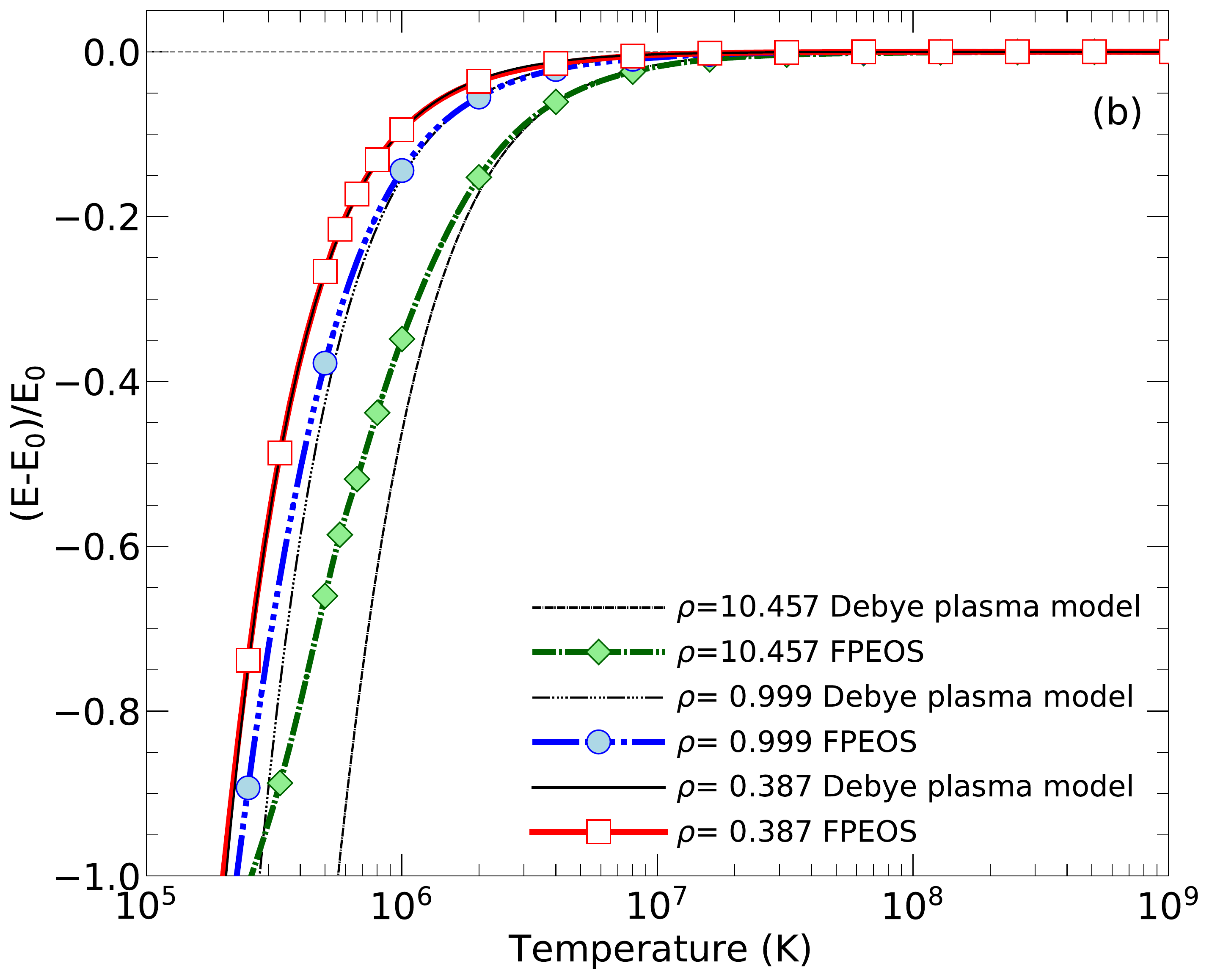}
    \caption{Internal energy of hot, dense helium is shown for a collection of isochores. Their densities are given in the captions in units of \gcc. For clarity, we shifted the individual curves in the upper panel by $\Delta$ and removed $E_0$, the energy of a noninteracting Fermi gas of electrons and classical nuclei. In the lower panel, we show that the results from our first-principles simulation converge to prediction from the Debye plasma models in the limit of high temperature where screening effects are the dominant type of the interaction.} 
\label{fig:E_helium}
\end{figure}

For every material, our PIMC and DFT-MD results can be combined into a single EOS table that can be smoothly interpolated. In Fig.~\ref{fig:E_helium}, we plot the internal energy of helium for a collection of isochores. To reduce the range of the Y axis, we removed the contribution from ideal Fermi gas of electrons and classical nuclei. In this figure, we also show that our results converge to the predictions of the Debye plasma model~\cite{Debye1923} in the limit of very high temperature. At lower temperatures, this model quickly fails because it does not include any bound states. 

\section{Results for Mixtures}

In Ref.~\cite{Militzer2020}, we demonstrated that ideal mixing approximation works well for temperatures above $2 \times 10^5\,$K and shock compression ratios greater than $\sim$3.2. The magnitude of nonideal mixing effects were found to be small and the shock Hugoniot curve of BN, B$_4$C, MgO, and MgSiO$_3$ could all be reproduced with high precision by mixing the EOSs of the elemental substances at constant pressure and temperature. The good agreement included the regimes of K and L shell ionization that lead to compression maxima on Hugoniot curves. This remarkable agreement is the basis for the mixture calculations that we implemented into our FPEOS database. Neglecting all inter-species interactions, the linear mixing approximation assumes all extensive properties of the mixtures can be derived by adding the contributions from the components 1 and 2 at the $P$ and $T$ conditions of interest as follows:
\begin{align}
V_{\rm mix} &= N_1 V_1 + N_2 V_2\;,\label{eq:Vmix}\\
m_{\rm mix} &= N_1 m_1 + N_2 m_2\;,\\
E_{\rm mix} &= N_1 E_1 + N_2 E_2\;,
\end{align}
 where all variables have been normalized per formula unit. $N_1$ and $N_2$ specify how many formula units of species 1 and 2 are contained in one unit of the mixture. The mass density of the mixture is given by $\rho_{\rm mix} = m_{\rm mix} / V_{\rm mix}$. 

\begin{figure}
    \centering
    \includegraphics[width=1.0\columnwidth]{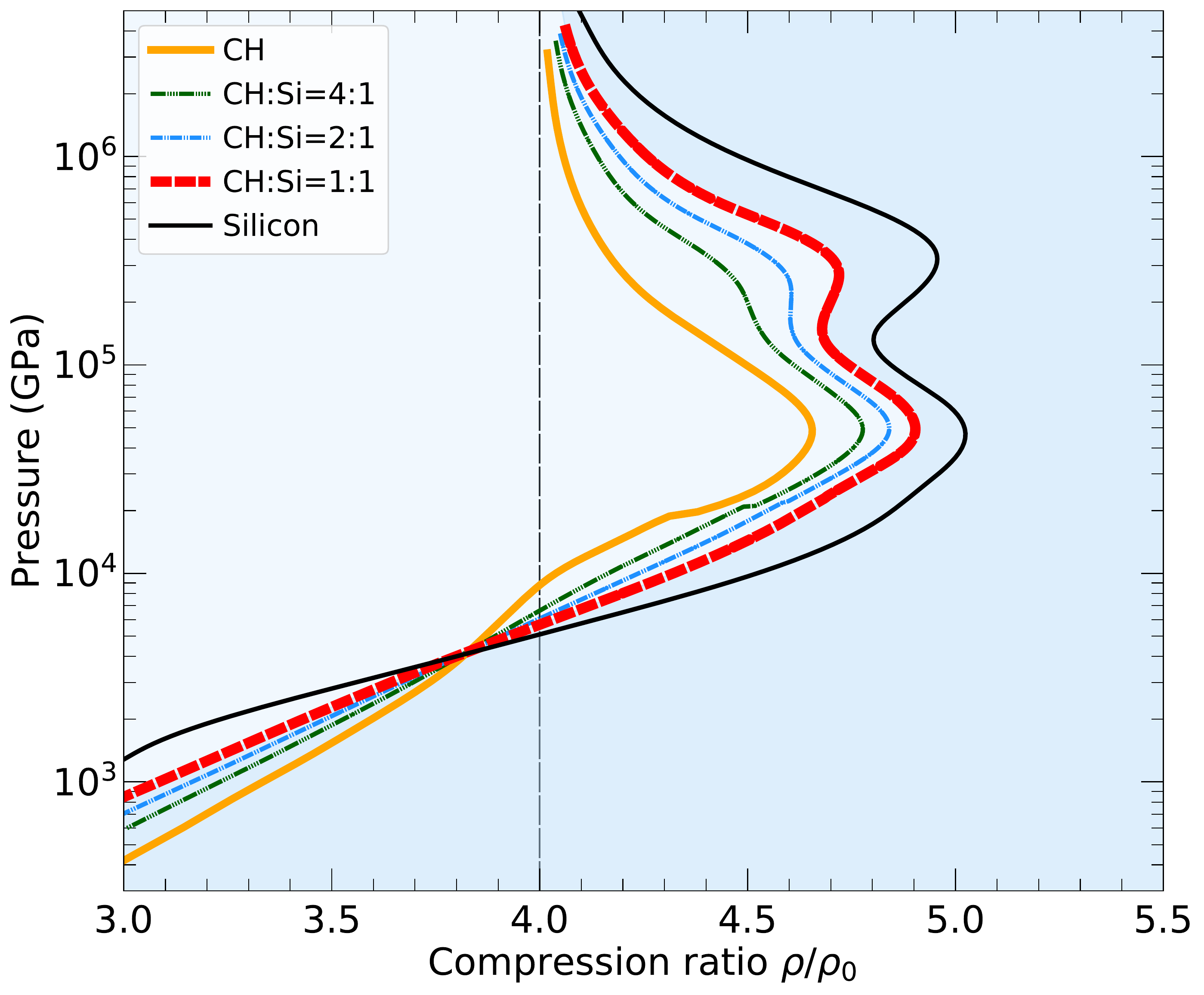}
    \caption{Shock Hugoniot curves of different mixtures of plastic (CH) and silicon. The silicon Hugoniot curve exhibits two compression maxima corresponding to the ionization of K and L shell electrons. Conversely, the CH curve shows only one maximum because the states of L shell electrons in carbon have merged with the conduction bands~\cite{ZhangCH2017}. We predict that two-maxima signature is preserved as long as the CH:Si concentration does not exceed a 2:1 ratio. With increasing CH concentration, the peak compression ratio decreases, as is expected from two endmember curves. } 
\label{fig:hugCH-Si}
\end{figure}

In Fig.~\ref{fig:hugCH-Si}, we compare the resulting shock Hugoniot curve of various mixtures of silicon and CH plastic. Plastics are typical coating materials of ICF capsules that may be doped with heavier elements to modify their behavior to absorb radiation. Elemental silicon was predicted to have two compression maxima that correspond to the conditions of K and L shell ionization~\cite{MilitzerDriver2015}. However, carbon and C-H mixtures were shown to have only one compression maximum~\cite{ZhangCH2017,ZhangCH2018,Doeppner2018,Kritcher2020} because the states of carbon's L shell electrons merged with the conduction band, which implies their excitations occur gradually and do not lead to a separate compression maximum. 

In Fig.~\ref{fig:hugCH-Si}, we show that the two-maxima signature of silicon is preserved as long as the CH:Si mixing ratio does not exceed 2:1. For higher ratios, silicon's K shell maximum disappears. With increasing the CH:Si mixing ratio, the shock compression overall decreases. The L shell maximum of silicon gradually transitions into the K shell maximum of carbon in CH without a significant change in pressure. 

\begin{figure}
    \centering
    \includegraphics[width=1.0\columnwidth]{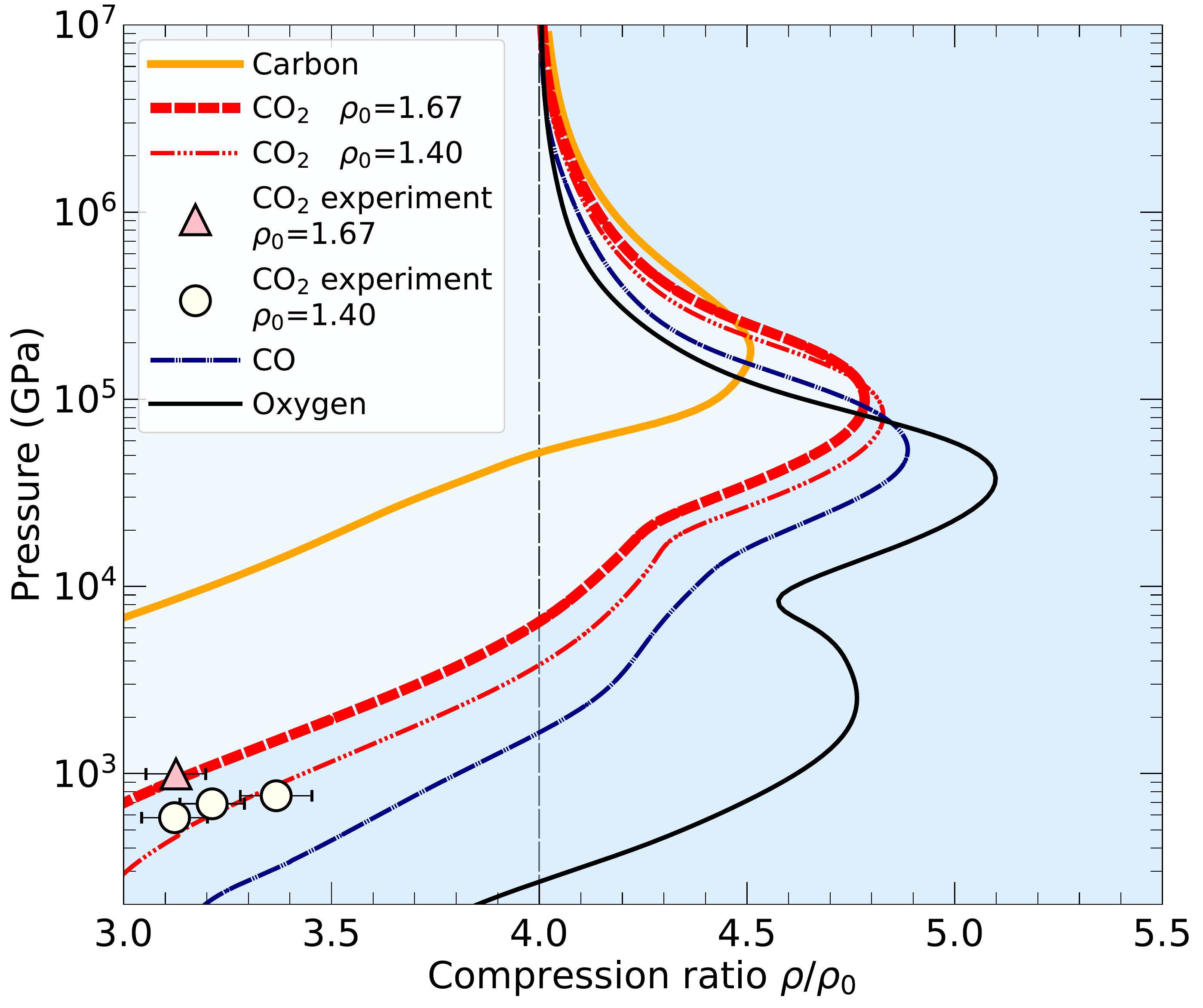}
    \caption{Shock Hugoniot curves of different mixtures of carbon and oxygen. Like CH in Fig.~\ref{fig:hugCH-Si}, the carbon Hugoniot curve exhibits only one compression maximum while oxygen, like silicon, exhibits two separate maxima from its L and K shell electrons. Neither CO nor CO$_2$ show two maxima, which means even a low-carbon concentration such as C:O=1:2 is sufficient to suppress the L shell compression maximum that one sees for elemental oxygen. 
     Our simulation results are in very good agreement with the experimental data from Ref.~\cite{Crandall2020}.
    } 
\label{fig:hugC-O}
\end{figure}

In Fig.~\ref{fig:hugC-O}, we show Hugoniot curves that we predict for different carbon-oxygen mixtures that make up the interiors of White Dwarf stars. As initial conditions for CO, we used its $\alpha$ structure with P2$_1$3 symmetry with $\rho_0$=1.0426 \gcc~to derive $E_0$ = $-$112.9115 Ha/CO. 

For CO$_2$, we used the Pa$\bar 3$ structure at $\rho_0$=1.40 and 1.67 \gcc~to respectively obtain $E_0$ = $-$188.1588 and $-$188.1574 Ha/CO$_2$.
As expected, the resulting Hugoniot curve falls in between those of elemental carbon and oxygen. While the oxygen curve shows two compression maxima, already a carbon content of C:O=1:2 appears to be  sufficient to eliminate the lower L shell maximum. We find excellent agreement with the revent shock wave experiments on CO$_2$ by Crandall et al.~\cite{Crandall2020}. All measurements agree with our predictions within error bars.

\begin{figure}
    \centering
    \includegraphics[width=1.0\columnwidth]{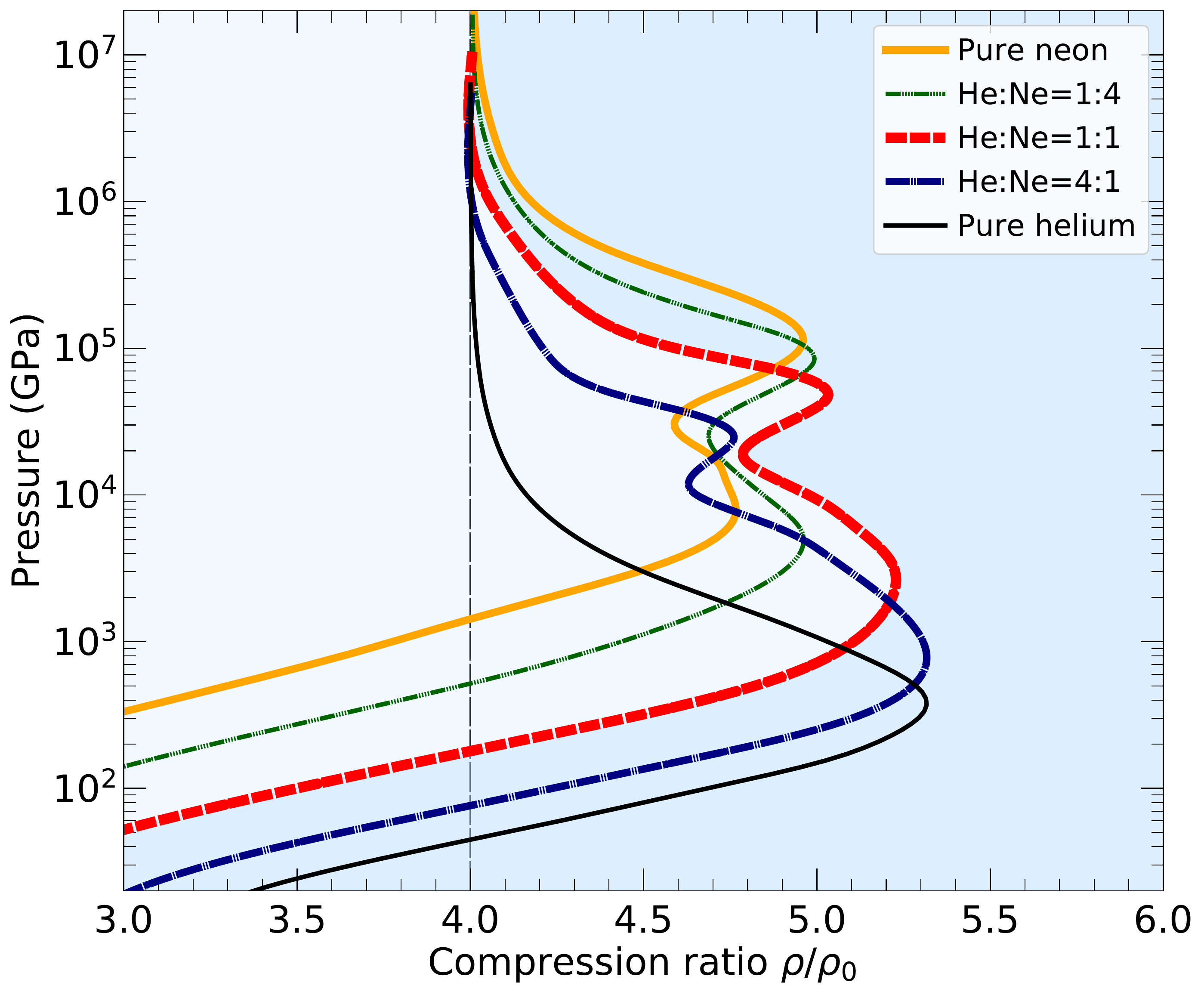}
    \caption{Shock Hugoniot curves of various mixtures of helium-neon mixtures. The helium curve only exhibits one very pronounced compression maximum because helium does not have an L shell. The neon curve shows two maxima because of the L and K shell ionization as like silicon in Fig.~\ref{fig:hugCH-Si} and oxygen in Fig.~\ref{fig:hugC-O}. With increasing neon concentration, the compression maxima shift to higher pressure (and temperature) because it takes more energy to ionize electrons from their respective shells. We find that already a low neon concentration of He:Ne=4:1 is sufficient for the Hugoniot curve to show two compression maxima.} 
\label{fig:hugHe-Ne}
\end{figure}

Neon is the most difficult material to transform into a metal, followed by helium~\cite{KM08}. In Fig.~\ref{fig:hugHe-Ne}, we study the shock properties of mixtures of the two inert gases. Neon exhibits two compression maxima while helium shows one at much lower pressure. Consequently, shock compression maxima shift down in pressure with increasing helium contents. A small neon contents as low as He:Ne=4:1 appears to be sufficient to cause two compression maxima.

\begin{figure}
    \centering
    \includegraphics[width=1.0\columnwidth]{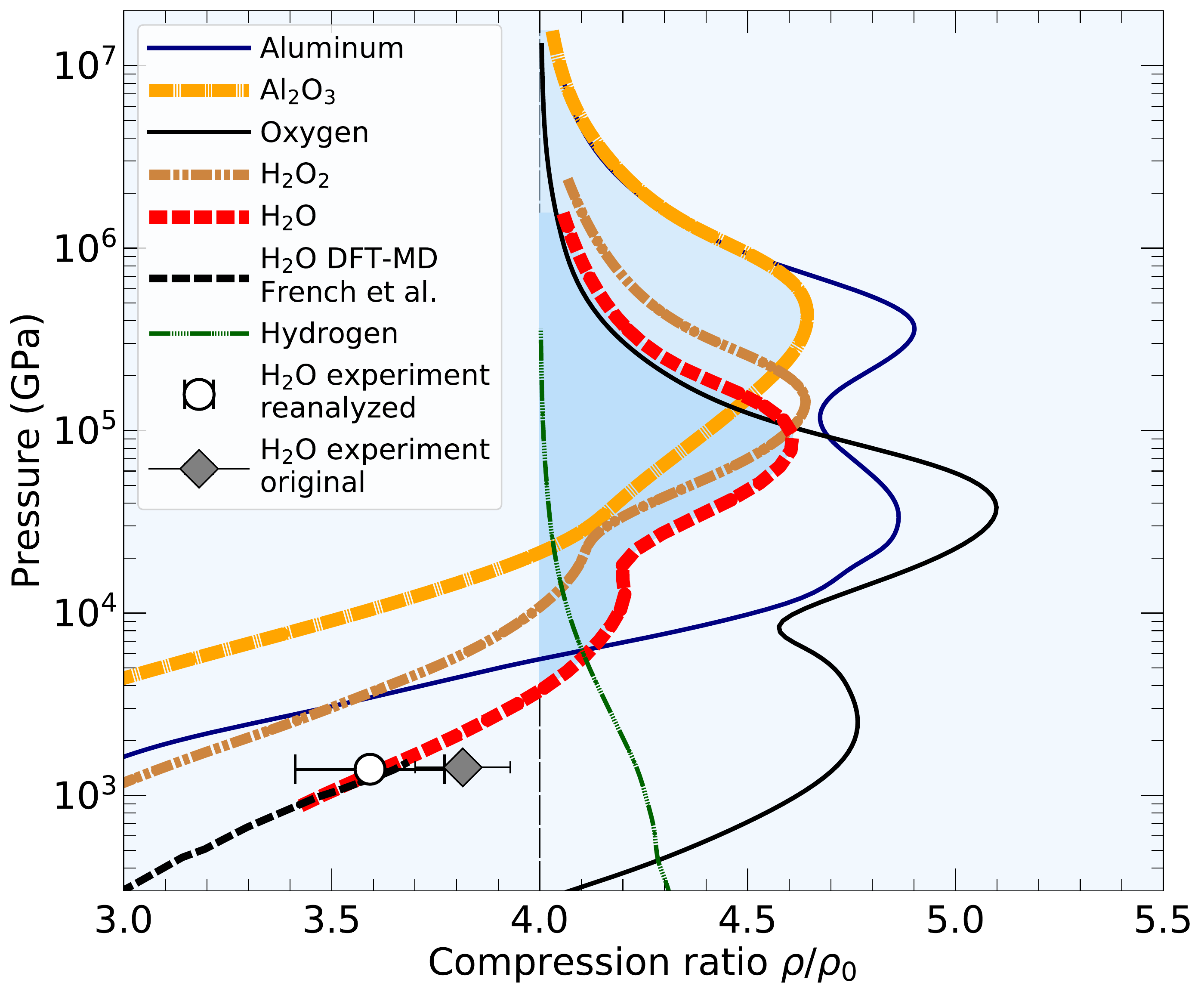}
    \caption{Shock Hugoniot curves of Al$_2$O$_3$, H$_2$O, and H$_2$O$_2$ are compared with those of the elemental constituents. For the lowest pressures, our H$_2$O curve converges to the predictions from French et al.~\cite{French2009}. Both theoretical predictions are in better agreement with the reanalyzed experimental results~\cite{Knudson12} than they are with the original measurements~\cite{Podurets1972}. H$_2$O maintains part of the two-peak signature of the oxygen Hugoniot curve, this influence is reduced in H$_2$O$_2$. Al$_2$O$_3$ shows only one compression maximum because the compression maxima of aluminum and oxygen are offset in pressure. }
\label{fig:Hug-mix}
\end{figure}

Finally, in Fig.~\ref{fig:Hug-mix} we compare the shock properties of water~\cite{French2009,Wilson2013,Mazevet2019}, hydrogen peroxide, and alumina with those of their elemental constituents. We predict water to exhibit two compression maxima that, despite a shift to higher pressures, are similar to the K and L shell ionization maxima of the oxygen. For compression ratios between 3.4 and 3.7, the Hugoniot curve that we derived with the linear mixing approximation is in very good agreement with fully interacting DFT-MD results of Ref. ~\cite{French2009}. This adds support to the prediction in Ref.~\cite{Militzer2020} that the linear mixing approximation works very well for compression ratios of 3.2 and larger. Both theoretical Hugoniot curves are in agreement with the reanalysis in Ref.~\cite{Knudson12} that shifted the experimental data point obtained by Podurets et al.~\cite{Podurets1972} to slightly lower densities.

We predict the shock Hugoniot curve of H$_2$O$_2$ to exhibit only a single compression maximum at $T$ = 3.784$\times 10^6\,$K, $P$=141$\,$600 GPa and $\rho/\rho_0$=4.639. Despite having a higher atomic oxygen fraction than H$_2$O, the lower L shell ionization appears only as a shoulder in the Hugoniot curve of H$_2$O$_2$, which is a consequence of its higher initial density, 1.713 \gcc. A density increase reduces the compression maxima along the Hugoniot curve because particles interact more strongly, which increases the pressure and thus reduces the compression ratio (see Figs.~\ref{fig:Hug_helium} and \ref{fig:hugC-O} as well as Ref.~\cite{Militzer2009}). If the initial density would be lowered to 1.35 \gcc~or less, the ionization of L shell electrons would again lead to a separate compression maximum. We set $E_0= -151.48932\,$Ha per formula unit (FU) in all Hugoniot calculations of H$_2$O$_2$.

For the computation of shock Hugoniot curve of alumina (Al$_2$O$_3$) were assume a corundum crystal structure and used $E_0=-708.807$ Ha/FU and $\rho_0=$3.9929 \gcc~for the initial conditions. The resulting Hugoniot curve only exhibits a single maximum, which is a surprise because oxygen and aluminum both show separate K and L shell maxima. However, these maxima are offset in pressure from one another and since both nuclei are present in this compound, their combined effects remove the L shell maximum. Furthermore the initial density of alumina is rather high, which reduces the magnitude of any compression maximum. 

\section{Results from Database Applications}

\begin{figure}[htb]
    \centering
    \includegraphics[width=1.0\columnwidth]{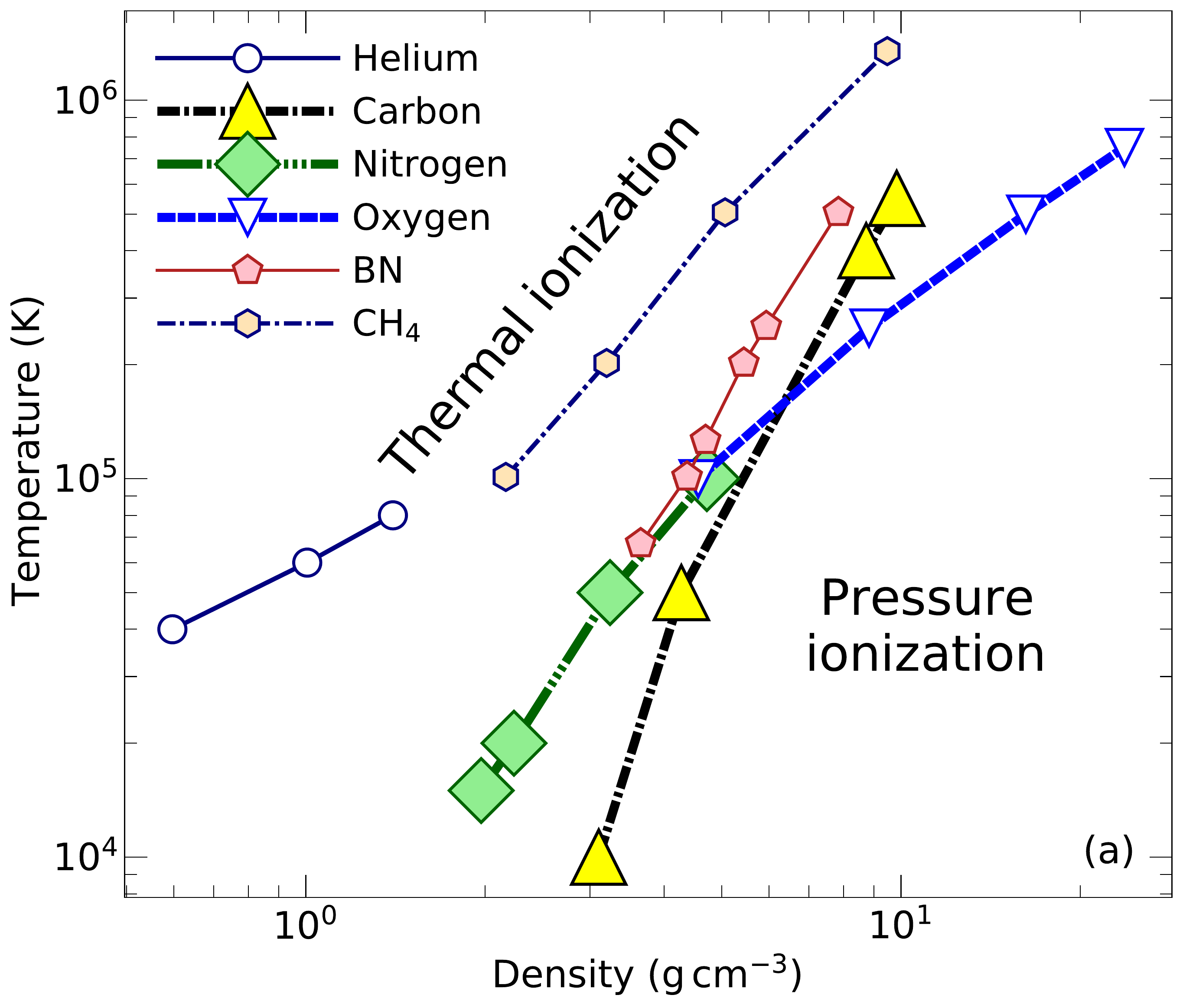}
    \includegraphics[width=1.0\columnwidth]{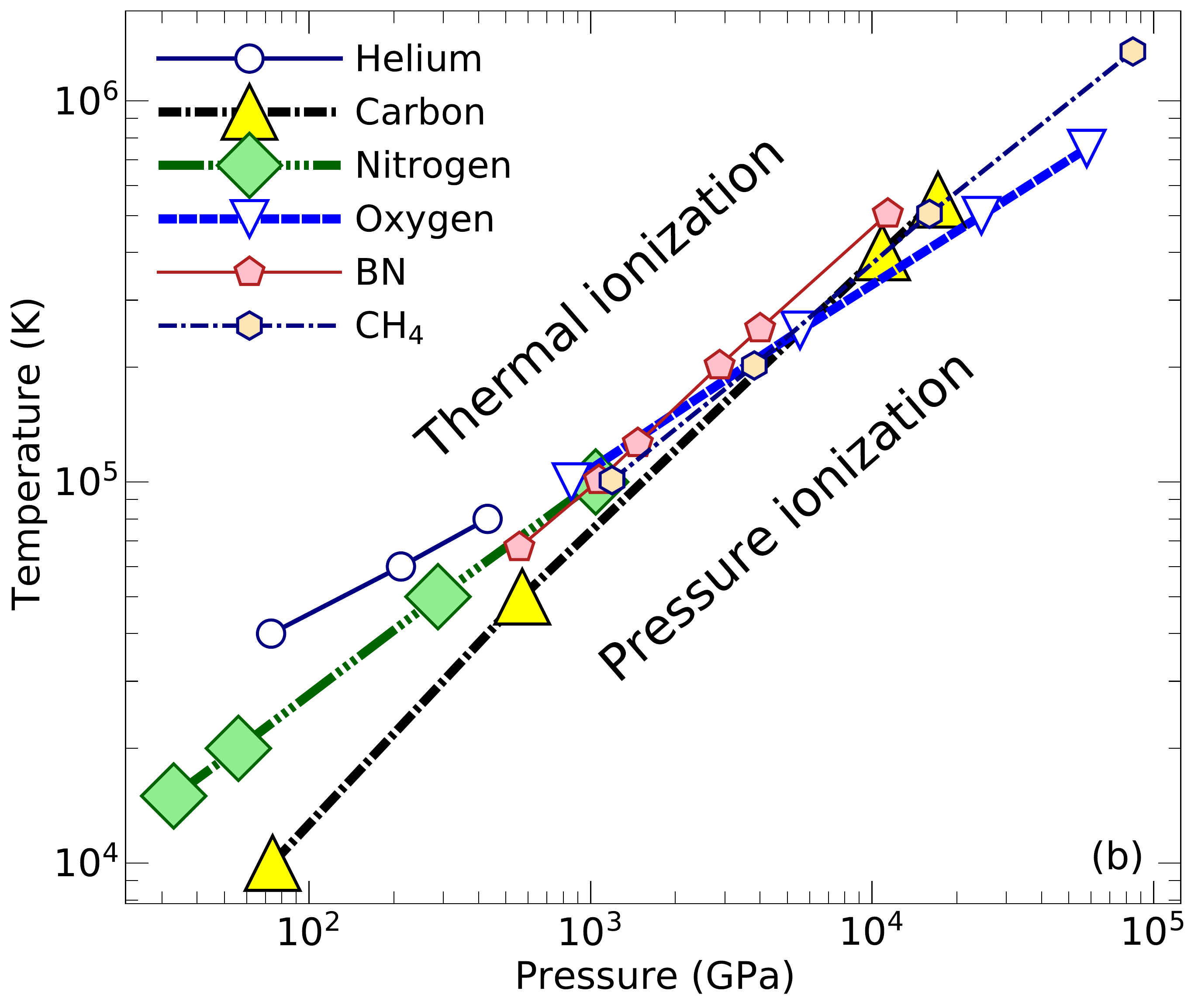}
    \caption{The dividing lines between the regimes of thermal and pressure ionization are shown in temperature-density and temperature-pressure spaces.}
\label{fig:energy_minima}
\end{figure}

In Ref.~\cite{GonzalezMilitzer2020}, it was shown that the regimes of pressure and thermal ionization can be distinguished from the slope, $\frac{\partial E}{\partial \rho}|_T$. At low density and high temperature, this slope is negative because with decreasing density, more and more free-particle states become available, more electrons become ionized, and as a result, the internal energy increases. This is called the thermal ionization regime, which is often described by the Saha ionization equilibrium~\cite{Ebeling1976}. Conversely, at high density the slope $\frac{\partial E}{\partial \rho}|_T$ becomes positive for two reasons. First, there is the confinement effect, which increases the kinetic energy of the free electrons and, second, the orbitals of the bound electrons hybridize and may even be pushed into the continuum of free-particle states, which is commonly referred to as pressure ionization. In Fig.~\ref{fig:energy_minima}, we employ the condition, $\frac{\partial E}{\partial \rho}|_T=0$, to distinguish between these two ionization regimes for six materials selected from our database. As expected, one finds low-Z materials like helium and CH$_4$ to switch from thermal to pressure ionization at a lower density compared to BN, nitrogen, carbon, and oxygen. Still, if one plots these transition lines in temperature-pressure space, they move much closer together (see second panel of Fig.~\ref{fig:energy_minima}). 
\begin{figure}
    \centering
    \includegraphics[width=1.0\columnwidth]{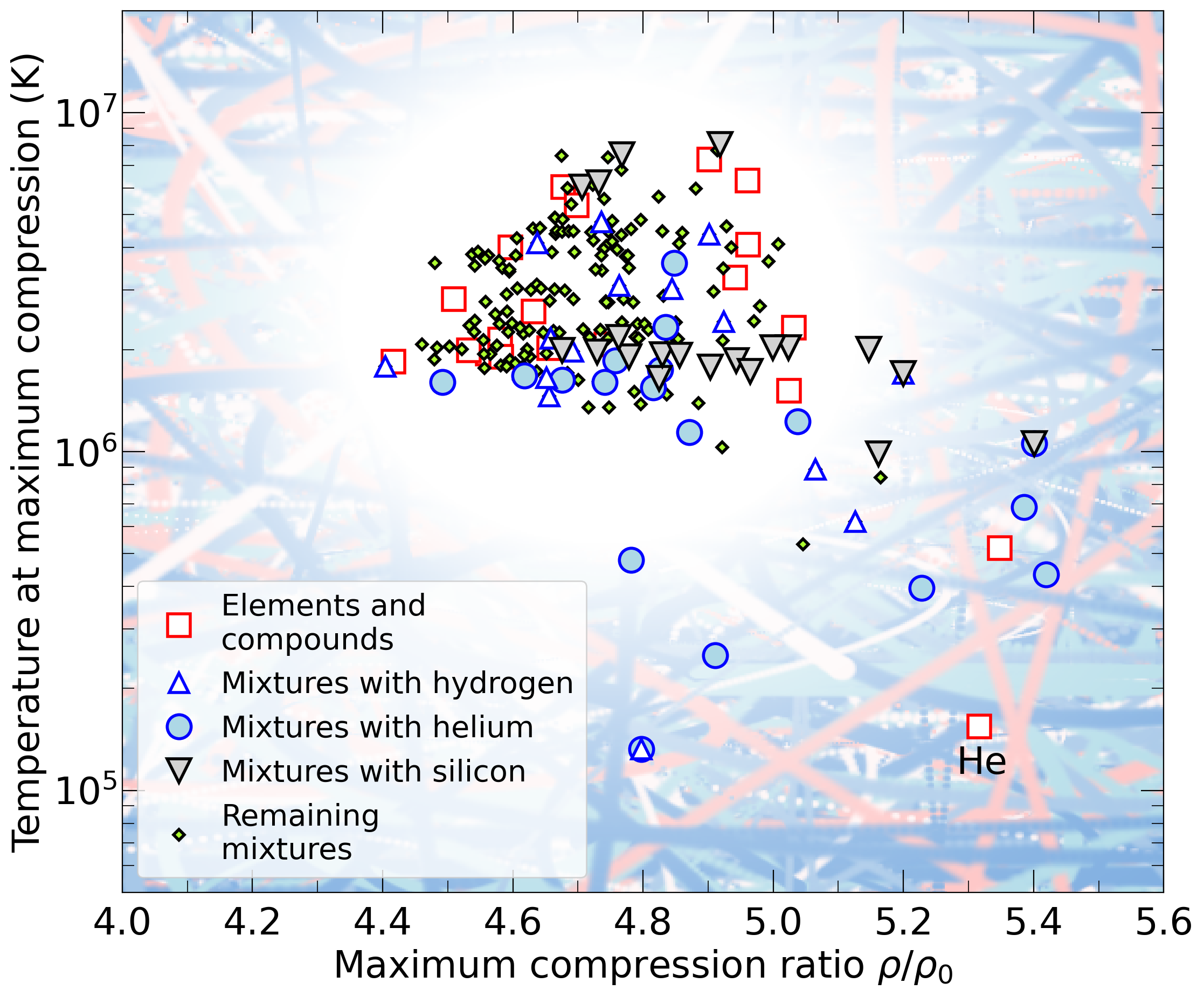}
    \caption{Conditions of the compression maximum on the shock Hugoniot curves of 21 elements and compounds as well as their mixtures. All materials show compression maxima larger than 4.3 that reflect the effects from electronic excitation. Low-Z materials like helium and helium-rich mixtures already reach their compression maxima for lower temperatures while silicon-rich mixtures require much higher temperatures to excite their K shell electrons. }
\label{fig:hug-mix-max}
\end{figure}

Furthermore, our FPEOS database enables us to efficiently compute the shock Hugoniot curves of all 21 compounds and 194 meaningful mixtures. In Fig.~\ref{fig:hug-mix-max}, we compare the conditions of shock compression maxima on all resulting Hugoniot curves for a 1:1 mixing ratio of formula units. These states may potentially be generated with laboratory experiments that either start with a chemical compound or by shocking a heterogeneous mixture of the two compounds in the mixture. Our calculations more accurately reflect that latter case because, unless noted otherwise, we derive $E_0$ and $V_0$ also from the linear mixing approximation for simplicity.

\begin{figure}
    \centering
    \includegraphics[width=1.0\columnwidth]{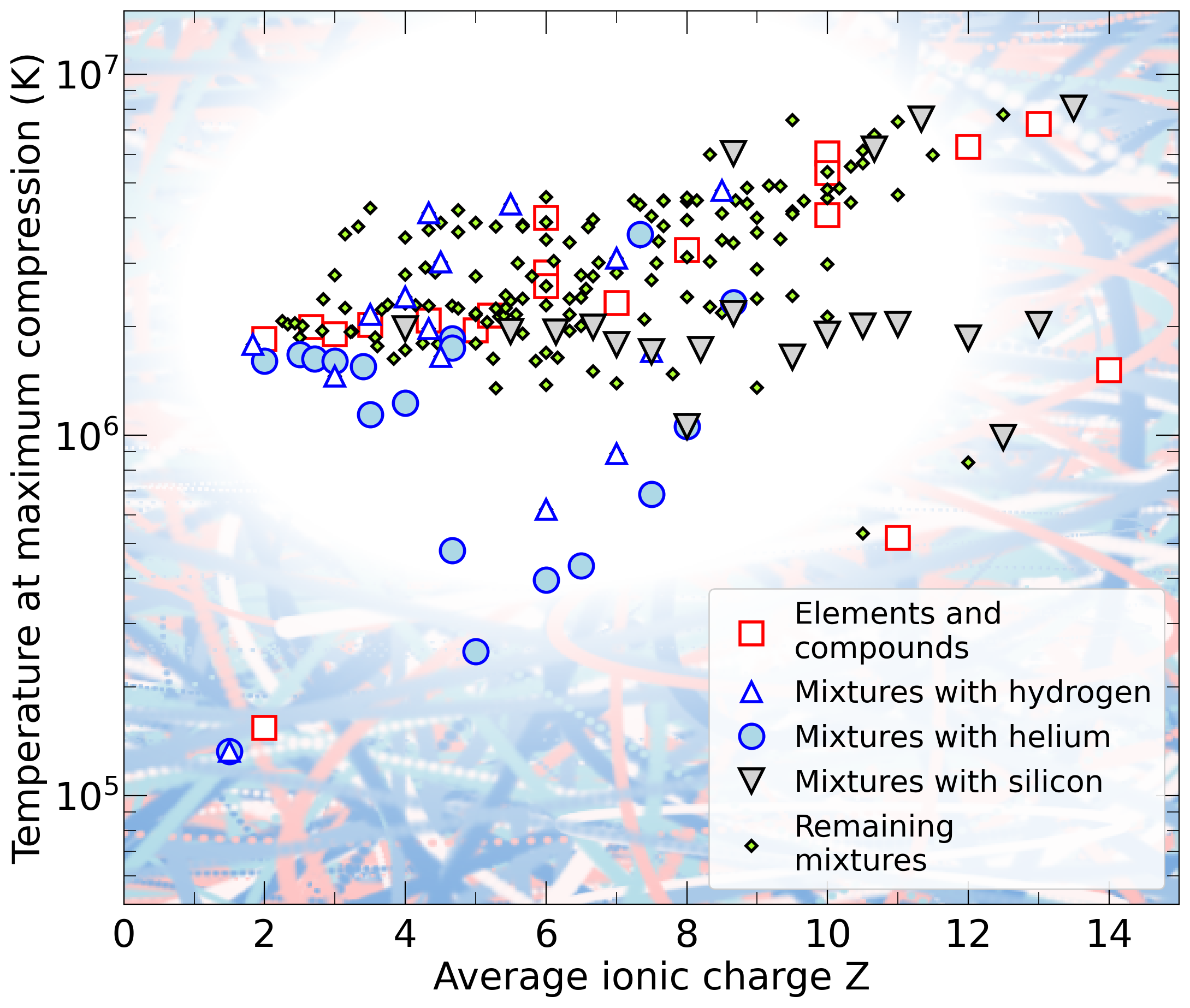}
    \caption{Temperature condition of the compression maximum on the shock Hugoniot curves of 21 elements and compounds as well as their mixtures is shown as a function of the average ionic charge $\left<Z\right>$ of these materials. While generally the compression maximum shifts towards higher temperature with increasing Z, the correlation is found to be not very strong.} 
\label{fig:Z-trend}
\end{figure}

We can identify a number of trends in Fig.~\ref{fig:hug-mix-max} but in general predicting the compression maxima of a specific mixture is not trivial~\cite{JDJohnson1997,Pain2007}. We find the mixtures of silicon exhibit a compression maximum at higher temperature, which is consistent with the ionization of the K shell electrons. Mixtures of helium tend to exhibit a compression maximum at lower temperature. However, mixtures with hydrogen do not follow this trend because it may be the other element in the mixture with hydrogen that is responsible for introducing the compression maximum. 

To study this trend, we study how strongly the temperature of shock compression maximum correlates with the average nuclei charge $\left< Z \right>$ of the mixture in Fig.~\ref{fig:Z-trend}. As expected, one finds some support for the trend of $T_\text{max}$ to increase with $\left< Z \right>$ but the correlation is not very strong. There are many mixtures with silicon that have $T_\text{max} \sim 2 \times 10^6\,$K but there are also several mixture with helium that have a similar $T_\text{max}$.

\begin{figure}
    \centering
    \includegraphics[width=1.0\columnwidth]{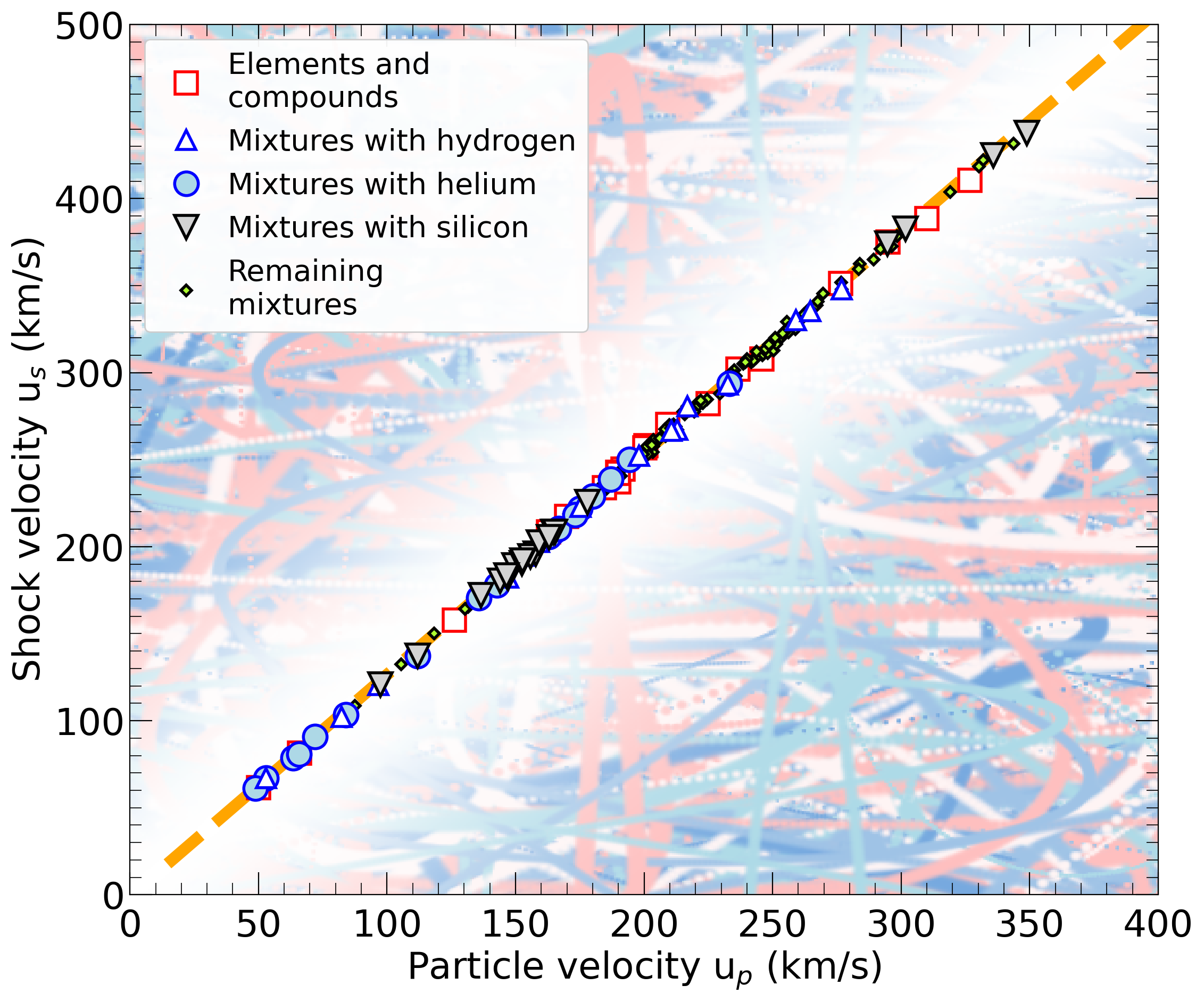}
    \caption{The particle and shock velocities are shown for the points of maximum compression along shock Hugoniot curves of 21 elements and compounds as well as their mixtures. With good accuracy, the dataset can be represented by the linear relationship of $u_s = 1.2727 \times u_p - 0.8588$ km/s (dashed orange line). The largest positive deviation is seen for a Si:Al mixture, in which case the fit underpredicts $u_s$ by 5.2 km/s, or 1.2\%. The largest negative deviation is found for a H:CH$_4$ mixture, in which case the fit overpredicts $u_s$ by 5.4 km/s, or 1.9\%. } 
\label{fig:up-us}
\end{figure}

\begin{figure}
    \centering
    \includegraphics[width=1.0\columnwidth]{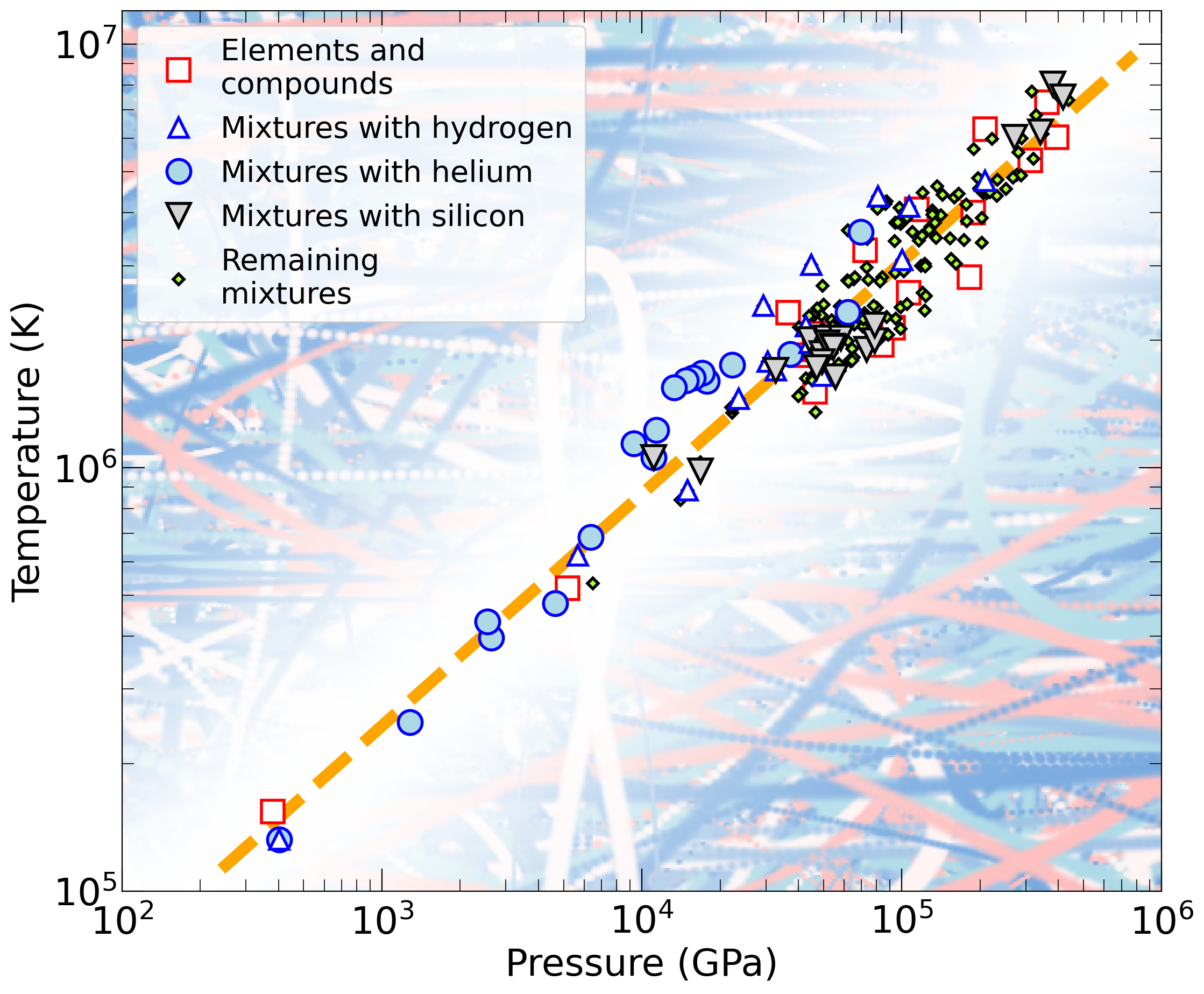}
    \caption{Pressure-temperature conditions for the points of maximum compression along shock Hugoniot curves of 21 elements and compounds as well as their mixtures. As expected, pressure and temperature are highly correlated, which can be represented by the fit log($T$/K) = 0.5744 $\times$ log($P$/GPa) + 3.7469. However, specific materials deviate substantially from this fit. The temperature of the maximum compression point of a H:Ne Hugoniot curve is 38\% lower than this fit would imply. Conversely, the temperature of a B:Al mixture is 49\% higher than predicted by the fit. } 
\label{fig:EOS_ranges}
\end{figure}

In Fig.~\ref{fig:up-us}, we converted the conditions of maximal shock compression into a $u_p$-$u_s$ plot. The shock and particle velocities were derived from $u_p=\sqrt{\xi \eta / m}$ and $u_s=\sqrt{\xi / (\eta m) }$ where $\xi = (P_1-P_0)V_0$, $\eta = 1-V_1/V_0$, and $m$ is the mass of one formula unit. We find that shock and particle velocities at maximum compression very closely follow the linear relationship, $u_s^{\rm max} = 1.2727 \times u_p^{\rm max} - 0.8588$ km/s, over a wide $u_p$ range from 50 to 350 km/s. The largest deviations from this trend are only +1.2\% and $-1.9$\%. This relationship, that we derived for the different compression maxima, shares similarities with the linear $u_p$-$u_s$ relationships that have been constructed for individual materials~\cite{ZhangSodium2017,Gonzalez2020} or groups of materials like fluid metals~\cite{Ozaki2016}. For a very high particle velocities of $\sim$400 km/s, the shock velocity has been found to rise faster than linear~\cite{ZhangSodium2017,Gonzalez2020} but the corresponding pressures and temperatures ($\sim10^6$ GPa and $\sim10^7$ K) cannot yet be reached in present-day planar shock experiments.

When we plot the temperature-pressure conditions of all computed Hugoniot maxima in log-log space in Fig.~\ref{fig:EOS_ranges}, we also find a linear trend but the correlation is weaker. Deviation can be as large as +49\% and $-$38\%.  

\begin{figure}
    \centering
    \includegraphics[width=1.0\columnwidth]{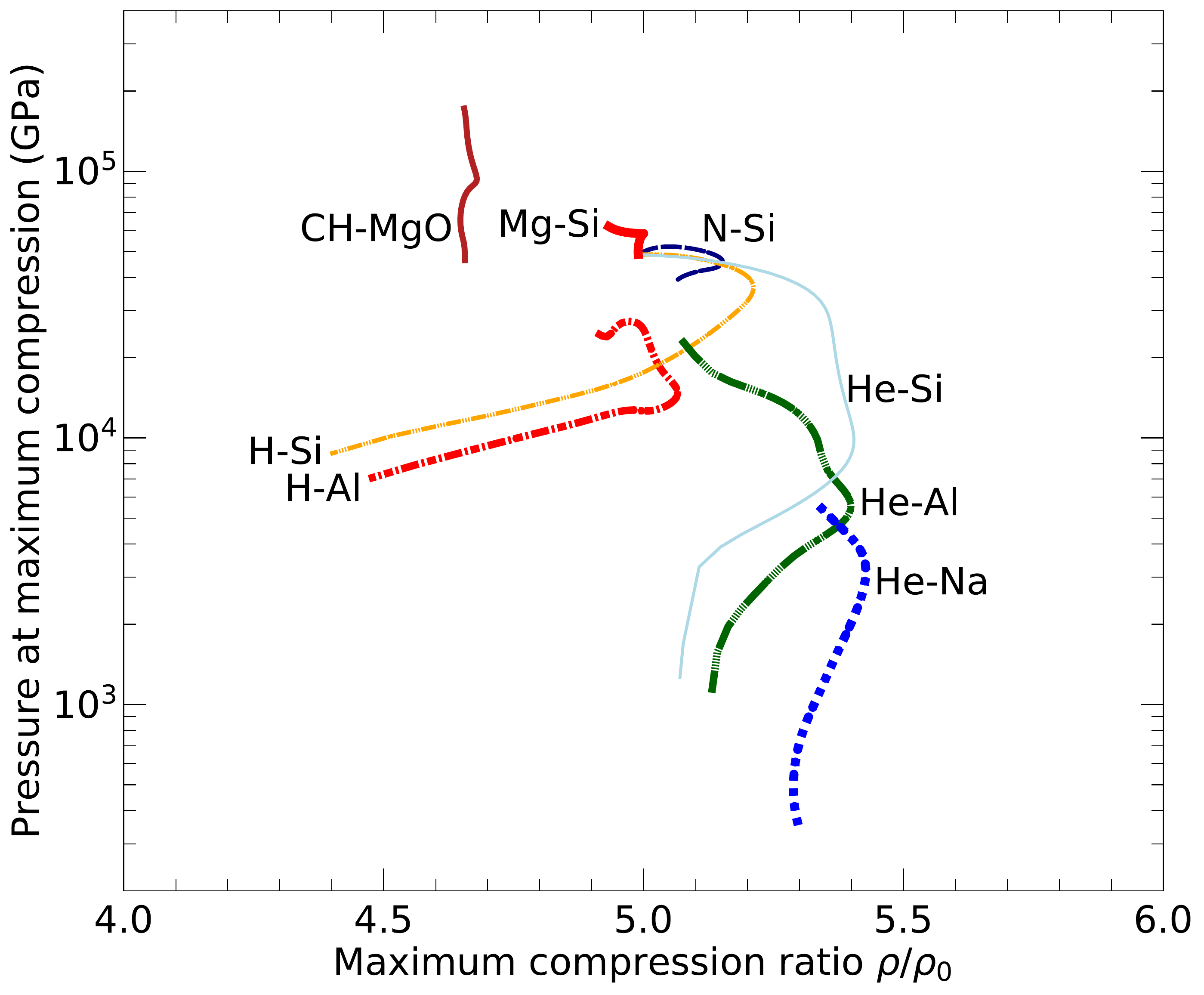}
    \caption{Conditions of the compression maximum on the shock Hugoniot curves of eight cases for which the mixtures exhibit a higher compression ratio than its endmembers. The lines emerge because a range of mixing ratios were considered.}
\label{fig:max-mix}
\end{figure}

In general, one expects the maximal shock compression ratio of a mixture to fall in between the maximal ratios of its two endmembers. However, there are exceptions because Eq.~\eqref{eq:hug} is nonlinear. So we combed through our database and found eight mixtures that exhibit higher shock compression ratios than their endmembers. In Fig.~\ref{fig:max-mix}, we plot how their maximal compression ratio and corresponding temperature vary as a function of mixing ratio. The strongest nonlinear behavior show mixtures of hydrogen, helium, and in one case nitrogen with heavier elements. The addition of a light element effectively lowers the initial density, which then increases the shock compression ratio as we have seen in Fig.~\ref{fig:Hug_helium}. This explanation, in principle, also applies to the two remaining, less intuitive cases: the Mg-Si mixture, where silicon has a low initial density, and the CH-MgO mixtures, where the introduction of CH leads to a reduction in density.

\section{Conclusions}

By assembling results from $\sim$5000 FP computer simulations of 21 elements and compounds, we have constructed a general-purpose FPEOS database for computation of matter at extreme conditions. It is our goal to make the calculations of shock Hugoniot curves and a ramp compression paths so efficient -- without compromising the precision of PIMC and DFT-MD methods -- that they become routine in the design and the analysis of WDM experiments. We thus provided our EOS tables as well the C++ and Python codes for the interpolation and the generation of various thermodynamic functions as supplementary material~\cite{SupplementalMaterial}.  

By invoking the linear mixing approximation at constant pressure and temperature, we first studied a selected number of binary mixtures, computed their shock Hugoniot curves, and related the resulting compression maxima to the ionization of L and K shell electrons. Then we applied our database to study the behavior of these maxima in 194 mixtures and identified trends in pressure, temperature, particle and shock velocity. Finally we identified eight unusual mixtures that should exhibit a higher shock compression ratio than their respective endmembers. 

\begin{acknowledgments}
  This work was in part supported by the National Science
  Foundation-Department of Energy (DOE) partnership for plasma science and engineering (grant DE-SC0016248) and by the DOE-National Nuclear Security Administration (grant DE-NA0003842). Computational support was provided by the Blue Waters computing project (NSF ACI 1640776) and the National Energy Research Scientific Computing Center (NERSC). 
  
  KPD and SZ acknowledge that this work was performed under the auspices of the U.S. Department of Energy by Lawrence Livermore National Laboratory under Contract DE-AC52-07NA27344, with the following disclaimer: This document was prepared as an account of work sponsored by an agency of the United States government. Neither the United States government nor Lawrence Livermore National Security, LLC, nor any of their employees makes any warranty, expressed or implied, or assumes any legal liability or responsibility for the accuracy, completeness, or usefulness of any information, apparatus, product, or process disclosed, or represents that its use would not infringe privately owned rights. Reference herein to any specific commercial product, process, or service by trade name, trademark, manufacturer, or otherwise does not necessarily constitute or imply its endorsement, recommendation, or favoring by the United States government or Lawrence Livermore National Security, LLC. The views and opinions of authors expressed herein do not necessarily state or reflect those of the United States government or Lawrence Livermore National Security, LLC, and shall not be used for advertising or product endorsement purposes. \end{acknowledgments}


\begin{thebibliography}{136}%
\makeatletter
\providecommand \@ifxundefined [1]{%
 \@ifx{#1\undefined}
}%
\providecommand \@ifnum [1]{%
 \ifnum #1\expandafter \@firstoftwo
 \else \expandafter \@secondoftwo
 \fi
}%
\providecommand \@ifx [1]{%
 \ifx #1\expandafter \@firstoftwo
 \else \expandafter \@secondoftwo
 \fi
}%
\providecommand \natexlab [1]{#1}%
\providecommand \enquote  [1]{``#1''}%
\providecommand \bibnamefont  [1]{#1}%
\providecommand \bibfnamefont [1]{#1}%
\providecommand \citenamefont [1]{#1}%
\providecommand \href@noop [0]{\@secondoftwo}%
\providecommand \href [0]{\begingroup \@sanitize@url \@href}%
\providecommand \@href[1]{\@@startlink{#1}\@@href}%
\providecommand \@@href[1]{\endgroup#1\@@endlink}%
\providecommand \@sanitize@url [0]{\catcode `\\12\catcode `\$12\catcode
  `\&12\catcode `\#12\catcode `\^12\catcode `\_12\catcode `\%12\relax}%
\providecommand \@@startlink[1]{}%
\providecommand \@@endlink[0]{}%
\providecommand \url  [0]{\begingroup\@sanitize@url \@url }%
\providecommand \@url [1]{\endgroup\@href {#1}{\urlprefix }}%
\providecommand \urlprefix  [0]{URL }%
\providecommand \Eprint [0]{\href }%
\providecommand \doibase [0]{https://doi.org/}%
\providecommand \selectlanguage [0]{\@gobble}%
\providecommand \bibinfo  [0]{\@secondoftwo}%
\providecommand \bibfield  [0]{\@secondoftwo}%
\providecommand \translation [1]{[#1]}%
\providecommand \BibitemOpen [0]{}%
\providecommand \bibitemStop [0]{}%
\providecommand \bibitemNoStop [0]{.\EOS\space}%
\providecommand \EOS [0]{\spacefactor3000\relax}%
\providecommand \BibitemShut  [1]{\csname bibitem#1\endcsname}%
\let\auto@bib@innerbib\@empty
\bibitem [{\citenamefont {Betti}(2009)}]{FESAC2009}%
  \BibitemOpen
  \bibinfo {editor} {\bibfnamefont {R.}~\bibnamefont {Betti}},\ ed.,\
  \href@noop {} {\emph {\bibinfo {title} {Advancing the Science of High Energy
  Density Laboratory Plasmas}}}\ (\bibinfo {organization} {Office of Fusion
  Energy Science (OFES)/Fusion Energy Science Advisory Committee (FESAC)},\
  \bibinfo {address} {Washington D.C.},\ \bibinfo {year} {2009})\BibitemShut
  {NoStop}%
\bibitem [{\citenamefont {Rosner}\ \emph {et~al.}(2010)\citenamefont {Rosner},
  \citenamefont {Hammer},\ and\ \citenamefont {Rothman}}]{HEDLP2009}%
  \BibitemOpen
  \bibinfo {editor} {\bibfnamefont {R.}~\bibnamefont {Rosner}}, \bibinfo
  {editor} {\bibfnamefont {D.}~\bibnamefont {Hammer}},\ and\ \bibinfo {editor}
  {\bibfnamefont {T.}~\bibnamefont {Rothman}},\ eds.,\ \href@noop {} {\emph
  {\bibinfo {title} {Basic Research Needs for high energy density laboratory
  physics}}}\ (\bibinfo {organization} {U.S. Department of Energy},\ \bibinfo
  {address} {Washington D.C.},\ \bibinfo {year} {2010})\BibitemShut {NoStop}%
\bibitem [{\citenamefont {Graziani}\ \emph
  {et~al.}(2014{\natexlab{a}})\citenamefont {Graziani}, \citenamefont
  {Desjarlais}, \citenamefont {Redmer},\ and\ \citenamefont {Trickey}}]{FCWDM}%
  \BibitemOpen
  \bibinfo {editor} {\bibfnamefont {F.}~\bibnamefont {Graziani}}, \bibinfo
  {editor} {\bibfnamefont {M.~P.}\ \bibnamefont {Desjarlais}}, \bibinfo
  {editor} {\bibfnamefont {R.}~\bibnamefont {Redmer}},\ and\ \bibinfo {editor}
  {\bibfnamefont {S.~B.}\ \bibnamefont {Trickey}},\ eds.,\ \href@noop {} {\emph
  {\bibinfo {title} {Frontiers and Challenges in Warm Dense Matter}}}\
  (\bibinfo  {publisher} {Springer International Publishing},\ \bibinfo {year}
  {2014})\BibitemShut {NoStop}%
\bibitem [{\citenamefont {Seidl}\ \emph {et~al.}(2009)\citenamefont {Seidl},
  \citenamefont {Anders}, \citenamefont {Bieniosek}, \citenamefont {Barnard},
  \citenamefont {Calanog}, \citenamefont {Chen}, \citenamefont {Cohen},
  \citenamefont {Coleman}, \citenamefont {Dorf}, \citenamefont {Gilson} \emph
  {et~al.}}]{Seidl2009}%
  \BibitemOpen
  \bibfield  {author} {\bibinfo {author} {\bibfnamefont {P.}~\bibnamefont
  {Seidl}}, \bibinfo {author} {\bibfnamefont {A.}~\bibnamefont {Anders}},
  \bibinfo {author} {\bibfnamefont {F.}~\bibnamefont {Bieniosek}}, \bibinfo
  {author} {\bibfnamefont {J.}~\bibnamefont {Barnard}}, \bibinfo {author}
  {\bibfnamefont {J.}~\bibnamefont {Calanog}}, \bibinfo {author} {\bibfnamefont
  {A.}~\bibnamefont {Chen}}, \bibinfo {author} {\bibfnamefont {R.}~\bibnamefont
  {Cohen}}, \bibinfo {author} {\bibfnamefont {J.}~\bibnamefont {Coleman}},
  \bibinfo {author} {\bibfnamefont {M.}~\bibnamefont {Dorf}}, \bibinfo {author}
  {\bibfnamefont {E.}~\bibnamefont {Gilson}}, \emph {et~al.},\ }\bibfield
  {title} {\bibinfo {title} {Progress in beam focusing and compression for
  warm-dense matter experiments},\ }\href@noop {} {\bibfield  {journal}
  {\bibinfo  {journal} {Nuclear Instruments and Methods in Physics Research
  Section A: Accelerators, Spectrometers, Detectors and Associated Equipment}\
  }\textbf {\bibinfo {volume} {606}},\ \bibinfo {pages} {75} (\bibinfo {year}
  {2009})}\BibitemShut {NoStop}%
\bibitem [{\citenamefont {Hammel}\ \emph {et~al.}(2010)\citenamefont {Hammel},
  \citenamefont {Haan}, \citenamefont {Clark}, \citenamefont {Edwards},
  \citenamefont {Langer}, \citenamefont {Marinak}, \citenamefont {Patel},
  \citenamefont {Salmonson},\ and\ \citenamefont {Scott}}]{Hammel2010}%
  \BibitemOpen
  \bibfield  {author} {\bibinfo {author} {\bibfnamefont {B.}~\bibnamefont
  {Hammel}}, \bibinfo {author} {\bibfnamefont {S.}~\bibnamefont {Haan}},
  \bibinfo {author} {\bibfnamefont {D.}~\bibnamefont {Clark}}, \bibinfo
  {author} {\bibfnamefont {M.}~\bibnamefont {Edwards}}, \bibinfo {author}
  {\bibfnamefont {S.}~\bibnamefont {Langer}}, \bibinfo {author} {\bibfnamefont
  {M.}~\bibnamefont {Marinak}}, \bibinfo {author} {\bibfnamefont
  {M.}~\bibnamefont {Patel}}, \bibinfo {author} {\bibfnamefont
  {J.}~\bibnamefont {Salmonson}},\ and\ \bibinfo {author} {\bibfnamefont
  {H.}~\bibnamefont {Scott}},\ }\bibfield  {title} {\bibinfo {title} {High-mode
  rayleigh-taylor growth in nif ignition capsules},\ }\href@noop {} {\bibfield
  {journal} {\bibinfo  {journal} {High Energy Density Physics}\ }\textbf
  {\bibinfo {volume} {6}},\ \bibinfo {pages} {171 } (\bibinfo {year} {2010})},\
  \bibinfo {note} {iCHED 2009 - 2nd International Conference on High Energy
  Density Physics}\BibitemShut {NoStop}%
\bibitem [{\citenamefont {Lindl}\ \emph {et~al.}(2014)\citenamefont {Lindl},
  \citenamefont {Landen}, \citenamefont {Edwards}, \citenamefont {Moses},\ and\
  \citenamefont {team}}]{Lindl2014}%
  \BibitemOpen
  \bibfield  {author} {\bibinfo {author} {\bibfnamefont {J.}~\bibnamefont
  {Lindl}}, \bibinfo {author} {\bibfnamefont {O.}~\bibnamefont {Landen}},
  \bibinfo {author} {\bibfnamefont {J.}~\bibnamefont {Edwards}}, \bibinfo
  {author} {\bibfnamefont {E.}~\bibnamefont {Moses}},\ and\ \bibinfo {author}
  {\bibfnamefont {N.}~\bibnamefont {team}},\ }\bibfield  {title} {\bibinfo
  {title} {Review of the national ignition campaign 2009-2012},\ }\href@noop {}
  {\bibfield  {journal} {\bibinfo  {journal} {Physics of Plasmas}\ }\textbf
  {\bibinfo {volume} {21}},\ \bibinfo {pages} {020501} (\bibinfo {year}
  {2014})}\BibitemShut {NoStop}%
\bibitem [{\citenamefont {Miyanishi}\ \emph {et~al.}(2015)\citenamefont
  {Miyanishi}, \citenamefont {Tange}, \citenamefont {Ozaki}, \citenamefont
  {Kimura}, \citenamefont {Sano}, \citenamefont {Sakawa}, \citenamefont
  {Tsuchiya},\ and\ \citenamefont {Kodama}}]{Miyanishi2015}%
  \BibitemOpen
  \bibfield  {author} {\bibinfo {author} {\bibfnamefont {K.}~\bibnamefont
  {Miyanishi}}, \bibinfo {author} {\bibfnamefont {Y.}~\bibnamefont {Tange}},
  \bibinfo {author} {\bibfnamefont {N.}~\bibnamefont {Ozaki}}, \bibinfo
  {author} {\bibfnamefont {T.}~\bibnamefont {Kimura}}, \bibinfo {author}
  {\bibfnamefont {T.}~\bibnamefont {Sano}}, \bibinfo {author} {\bibfnamefont
  {Y.}~\bibnamefont {Sakawa}}, \bibinfo {author} {\bibfnamefont
  {T.}~\bibnamefont {Tsuchiya}},\ and\ \bibinfo {author} {\bibfnamefont
  {R.}~\bibnamefont {Kodama}},\ }\bibfield  {title} {\bibinfo {title}
  {Laser-shock compression of magnesium oxide in the warm-dense-matter
  regime},\ }\href@noop {} {\bibfield  {journal} {\bibinfo  {journal} {Phys.
  Rev. E}\ }\textbf {\bibinfo {volume} {92}},\ \bibinfo {pages} {023103}
  (\bibinfo {year} {2015})}\BibitemShut {NoStop}%
\bibitem [{\citenamefont {Betti}\ and\ \citenamefont
  {Hurricane}(2016)}]{Betti2016}%
  \BibitemOpen
  \bibfield  {author} {\bibinfo {author} {\bibfnamefont {R.}~\bibnamefont
  {Betti}}\ and\ \bibinfo {author} {\bibfnamefont {O.}~\bibnamefont
  {Hurricane}},\ }\bibfield  {title} {\bibinfo {title} {Inertial-confinement
  fusion with lasers},\ }\href@noop {} {\bibfield  {journal} {\bibinfo
  {journal} {Nature Physics}\ }\textbf {\bibinfo {volume} {12}},\ \bibinfo
  {pages} {435} (\bibinfo {year} {2016})}\BibitemShut {NoStop}%
\bibitem [{\citenamefont {Gaffney}\ \emph {et~al.}(2018)\citenamefont
  {Gaffney}, \citenamefont {Hu}, \citenamefont {Arnault}, \citenamefont
  {Becker}, \citenamefont {Benedict}, \citenamefont {Boehly}, \citenamefont
  {Celliers}, \citenamefont {Ceperley}, \citenamefont {{\v{C}}ert{\'\i}k},
  \citenamefont {Cl{\'e}rouin} \emph {et~al.}}]{Gaffney2018}%
  \BibitemOpen
  \bibfield  {author} {\bibinfo {author} {\bibfnamefont {J.}~\bibnamefont
  {Gaffney}}, \bibinfo {author} {\bibfnamefont {S.}~\bibnamefont {Hu}},
  \bibinfo {author} {\bibfnamefont {P.}~\bibnamefont {Arnault}}, \bibinfo
  {author} {\bibfnamefont {A.}~\bibnamefont {Becker}}, \bibinfo {author}
  {\bibfnamefont {L.}~\bibnamefont {Benedict}}, \bibinfo {author}
  {\bibfnamefont {T.}~\bibnamefont {Boehly}}, \bibinfo {author} {\bibfnamefont
  {P.}~\bibnamefont {Celliers}}, \bibinfo {author} {\bibfnamefont
  {D.}~\bibnamefont {Ceperley}}, \bibinfo {author} {\bibfnamefont
  {O.}~\bibnamefont {{\v{C}}ert{\'\i}k}}, \bibinfo {author} {\bibfnamefont
  {J.}~\bibnamefont {Cl{\'e}rouin}}, \emph {et~al.},\ }\bibfield  {title}
  {\bibinfo {title} {A review of equation-of-state models for inertial
  confinement fusion materials},\ }\href@noop {} {\bibfield  {journal}
  {\bibinfo  {journal} {High Energy Density Physics}\ }\textbf {\bibinfo
  {volume} {28}},\ \bibinfo {pages} {7} (\bibinfo {year} {2018})}\BibitemShut
  {NoStop}%
\bibitem [{\citenamefont {Zeldovich}\ and\ \citenamefont
  {Raizer}(1968)}]{Ze66}%
  \BibitemOpen
  \bibfield  {author} {\bibinfo {author} {\bibfnamefont {Y.~B.}\ \bibnamefont
  {Zeldovich}}\ and\ \bibinfo {author} {\bibfnamefont {Y.~P.}\ \bibnamefont
  {Raizer}},\ }\href@noop {} {\emph {\bibinfo {title} {Elements of Gasdynamics
  and the Classical Theory of Shock Waves}}}\ (\bibinfo  {publisher} {Academic
  Press},\ \bibinfo {address} {New York},\ \bibinfo {year} {1968})\BibitemShut
  {NoStop}%
\bibitem [{\citenamefont {Remington}\ \emph {et~al.}(2006)\citenamefont
  {Remington}, \citenamefont {Drake},\ and\ \citenamefont
  {Ryutov}}]{Remington2006}%
  \BibitemOpen
  \bibfield  {author} {\bibinfo {author} {\bibfnamefont {B.~A.}\ \bibnamefont
  {Remington}}, \bibinfo {author} {\bibfnamefont {R.~P.}\ \bibnamefont
  {Drake}},\ and\ \bibinfo {author} {\bibfnamefont {D.~D.}\ \bibnamefont
  {Ryutov}},\ }\bibfield  {title} {\bibinfo {title} {Experimental astrophysics
  with high power lasers and $z$ pinches},\ }\href
  {https://doi.org/10.1103/RevModPhys.78.755} {\bibfield  {journal} {\bibinfo
  {journal} {Rev. Mod. Phys.}\ }\textbf {\bibinfo {volume} {78}},\ \bibinfo
  {pages} {755} (\bibinfo {year} {2006})}\BibitemShut {NoStop}%
\bibitem [{\citenamefont {Fortov}(2009)}]{Fortov2009}%
  \BibitemOpen
  \bibfield  {author} {\bibinfo {author} {\bibfnamefont {V.}~\bibnamefont
  {Fortov}},\ }\bibfield  {title} {\bibinfo {title} {Extreme states of matter
  on earth and in space},\ }\href@noop {} {\bibfield  {journal} {\bibinfo
  {journal} {Phys. Uspekhi}\ }\textbf {\bibinfo {volume} {52}},\ \bibinfo
  {pages} {615} (\bibinfo {year} {2009})}\BibitemShut {NoStop}%
\bibitem [{\citenamefont {Saumon}\ \emph {et~al.}(1995)\citenamefont {Saumon},
  \citenamefont {Chabrier},\ and\ \citenamefont {Horn}}]{SC95}%
  \BibitemOpen
  \bibfield  {author} {\bibinfo {author} {\bibfnamefont {D.}~\bibnamefont
  {Saumon}}, \bibinfo {author} {\bibfnamefont {G.}~\bibnamefont {Chabrier}},\
  and\ \bibinfo {author} {\bibfnamefont {H.~M.~V.}\ \bibnamefont {Horn}},\
  }\href@noop {} {\bibfield  {journal} {\bibinfo  {journal} {Astrophys. J.
  Suppl.}\ }\textbf {\bibinfo {volume} {99}},\ \bibinfo {pages} {713} (\bibinfo
  {year} {1995})}\BibitemShut {NoStop}%
\bibitem [{\citenamefont {Chabrier}\ and\ \citenamefont
  {Baraffe}(2000)}]{ChabrierBaraffe2000}%
  \BibitemOpen
  \bibfield  {author} {\bibinfo {author} {\bibfnamefont {G.}~\bibnamefont
  {Chabrier}}\ and\ \bibinfo {author} {\bibfnamefont {I.}~\bibnamefont
  {Baraffe}},\ }\bibfield  {title} {\bibinfo {title} {Theory of low-mass stars
  and substellar objects},\ }\href@noop {} {\bibfield  {journal} {\bibinfo
  {journal} {Annual Review of Astronomy and Astrophysics}\ }\textbf {\bibinfo
  {volume} {38}},\ \bibinfo {pages} {337} (\bibinfo {year} {2000})}\BibitemShut
  {NoStop}%
\bibitem [{\citenamefont {Ebeling}\ \emph {et~al.}(1976)\citenamefont
  {Ebeling}, \citenamefont {Kraeft},\ and\ \citenamefont
  {Kremp}}]{Ebeling1976}%
  \BibitemOpen
  \bibfield  {author} {\bibinfo {author} {\bibfnamefont {W.}~\bibnamefont
  {Ebeling}}, \bibinfo {author} {\bibfnamefont {W.}~\bibnamefont {Kraeft}},\
  and\ \bibinfo {author} {\bibfnamefont {D.}~\bibnamefont {Kremp}},\ }\bibfield
   {title} {\bibinfo {title} {Theory of bound states and ionization equilibrium
  in plasmas and solids},\ }in\ \href@noop {} {\emph {\bibinfo {booktitle}
  {Ergebnisse der Plasmaphysik und der Gaselektronik}}},\ \bibinfo {series and
  number} {Band 5}\ (\bibinfo  {publisher} {Akademie-Verlag},\ \bibinfo
  {address} {Berlin},\ \bibinfo {year} {1976})\BibitemShut {NoStop}%
\bibitem [{\citenamefont {Ebeling}\ and\ \citenamefont
  {Richert}(1985)}]{ER85a}%
  \BibitemOpen
  \bibfield  {author} {\bibinfo {author} {\bibfnamefont {W.}~\bibnamefont
  {Ebeling}}\ and\ \bibinfo {author} {\bibfnamefont {W.}~\bibnamefont
  {Richert}},\ }\href@noop {} {\bibfield  {journal} {\bibinfo  {journal} {Phys.
  Stat. Sol.}\ }\textbf {\bibinfo {volume} {{128}}},\ \bibinfo {pages} {467}
  (\bibinfo {year} {1985})}\BibitemShut {NoStop}%
\bibitem [{\citenamefont {Rogers}(1986)}]{Ro86}%
  \BibitemOpen
  \bibfield  {author} {\bibinfo {author} {\bibfnamefont {F.}~\bibnamefont
  {Rogers}},\ }\href@noop {} {\bibfield  {journal} {\bibinfo  {journal}
  {Astrophys. J.}\ }\textbf {\bibinfo {volume} {310}},\ \bibinfo {pages} {723}
  (\bibinfo {year} {1986})}\BibitemShut {NoStop}%
\bibitem [{\citenamefont {Rogers}(1990)}]{Ro90}%
  \BibitemOpen
  \bibfield  {author} {\bibinfo {author} {\bibfnamefont {F.}~\bibnamefont
  {Rogers}},\ }\bibfield  {title} {\bibinfo {title} {A distribution function
  approach for effective occupation numbers and the equation of state of
  hydrogen plasmas},\ }\href@noop {} {\bibfield  {journal} {\bibinfo  {journal}
  {Astrophys. J.}\ }\textbf {\bibinfo {volume} {352}},\ \bibinfo {pages} {689}
  (\bibinfo {year} {1990})}\BibitemShut {NoStop}%
\bibitem [{\citenamefont {Potekhin}\ and\ \citenamefont
  {Chabrier}(2000)}]{PotekhinChabrier2000}%
  \BibitemOpen
  \bibfield  {author} {\bibinfo {author} {\bibfnamefont {A.~Y.}\ \bibnamefont
  {Potekhin}}\ and\ \bibinfo {author} {\bibfnamefont {G.}~\bibnamefont
  {Chabrier}},\ }\href@noop {} {\bibfield  {journal} {\bibinfo  {journal}
  {Phys. Rev. E}\ }\textbf {\bibinfo {volume} {62}},\ \bibinfo {pages} {8554}
  (\bibinfo {year} {2000})}\BibitemShut {NoStop}%
\bibitem [{\citenamefont {Rozsnyai}\ \emph {et~al.}(2001)\citenamefont
  {Rozsnyai}, \citenamefont {Albritton}, \citenamefont {Young}, \citenamefont
  {Sonnad},\ and\ \citenamefont {Liberman}}]{Rozsnyai2001}%
  \BibitemOpen
  \bibfield  {author} {\bibinfo {author} {\bibfnamefont {B.~F.}\ \bibnamefont
  {Rozsnyai}}, \bibinfo {author} {\bibfnamefont {J.~R.}\ \bibnamefont
  {Albritton}}, \bibinfo {author} {\bibfnamefont {D.~A.}\ \bibnamefont
  {Young}}, \bibinfo {author} {\bibfnamefont {V.~N.}\ \bibnamefont {Sonnad}},\
  and\ \bibinfo {author} {\bibfnamefont {D.~A.}\ \bibnamefont {Liberman}},\
  }\bibfield  {title} {\bibinfo {title} {Theory and experiment for ultrahigh
  pressure shock hugoniots},\ }\href@noop {} {\bibfield  {journal} {\bibinfo
  {journal} {Physics Letters A}\ }\textbf {\bibinfo {volume} {291}},\ \bibinfo
  {pages} {226} (\bibinfo {year} {2001})}\BibitemShut {NoStop}%
\bibitem [{\citenamefont {Debye}\ and\ \citenamefont
  {H{\"u}ckel}(1923)}]{Debye1923}%
  \BibitemOpen
  \bibfield  {author} {\bibinfo {author} {\bibfnamefont {P.}~\bibnamefont
  {Debye}}\ and\ \bibinfo {author} {\bibfnamefont {E.}~\bibnamefont
  {H{\"u}ckel}},\ }\bibfield  {title} {\bibinfo {title} {Zur theorie der
  elektrolyte},\ }\href@noop {} {\bibfield  {journal} {\bibinfo  {journal}
  {Phys. Z}\ }\textbf {\bibinfo {volume} {24}},\ \bibinfo {pages} {185}
  (\bibinfo {year} {1923})}\BibitemShut {NoStop}%
\bibitem [{\citenamefont {Surh}\ \emph {et~al.}(2001)\citenamefont {Surh},
  \citenamefont {Barbee~III},\ and\ \citenamefont {Yang}}]{Surh01}%
  \BibitemOpen
  \bibfield  {author} {\bibinfo {author} {\bibfnamefont {M.~P.}\ \bibnamefont
  {Surh}}, \bibinfo {author} {\bibfnamefont {T.~W.}\ \bibnamefont
  {Barbee~III}},\ and\ \bibinfo {author} {\bibfnamefont {L.~H.}\ \bibnamefont
  {Yang}},\ }\bibfield  {title} {\bibinfo {title} {First principles molecular
  dynamics of dense plasmas},\ }\href@noop {} {\bibfield  {journal} {\bibinfo
  {journal} {Physical review letters}\ }\textbf {\bibinfo {volume} {86}},\
  \bibinfo {pages} {5958} (\bibinfo {year} {2001})}\BibitemShut {NoStop}%
\bibitem [{\citenamefont {Nellis}\ \emph {et~al.}(1991)\citenamefont {Nellis},
  \citenamefont {Radousky}, \citenamefont {Hamilton}, \citenamefont {Mitchell},
  \citenamefont {Holmes}, \citenamefont {Christianson},\ and\ \citenamefont
  {Van~Thiel}}]{Nellis1991}%
  \BibitemOpen
  \bibfield  {author} {\bibinfo {author} {\bibfnamefont {W.}~\bibnamefont
  {Nellis}}, \bibinfo {author} {\bibfnamefont {H.}~\bibnamefont {Radousky}},
  \bibinfo {author} {\bibfnamefont {D.}~\bibnamefont {Hamilton}}, \bibinfo
  {author} {\bibfnamefont {A.}~\bibnamefont {Mitchell}}, \bibinfo {author}
  {\bibfnamefont {N.}~\bibnamefont {Holmes}}, \bibinfo {author} {\bibfnamefont
  {K.}~\bibnamefont {Christianson}},\ and\ \bibinfo {author} {\bibfnamefont
  {M.}~\bibnamefont {Van~Thiel}},\ }\bibfield  {title} {\bibinfo {title}
  {Equation-of-state, shock-temperature, and electrical-conductivity data of
  dense fluid nitrogen in the region of the dissociative phase transition},\
  }\href@noop {} {\bibfield  {journal} {\bibinfo  {journal} {The Journal of
  chemical physics}\ }\textbf {\bibinfo {volume} {94}},\ \bibinfo {pages}
  {2244} (\bibinfo {year} {1991})}\BibitemShut {NoStop}%
\bibitem [{\citenamefont {Weir}\ \emph {et~al.}(1996)\citenamefont {Weir},
  \citenamefont {Mitchell},\ and\ \citenamefont {Nellis}}]{Weir1996}%
  \BibitemOpen
  \bibfield  {author} {\bibinfo {author} {\bibfnamefont {S.}~\bibnamefont
  {Weir}}, \bibinfo {author} {\bibfnamefont {A.}~\bibnamefont {Mitchell}},\
  and\ \bibinfo {author} {\bibfnamefont {W.}~\bibnamefont {Nellis}},\
  }\bibfield  {title} {\bibinfo {title} {{Metallization of fluid molecular
  hydrogen at 140 GPa (1.4 Mbar)}},\ }\href@noop {} {\bibfield  {journal}
  {\bibinfo  {journal} {Physical review letters}\ }\textbf {\bibinfo {volume}
  {76}},\ \bibinfo {pages} {1860} (\bibinfo {year} {1996})}\BibitemShut
  {NoStop}%
\bibitem [{\citenamefont {Knudson}\ \emph
  {et~al.}(2012{\natexlab{a}})\citenamefont {Knudson}, \citenamefont
  {Desjarlais}, \citenamefont {Lemke}, \citenamefont {Mattsson}, \citenamefont
  {French}, \citenamefont {Nettelmann},\ and\ \citenamefont
  {Redmer}}]{Knudson2012}%
  \BibitemOpen
  \bibfield  {author} {\bibinfo {author} {\bibfnamefont {M.~D.}\ \bibnamefont
  {Knudson}}, \bibinfo {author} {\bibfnamefont {M.~P.}\ \bibnamefont
  {Desjarlais}}, \bibinfo {author} {\bibfnamefont {R.}~\bibnamefont {Lemke}},
  \bibinfo {author} {\bibfnamefont {T.}~\bibnamefont {Mattsson}}, \bibinfo
  {author} {\bibfnamefont {M.}~\bibnamefont {French}}, \bibinfo {author}
  {\bibfnamefont {N.}~\bibnamefont {Nettelmann}},\ and\ \bibinfo {author}
  {\bibfnamefont {R.}~\bibnamefont {Redmer}},\ }\bibfield  {title} {\bibinfo
  {title} {{Probing the interiors of the ice giants: Shock compression of water
  to 700 GPa and 3.8 g/cm$^3$}},\ }\href@noop {} {\bibfield  {journal}
  {\bibinfo  {journal} {Physical Review Letters}\ }\textbf {\bibinfo {volume}
  {108}},\ \bibinfo {pages} {091102} (\bibinfo {year}
  {2012}{\natexlab{a}})}\BibitemShut {NoStop}%
\bibitem [{\citenamefont {Gomez}\ \emph {et~al.}(2020)\citenamefont {Gomez},
  \citenamefont {Slutz}, \citenamefont {Jennings}, \citenamefont {Ampleford},
  \citenamefont {Weis}, \citenamefont {Myers}, \citenamefont {Yager-Elorriaga},
  \citenamefont {Hahn}, \citenamefont {Hansen}, \citenamefont {Harding} \emph
  {et~al.}}]{Gomez2020}%
  \BibitemOpen
  \bibfield  {author} {\bibinfo {author} {\bibfnamefont {M.}~\bibnamefont
  {Gomez}}, \bibinfo {author} {\bibfnamefont {S.}~\bibnamefont {Slutz}},
  \bibinfo {author} {\bibfnamefont {C.}~\bibnamefont {Jennings}}, \bibinfo
  {author} {\bibfnamefont {D.}~\bibnamefont {Ampleford}}, \bibinfo {author}
  {\bibfnamefont {M.}~\bibnamefont {Weis}}, \bibinfo {author} {\bibfnamefont
  {C.}~\bibnamefont {Myers}}, \bibinfo {author} {\bibfnamefont
  {D.}~\bibnamefont {Yager-Elorriaga}}, \bibinfo {author} {\bibfnamefont
  {K.}~\bibnamefont {Hahn}}, \bibinfo {author} {\bibfnamefont {S.}~\bibnamefont
  {Hansen}}, \bibinfo {author} {\bibfnamefont {E.}~\bibnamefont {Harding}},
  \emph {et~al.},\ }\bibfield  {title} {\bibinfo {title} {Performance scaling
  in magnetized liner inertial fusion experiments},\ }\href@noop {} {\bibfield
  {journal} {\bibinfo  {journal} {Physical Review Letters}\ }\textbf {\bibinfo
  {volume} {125}},\ \bibinfo {pages} {155002} (\bibinfo {year}
  {2020})}\BibitemShut {NoStop}%
\bibitem [{\citenamefont {Graziani}\ \emph
  {et~al.}(2014{\natexlab{b}})\citenamefont {Graziani}, \citenamefont
  {Desjarlais}, \citenamefont {Redmer},\ and\ \citenamefont
  {Trickey}}]{Graziani2014Book}%
  \BibitemOpen
  \bibfield  {author} {\bibinfo {author} {\bibfnamefont {F.}~\bibnamefont
  {Graziani}}, \bibinfo {author} {\bibfnamefont {M.~P.}\ \bibnamefont
  {Desjarlais}}, \bibinfo {author} {\bibfnamefont {R.}~\bibnamefont {Redmer}},\
  and\ \bibinfo {author} {\bibfnamefont {S.~B.}\ \bibnamefont {Trickey}},\
  }\href@noop {} {\emph {\bibinfo {title} {Frontiers and challenges in warm
  dense matter}}},\ Vol.~\bibinfo {volume} {96}\ (\bibinfo  {publisher}
  {Springer Science \& Business},\ \bibinfo {year} {2014})\BibitemShut
  {NoStop}%
\bibitem [{\citenamefont {Hubbard}(1984)}]{hubbard_planets}%
  \BibitemOpen
  \bibfield  {author} {\bibinfo {author} {\bibfnamefont {W.~B.}\ \bibnamefont
  {Hubbard}},\ }\href@noop {} {\emph {\bibinfo {title} {Planetary Interiors}}}\
  (\bibinfo  {publisher} {University of Arizona Press},\ \bibinfo {address}
  {Tucson, AZ},\ \bibinfo {year} {1984})\BibitemShut {NoStop}%
\bibitem [{\citenamefont {Vorberger}\ \emph {et~al.}(2007)\citenamefont
  {Vorberger}, \citenamefont {Tamblyn}, \citenamefont {Militzer},\ and\
  \citenamefont {Bonev}}]{Vorberger2007}%
  \BibitemOpen
  \bibfield  {author} {\bibinfo {author} {\bibfnamefont {J.}~\bibnamefont
  {Vorberger}}, \bibinfo {author} {\bibfnamefont {I.}~\bibnamefont {Tamblyn}},
  \bibinfo {author} {\bibfnamefont {B.}~\bibnamefont {Militzer}},\ and\
  \bibinfo {author} {\bibfnamefont {S.~A.}\ \bibnamefont {Bonev}},\ }\bibfield
  {title} {\bibinfo {title} {Hydrogen-helium mixtures in the interiors of giant
  planets},\ }\href {https://doi.org/10.1103/PhysRevB.75.024206} {\bibfield
  {journal} {\bibinfo  {journal} {Phys. Rev. B}\ }\textbf {\bibinfo {volume}
  {75}},\ \bibinfo {pages} {024206} (\bibinfo {year} {2007})}\BibitemShut
  {NoStop}%
\bibitem [{\citenamefont {Militzer}\ and\ \citenamefont
  {Hubbard}(2013)}]{Militzer2013b}%
  \BibitemOpen
  \bibfield  {author} {\bibinfo {author} {\bibfnamefont {B.}~\bibnamefont
  {Militzer}}\ and\ \bibinfo {author} {\bibfnamefont {W.~B.}\ \bibnamefont
  {Hubbard}},\ }\bibfield  {title} {\bibinfo {title} {{Ab Initio Equation of
  State for Hydrogen-Helium Mixtures With Recalibration of the Giant-Planet
  Mass-Radius Relation}},\ }\href {https://doi.org/10.1088/0004-637X/774/2/148}
  {\bibfield  {journal} {\bibinfo  {journal} {Astrophys. J.}\ }\textbf
  {\bibinfo {volume} {774}},\ \bibinfo {pages} {148} (\bibinfo {year}
  {2013})}\BibitemShut {NoStop}%
\bibitem [{\citenamefont {Militzer}(2013)}]{Militzer2013c}%
  \BibitemOpen
  \bibfield  {author} {\bibinfo {author} {\bibfnamefont {B.}~\bibnamefont
  {Militzer}},\ }\bibfield  {title} {\bibinfo {title} {{Equation of state
  calculations of hydrogen-helium mixtures in solar and extrasolar giant
  planets}},\ }\href {https://doi.org/10.1103/PhysRevB.87.014202} {\bibfield
  {journal} {\bibinfo  {journal} {Physical Review B - Condensed Matter and
  Materials Physics}\ }\textbf {\bibinfo {volume} {87}},\ \bibinfo {pages}
  {014202} (\bibinfo {year} {2013})}\BibitemShut {NoStop}%
\bibitem [{\citenamefont {Wahl}\ \emph
  {et~al.}(2017{\natexlab{a}})\citenamefont {Wahl}, \citenamefont {Hubbard},
  \citenamefont {Militzer}, \citenamefont {Guillot}, \citenamefont {Miguel},
  \citenamefont {Movshovitz}, \citenamefont {Kaspi}, \citenamefont {Helled},
  \citenamefont {Reese}, \citenamefont {Galanti}, \citenamefont {Levin},
  \citenamefont {Connerney},\ and\ \citenamefont {Bolton}}]{Wahl2017}%
  \BibitemOpen
  \bibfield  {author} {\bibinfo {author} {\bibfnamefont {S.~M.}\ \bibnamefont
  {Wahl}}, \bibinfo {author} {\bibfnamefont {W.}~\bibnamefont {Hubbard}},
  \bibinfo {author} {\bibfnamefont {B.}~\bibnamefont {Militzer}}, \bibinfo
  {author} {\bibfnamefont {T.}~\bibnamefont {Guillot}}, \bibinfo {author}
  {\bibfnamefont {Y.}~\bibnamefont {Miguel}}, \bibinfo {author} {\bibfnamefont
  {N.}~\bibnamefont {Movshovitz}}, \bibinfo {author} {\bibfnamefont
  {Y.}~\bibnamefont {Kaspi}}, \bibinfo {author} {\bibfnamefont
  {R.}~\bibnamefont {Helled}}, \bibinfo {author} {\bibfnamefont
  {D.}~\bibnamefont {Reese}}, \bibinfo {author} {\bibfnamefont
  {E.}~\bibnamefont {Galanti}}, \bibinfo {author} {\bibfnamefont
  {S.}~\bibnamefont {Levin}}, \bibinfo {author} {\bibfnamefont {J.~E.}\
  \bibnamefont {Connerney}},\ and\ \bibinfo {author} {\bibfnamefont {S.~J.}\
  \bibnamefont {Bolton}},\ }\bibfield  {title} {\bibinfo {title} {{Comparing
  Jupiter interior structure models to Juno gravity measurements and the role
  of a dilute core}},\ }\href {https://doi.org/10.1002/2017GL073160} {\bibfield
   {journal} {\bibinfo  {journal} {Geophysical Research Letters}\ }\textbf
  {\bibinfo {volume} {44}},\ \bibinfo {pages} {4649} (\bibinfo {year}
  {2017}{\natexlab{a}})},\ \Eprint {https://arxiv.org/abs/1707.01997}
  {1707.01997} \BibitemShut {NoStop}%
\bibitem [{\citenamefont {Chabrier}\ \emph {et~al.}(2019)\citenamefont
  {Chabrier}, \citenamefont {Mazevet},\ and\ \citenamefont
  {Soubiran}}]{CMS_EOS}%
  \BibitemOpen
  \bibfield  {author} {\bibinfo {author} {\bibfnamefont {G.}~\bibnamefont
  {Chabrier}}, \bibinfo {author} {\bibfnamefont {S.}~\bibnamefont {Mazevet}},\
  and\ \bibinfo {author} {\bibfnamefont {F.}~\bibnamefont {Soubiran}},\
  }\bibfield  {title} {\bibinfo {title} {{A New Equation of State for Dense
  Hydrogen–Helium Mixtures}},\ }\href@noop {} {\bibfield  {journal} {\bibinfo
   {journal} {Astrophys. J.}\ }\textbf {\bibinfo {volume} {872}},\ \bibinfo
  {pages} {51} (\bibinfo {year} {2019})}\BibitemShut {NoStop}%
\bibitem [{\citenamefont {Militzer}\ \emph
  {et~al.}(2019{\natexlab{a}})\citenamefont {Militzer}, \citenamefont {Wahl},\
  and\ \citenamefont {Hubbard}}]{Militzer2019b}%
  \BibitemOpen
  \bibfield  {author} {\bibinfo {author} {\bibfnamefont {B.}~\bibnamefont
  {Militzer}}, \bibinfo {author} {\bibfnamefont {S.}~\bibnamefont {Wahl}},\
  and\ \bibinfo {author} {\bibfnamefont {W.}~\bibnamefont {Hubbard}},\
  }\bibfield  {title} {\bibinfo {title} {Models of saturn's interior
  constructed with an accelerated concentric maclaurin spheroid method},\
  }\href@noop {} {\bibfield  {journal} {\bibinfo  {journal} {The Astrophysical
  Journal}\ }\textbf {\bibinfo {volume} {879}},\ \bibinfo {pages} {78}
  (\bibinfo {year} {2019}{\natexlab{a}})}\BibitemShut {NoStop}%
\bibitem [{\citenamefont {Wilson}\ and\ \citenamefont
  {Militzer}(2012)}]{Wilson2012}%
  \BibitemOpen
  \bibfield  {author} {\bibinfo {author} {\bibfnamefont {H.~F.}\ \bibnamefont
  {Wilson}}\ and\ \bibinfo {author} {\bibfnamefont {B.}~\bibnamefont
  {Militzer}},\ }\bibfield  {title} {\bibinfo {title} {{Rocky Core Solubility
  in Jupiter and Giant Exoplanets}},\ }\href
  {https://doi.org/10.1103/PhysRevLett.108.111101} {\bibfield  {journal}
  {\bibinfo  {journal} {Physical Review Letters}\ }\textbf {\bibinfo {volume}
  {108}},\ \bibinfo {pages} {111101} (\bibinfo {year} {2012})}\BibitemShut
  {NoStop}%
\bibitem [{\citenamefont {Wahl}\ \emph {et~al.}(2013)\citenamefont {Wahl},
  \citenamefont {Wilson},\ and\ \citenamefont {Militzer}}]{Wahl2013a}%
  \BibitemOpen
  \bibfield  {author} {\bibinfo {author} {\bibfnamefont {S.~M.}\ \bibnamefont
  {Wahl}}, \bibinfo {author} {\bibfnamefont {H.~F.}\ \bibnamefont {Wilson}},\
  and\ \bibinfo {author} {\bibfnamefont {B.}~\bibnamefont {Militzer}},\
  }\bibfield  {title} {\bibinfo {title} {{Solubility of Iron in Metallic
  Hydrogen and Stability of Dense Cores in Giant Planets}},\ }\href
  {https://doi.org/10.1088/0004-637X/773/2/95} {\bibfield  {journal} {\bibinfo
  {journal} {The Astrophysical Journal}\ }\textbf {\bibinfo {volume} {773}},\
  \bibinfo {pages} {95} (\bibinfo {year} {2013})}\BibitemShut {NoStop}%
\bibitem [{\citenamefont {Becker}\ \emph {et~al.}(2014)\citenamefont {Becker},
  \citenamefont {Lorenzen}, \citenamefont {Fortney}, \citenamefont
  {Nettelmann}, \citenamefont {Schöttler},\ and\ \citenamefont
  {Redmer}}]{REOS3}%
  \BibitemOpen
  \bibfield  {author} {\bibinfo {author} {\bibfnamefont {A.}~\bibnamefont
  {Becker}}, \bibinfo {author} {\bibfnamefont {W.}~\bibnamefont {Lorenzen}},
  \bibinfo {author} {\bibfnamefont {J.}~\bibnamefont {Fortney}}, \bibinfo
  {author} {\bibfnamefont {N.}~\bibnamefont {Nettelmann}}, \bibinfo {author}
  {\bibfnamefont {M.}~\bibnamefont {Schöttler}},\ and\ \bibinfo {author}
  {\bibfnamefont {R.}~\bibnamefont {Redmer}},\ }\bibfield  {title} {\bibinfo
  {title} {{Ab initio equations of states for hydrogen (H-REOS.3) and helium
  (He-REOS.3) and their implications for the interiors of Brown Dwarfs}},\
  }\href@noop {} {\bibfield  {journal} {\bibinfo  {journal} {Astrophys. J.
  Suppl. S.}\ }\textbf {\bibinfo {volume} {215}},\ \bibinfo {pages} {21}
  (\bibinfo {year} {2014})}\BibitemShut {NoStop}%
\bibitem [{\citenamefont {Soubiran}\ and\ \citenamefont
  {Militzer}(2016)}]{Soubiran2016}%
  \BibitemOpen
  \bibfield  {author} {\bibinfo {author} {\bibfnamefont {F.}~\bibnamefont
  {Soubiran}}\ and\ \bibinfo {author} {\bibfnamefont {B.}~\bibnamefont
  {Militzer}},\ }\bibfield  {title} {\bibinfo {title} {{The properties of heavy
  elements in giant planet envelopes}},\ }\href
  {http://arxiv.org/abs/1606.04162} {\bibfield  {journal} {\bibinfo  {journal}
  {Astrophys. J., \textit{soumis}}\ } (\bibinfo {year} {2016})},\ \Eprint
  {https://arxiv.org/abs/1606.04162} {1606.04162} \BibitemShut {NoStop}%
\bibitem [{\citenamefont {Gonz{\'{a}}lez-Cataldo}\ \emph
  {et~al.}(2014)\citenamefont {Gonz{\'{a}}lez-Cataldo}, \citenamefont
  {Wilson},\ and\ \citenamefont {Militzer}}]{Gonzalez-Cataldo2014}%
  \BibitemOpen
  \bibfield  {author} {\bibinfo {author} {\bibfnamefont {F.}~\bibnamefont
  {Gonz{\'{a}}lez-Cataldo}}, \bibinfo {author} {\bibfnamefont {H.~F.}\
  \bibnamefont {Wilson}},\ and\ \bibinfo {author} {\bibfnamefont
  {B.}~\bibnamefont {Militzer}},\ }\bibfield  {title} {\bibinfo {title} {{Ab
  Initio Free Energy Calculations of the Solubility of Silica in Metallic
  Hydrogen and Application To Giant Planet Cores}},\ }\href
  {https://doi.org/10.1088/0004-637X/787/1/79} {\bibfield  {journal} {\bibinfo
  {journal} {The Astrophysical Journal}\ }\textbf {\bibinfo {volume} {787}},\
  \bibinfo {pages} {79} (\bibinfo {year} {2014})}\BibitemShut {NoStop}%
\bibitem [{\citenamefont {Gonz{\'{a}}lez-Cataldo}\ \emph
  {et~al.}(2016)\citenamefont {Gonz{\'{a}}lez-Cataldo}, \citenamefont {Davis},\
  and\ \citenamefont {Guti{\'{e}}rrez}}]{Gonzalez-Cataldo2016}%
  \BibitemOpen
  \bibfield  {author} {\bibinfo {author} {\bibfnamefont {F.}~\bibnamefont
  {Gonz{\'{a}}lez-Cataldo}}, \bibinfo {author} {\bibfnamefont {S.}~\bibnamefont
  {Davis}},\ and\ \bibinfo {author} {\bibfnamefont {G.}~\bibnamefont
  {Guti{\'{e}}rrez}},\ }\bibfield  {title} {\bibinfo {title} {{Melting curve of
  SiO$_2$ at multimegabar pressures: implications for gas giants and
  super-Earths}},\ }\href {http://dx.doi.org/10.1038/srep26537
  http://www.nature.com/articles/srep26537} {\bibfield  {journal} {\bibinfo
  {journal} {Scientific Reports}\ }\textbf {\bibinfo {volume} {6}},\ \bibinfo
  {pages} {26537} (\bibinfo {year} {2016})}\BibitemShut {NoStop}%
\bibitem [{\citenamefont {Soubiran}\ \emph {et~al.}(2017)\citenamefont
  {Soubiran}, \citenamefont {Militzer}, \citenamefont {Driver},\ and\
  \citenamefont {Zhang}}]{Soubiran2017}%
  \BibitemOpen
  \bibfield  {author} {\bibinfo {author} {\bibfnamefont {F.}~\bibnamefont
  {Soubiran}}, \bibinfo {author} {\bibfnamefont {B.}~\bibnamefont {Militzer}},
  \bibinfo {author} {\bibfnamefont {K.~P.}\ \bibnamefont {Driver}},\ and\
  \bibinfo {author} {\bibfnamefont {S.}~\bibnamefont {Zhang}},\ }\bibfield
  {title} {\bibinfo {title} {{Properties of hydrogen, helium, and silicon
  dioxide mixtures in giant planet interiors}},\ }\href
  {https://doi.org/10.1063/1.4978618} {\bibfield  {journal} {\bibinfo
  {journal} {Physics of Plasmas}\ }\textbf {\bibinfo {volume} {24}},\ \bibinfo
  {pages} {041401} (\bibinfo {year} {2017})}\BibitemShut {NoStop}%
\bibitem [{\citenamefont {Saumon}\ and\ \citenamefont
  {Guillot}(2004)}]{Saumon2004}%
  \BibitemOpen
  \bibfield  {author} {\bibinfo {author} {\bibfnamefont {D.}~\bibnamefont
  {Saumon}}\ and\ \bibinfo {author} {\bibfnamefont {T.}~\bibnamefont
  {Guillot}},\ }\bibfield  {title} {\bibinfo {title} {{Shock Compression of
  Deuterium and the Interiors of Jupiter and Saturn}},\ }\href
  {https://doi.org/10.1086/421257} {\bibfield  {journal} {\bibinfo  {journal}
  {The Astrophysical Journal}\ }\textbf {\bibinfo {volume} {609}},\ \bibinfo
  {pages} {1170} (\bibinfo {year} {2004})}\BibitemShut {NoStop}%
\bibitem [{\citenamefont {Baraffe}\ \emph {et~al.}(2014)\citenamefont
  {Baraffe}, \citenamefont {Chabrier}, \citenamefont {Fortney},\ and\
  \citenamefont {Sotin}}]{Baraffe2014}%
  \BibitemOpen
  \bibfield  {author} {\bibinfo {author} {\bibfnamefont {I.}~\bibnamefont
  {Baraffe}}, \bibinfo {author} {\bibfnamefont {G.}~\bibnamefont {Chabrier}},
  \bibinfo {author} {\bibfnamefont {J.}~\bibnamefont {Fortney}},\ and\ \bibinfo
  {author} {\bibfnamefont {C.}~\bibnamefont {Sotin}},\ }\bibfield  {title}
  {\bibinfo {title} {{Planetary Internal Structures}},\ }in\ \href
  {https://doi.org/10.2458/azu_uapress_9780816531240-ch033} {\emph {\bibinfo
  {booktitle} {Protostars and Planets VI}}}\ (\bibinfo  {publisher} {University
  of Arizona Press},\ \bibinfo {year} {2014})\ \Eprint
  {https://arxiv.org/abs/arXiv:1401.4738v1} {arXiv:1401.4738v1} \BibitemShut
  {NoStop}%
\bibitem [{\citenamefont {Militzer}\ \emph {et~al.}(2016)\citenamefont
  {Militzer}, \citenamefont {Soubiran}, \citenamefont {Wahl},\ and\
  \citenamefont {Hubbard}}]{Militzer2016b}%
  \BibitemOpen
  \bibfield  {author} {\bibinfo {author} {\bibfnamefont {B.}~\bibnamefont
  {Militzer}}, \bibinfo {author} {\bibfnamefont {F.}~\bibnamefont {Soubiran}},
  \bibinfo {author} {\bibfnamefont {S.~M.}\ \bibnamefont {Wahl}},\ and\
  \bibinfo {author} {\bibfnamefont {W.}~\bibnamefont {Hubbard}},\ }\bibfield
  {title} {\bibinfo {title} {{Understanding Jupiter's interior}},\ }\href
  {https://doi.org/10.1002/2016JE005080} {\bibfield  {journal} {\bibinfo
  {journal} {Journal of Geophysical Research: Planets}\ }\textbf {\bibinfo
  {volume} {121}},\ \bibinfo {pages} {1552} (\bibinfo {year} {2016})},\ \Eprint
  {https://arxiv.org/abs/1608.02685} {1608.02685} \BibitemShut {NoStop}%
\bibitem [{\citenamefont {Guillot}(1999)}]{Guillot1999}%
  \BibitemOpen
  \bibfield  {author} {\bibinfo {author} {\bibfnamefont {T.}~\bibnamefont
  {Guillot}},\ }\bibfield  {title} {\bibinfo {title} {{Interiors of Giant
  Planets Inside and Outside the Solar System}},\ }\href
  {https://doi.org/10.1126/science.286.5437.72} {\bibfield  {journal} {\bibinfo
   {journal} {Science}\ }\textbf {\bibinfo {volume} {286}},\ \bibinfo {pages}
  {72} (\bibinfo {year} {1999})}\BibitemShut {NoStop}%
\bibitem [{\citenamefont {{\it et al.}}(2011)}]{Borucki2011}%
  \BibitemOpen
  \bibfield  {author} {\bibinfo {author} {\bibfnamefont {W.~B.}\ \bibnamefont
  {{\it et al.}}},\ }\href@noop {} {\bibfield  {journal} {\bibinfo  {journal}
  {Astrophys. J}\ }\textbf {\bibinfo {volume} {736}},\ \bibinfo {pages} {19}
  (\bibinfo {year} {2011})}\BibitemShut {NoStop}%
\bibitem [{\citenamefont {Deming}\ and\ \citenamefont
  {Knutson}(2020)}]{Deming2020}%
  \BibitemOpen
  \bibfield  {author} {\bibinfo {author} {\bibfnamefont {D.}~\bibnamefont
  {Deming}}\ and\ \bibinfo {author} {\bibfnamefont {H.~A.}\ \bibnamefont
  {Knutson}},\ }\bibfield  {title} {\bibinfo {title} {Highlights of
  exoplanetary science from spitzer},\ }\href@noop {} {\bibfield  {journal}
  {\bibinfo  {journal} {Nature Astronomy}\ }\textbf {\bibinfo {volume} {4}},\
  \bibinfo {pages} {453} (\bibinfo {year} {2020})}\BibitemShut {NoStop}%
\bibitem [{\citenamefont {Madhusudhan}\ \emph {et~al.}(2011)\citenamefont
  {Madhusudhan}, \citenamefont {Harrington}, \citenamefont {Stevenson},
  \citenamefont {Nymeyer}, \citenamefont {Campo}, \citenamefont {Wheatley},
  \citenamefont {Deming}, \citenamefont {Blecic}, \citenamefont {Hardy},
  \citenamefont {Lust} \emph {et~al.}}]{Madhusudhan2011}%
  \BibitemOpen
  \bibfield  {author} {\bibinfo {author} {\bibfnamefont {N.}~\bibnamefont
  {Madhusudhan}}, \bibinfo {author} {\bibfnamefont {J.}~\bibnamefont
  {Harrington}}, \bibinfo {author} {\bibfnamefont {K.~B.}\ \bibnamefont
  {Stevenson}}, \bibinfo {author} {\bibfnamefont {S.}~\bibnamefont {Nymeyer}},
  \bibinfo {author} {\bibfnamefont {C.~J.}\ \bibnamefont {Campo}}, \bibinfo
  {author} {\bibfnamefont {P.~J.}\ \bibnamefont {Wheatley}}, \bibinfo {author}
  {\bibfnamefont {D.}~\bibnamefont {Deming}}, \bibinfo {author} {\bibfnamefont
  {J.}~\bibnamefont {Blecic}}, \bibinfo {author} {\bibfnamefont {R.~A.}\
  \bibnamefont {Hardy}}, \bibinfo {author} {\bibfnamefont {N.~B.}\ \bibnamefont
  {Lust}}, \emph {et~al.},\ }\bibfield  {title} {\bibinfo {title} {{A high C/O
  ratio and weak thermal inversion in the atmosphere of exoplanet WASP-12b}},\
  }\href@noop {} {\bibfield  {journal} {\bibinfo  {journal} {Nature}\ }\textbf
  {\bibinfo {volume} {469}},\ \bibinfo {pages} {64} (\bibinfo {year}
  {2011})}\BibitemShut {NoStop}%
\bibitem [{\citenamefont {Wagner}\ \emph {et~al.}(2012)\citenamefont {Wagner},
  \citenamefont {Tosi}, \citenamefont {Sohl}, \citenamefont {Rauer},\ and\
  \citenamefont {Spohn}}]{Wagner2012}%
  \BibitemOpen
  \bibfield  {author} {\bibinfo {author} {\bibfnamefont {F.~W.}\ \bibnamefont
  {Wagner}}, \bibinfo {author} {\bibfnamefont {N.}~\bibnamefont {Tosi}},
  \bibinfo {author} {\bibfnamefont {F.}~\bibnamefont {Sohl}}, \bibinfo {author}
  {\bibfnamefont {H.}~\bibnamefont {Rauer}},\ and\ \bibinfo {author}
  {\bibfnamefont {T.}~\bibnamefont {Spohn}},\ }\bibfield  {title} {\bibinfo
  {title} {{Rocky super-Earth interiors}},\ }\href@noop {} {\bibfield
  {journal} {\bibinfo  {journal} {Astronomy {\&} Astrophysics}\ }\textbf
  {\bibinfo {volume} {541}},\ \bibinfo {pages} {A103} (\bibinfo {year}
  {2012})}\BibitemShut {NoStop}%
\bibitem [{\citenamefont {Wilson}\ and\ \citenamefont
  {Militzer}(2014)}]{Wilson2014}%
  \BibitemOpen
  \bibfield  {author} {\bibinfo {author} {\bibfnamefont {H.~F.}\ \bibnamefont
  {Wilson}}\ and\ \bibinfo {author} {\bibfnamefont {B.}~\bibnamefont
  {Militzer}},\ }\bibfield  {title} {\bibinfo {title} {Interior phase
  transformations and mass-radius relationships of silicon-carbon planets},\
  }\href@noop {} {\bibfield  {journal} {\bibinfo  {journal} {The Astrophysical
  Journal}\ }\textbf {\bibinfo {volume} {793}},\ \bibinfo {pages} {34}
  (\bibinfo {year} {2014})}\BibitemShut {NoStop}%
\bibitem [{\citenamefont {Christensen-Dalsgaard}\ \emph
  {et~al.}(1996)\citenamefont {Christensen-Dalsgaard}, \citenamefont
  {D{\"a}ppen}, \citenamefont {Ajukov}, \citenamefont {Anderson}, \citenamefont
  {Antia}, \citenamefont {Basu}, \citenamefont {Baturin}, \citenamefont
  {Berthomieu}, \citenamefont {Chaboyer}, \citenamefont {Chitre} \emph
  {et~al.}}]{Christensen1996}%
  \BibitemOpen
  \bibfield  {author} {\bibinfo {author} {\bibfnamefont {J.}~\bibnamefont
  {Christensen-Dalsgaard}}, \bibinfo {author} {\bibfnamefont {W.}~\bibnamefont
  {D{\"a}ppen}}, \bibinfo {author} {\bibfnamefont {S.}~\bibnamefont {Ajukov}},
  \bibinfo {author} {\bibfnamefont {E.}~\bibnamefont {Anderson}}, \bibinfo
  {author} {\bibfnamefont {H.}~\bibnamefont {Antia}}, \bibinfo {author}
  {\bibfnamefont {S.}~\bibnamefont {Basu}}, \bibinfo {author} {\bibfnamefont
  {V.}~\bibnamefont {Baturin}}, \bibinfo {author} {\bibfnamefont
  {G.}~\bibnamefont {Berthomieu}}, \bibinfo {author} {\bibfnamefont
  {B.}~\bibnamefont {Chaboyer}}, \bibinfo {author} {\bibfnamefont
  {S.}~\bibnamefont {Chitre}}, \emph {et~al.},\ }\bibfield  {title} {\bibinfo
  {title} {The current state of solar modeling},\ }\href@noop {} {\bibfield
  {journal} {\bibinfo  {journal} {Science}\ }\textbf {\bibinfo {volume}
  {272}},\ \bibinfo {pages} {1286} (\bibinfo {year} {1996})}\BibitemShut
  {NoStop}%
\bibitem [{\citenamefont {Christensen-Dalsgaard}(2002)}]{Christensen2002}%
  \BibitemOpen
  \bibfield  {author} {\bibinfo {author} {\bibfnamefont {J.}~\bibnamefont
  {Christensen-Dalsgaard}},\ }\bibfield  {title} {\bibinfo {title}
  {Helioseismology},\ }\href@noop {} {\bibfield  {journal} {\bibinfo  {journal}
  {Reviews of Modern Physics}\ }\textbf {\bibinfo {volume} {74}},\ \bibinfo
  {pages} {1073} (\bibinfo {year} {2002})}\BibitemShut {NoStop}%
\bibitem [{\citenamefont {Schumacher}\ and\ \citenamefont
  {Sreenivasan}(2020)}]{Schumacher2020}%
  \BibitemOpen
  \bibfield  {author} {\bibinfo {author} {\bibfnamefont {J.}~\bibnamefont
  {Schumacher}}\ and\ \bibinfo {author} {\bibfnamefont {K.~R.}\ \bibnamefont
  {Sreenivasan}},\ }\bibfield  {title} {\bibinfo {title} {Colloquium: Unusual
  dynamics of convection in the sun},\ }\href@noop {} {\bibfield  {journal}
  {\bibinfo  {journal} {Reviews of Modern Physics}\ }\textbf {\bibinfo {volume}
  {92}},\ \bibinfo {pages} {041001} (\bibinfo {year} {2020})}\BibitemShut
  {NoStop}%
\bibitem [{\citenamefont {Aerts}(2019)}]{Aerts2019}%
  \BibitemOpen
  \bibfield  {author} {\bibinfo {author} {\bibfnamefont {C.}~\bibnamefont
  {Aerts}},\ }\bibfield  {title} {\bibinfo {title} {Probing the interior
  physics of stars through asteroseismology},\ }\href@noop {} {\bibfield
  {journal} {\bibinfo  {journal} {arXiv preprint arXiv:1912.12300}\ } (\bibinfo
  {year} {2019})}\BibitemShut {NoStop}%
\bibitem [{\citenamefont {Hedman}\ and\ \citenamefont
  {Nicholson}(2013)}]{HedmanNicholson2013}%
  \BibitemOpen
  \bibfield  {author} {\bibinfo {author} {\bibfnamefont {M.~M.}\ \bibnamefont
  {Hedman}}\ and\ \bibinfo {author} {\bibfnamefont {P.~D.}\ \bibnamefont
  {Nicholson}},\ }\bibfield  {title} {\bibinfo {title} {{KRONOSEISMOLOGY: USING
  DENSITY WAVES IN SATURN’S C RING TO PROBE THE PLANET’S INTERIOR}},\
  }\href@noop {} {\bibfield  {journal} {\bibinfo  {journal} {Astrophys. J.}\
  }\textbf {\bibinfo {volume} {146}},\ \bibinfo {pages} {12} (\bibinfo {year}
  {2013})}\BibitemShut {NoStop}%
\bibitem [{\citenamefont {Lambert}\ \emph {et~al.}(2006)\citenamefont
  {Lambert}, \citenamefont {Cl\'{e}rouin},\ and\ \citenamefont
  {Z\'{e}rah}}]{Lambert2006}%
  \BibitemOpen
  \bibfield  {author} {\bibinfo {author} {\bibfnamefont {F.}~\bibnamefont
  {Lambert}}, \bibinfo {author} {\bibfnamefont {J.}~\bibnamefont
  {Cl\'{e}rouin}},\ and\ \bibinfo {author} {\bibfnamefont {G.}~\bibnamefont
  {Z\'{e}rah}},\ }\bibfield  {title} {\bibinfo {title} {{Very-high-temperature
  molecular dynamics}},\ }\href@noop {} {\bibfield  {journal} {\bibinfo
  {journal} {Phys. Rev. E}\ }\textbf {\bibinfo {volume} {73}},\ \bibinfo
  {pages} {016403} (\bibinfo {year} {2006})}\BibitemShut {NoStop}%
\bibitem [{\citenamefont {Karasiev}\ \emph {et~al.}(2013)\citenamefont
  {Karasiev}, \citenamefont {Chakraborty}, \citenamefont {Shukruto},\ and\
  \citenamefont {Trickey}}]{Karasiev2013}%
  \BibitemOpen
  \bibfield  {author} {\bibinfo {author} {\bibfnamefont {V.~V.}\ \bibnamefont
  {Karasiev}}, \bibinfo {author} {\bibfnamefont {D.}~\bibnamefont
  {Chakraborty}}, \bibinfo {author} {\bibfnamefont {O.~A.}\ \bibnamefont
  {Shukruto}},\ and\ \bibinfo {author} {\bibfnamefont {S.}~\bibnamefont
  {Trickey}},\ }\bibfield  {title} {\bibinfo {title} {Nonempirical generalized
  gradient approximation free-energy functional for orbital-free simulations},\
  }\href@noop {} {\bibfield  {journal} {\bibinfo  {journal} {Physical Review
  B}\ }\textbf {\bibinfo {volume} {88}},\ \bibinfo {pages} {161108} (\bibinfo
  {year} {2013})}\BibitemShut {NoStop}%
\bibitem [{\citenamefont {Sjostrom}\ and\ \citenamefont
  {Daligault}(2014)}]{Sjostrom2014}%
  \BibitemOpen
  \bibfield  {author} {\bibinfo {author} {\bibfnamefont {T.}~\bibnamefont
  {Sjostrom}}\ and\ \bibinfo {author} {\bibfnamefont {J.}~\bibnamefont
  {Daligault}},\ }\bibfield  {title} {\bibinfo {title} {Fast and accurate
  quantum molecular dynamics of dense plasmas across temperature regimes},\
  }\href@noop {} {\bibfield  {journal} {\bibinfo  {journal} {Physical review
  letters}\ }\textbf {\bibinfo {volume} {113}},\ \bibinfo {pages} {155006}
  (\bibinfo {year} {2014})}\BibitemShut {NoStop}%
\bibitem [{\citenamefont {S.}\ \emph {et~al.}(2019)\citenamefont {S.},
  \citenamefont {A.}, \citenamefont {G.},\ and\ \citenamefont
  {Y.}}]{Mazevet2019}%
  \BibitemOpen
  \bibfield  {author} {\bibinfo {author} {\bibfnamefont {M.}~\bibnamefont
  {S.}}, \bibinfo {author} {\bibfnamefont {L.}~\bibnamefont {A.}}, \bibinfo
  {author} {\bibfnamefont {C.}~\bibnamefont {G.}},\ and\ \bibinfo {author}
  {\bibfnamefont {P.~A.}\ \bibnamefont {Y.}},\ }\href@noop {} {\bibfield
  {journal} {\bibinfo  {journal} {Astronomy {\&} Astrophysics}\ }\textbf
  {\bibinfo {volume} {612}},\ \bibinfo {pages} {A128} (\bibinfo {year}
  {2019})}\BibitemShut {NoStop}%
\bibitem [{\citenamefont {Sterne}\ \emph {et~al.}(2007)\citenamefont {Sterne},
  \citenamefont {Hansen}, \citenamefont {Wilson},\ and\ \citenamefont
  {Isaacs}}]{Sterne2007}%
  \BibitemOpen
  \bibfield  {author} {\bibinfo {author} {\bibfnamefont {P.}~\bibnamefont
  {Sterne}}, \bibinfo {author} {\bibfnamefont {S.}~\bibnamefont {Hansen}},
  \bibinfo {author} {\bibfnamefont {B.}~\bibnamefont {Wilson}},\ and\ \bibinfo
  {author} {\bibfnamefont {W.}~\bibnamefont {Isaacs}},\ }\bibfield  {title}
  {\bibinfo {title} {Equation of state, occupation probabilities and
  conductivities in the average atom purgatorio code},\ }\href@noop {}
  {\bibfield  {journal} {\bibinfo  {journal} {High Energy Density Physics}\
  }\textbf {\bibinfo {volume} {3}},\ \bibinfo {pages} {278} (\bibinfo {year}
  {2007})}\BibitemShut {NoStop}%
\bibitem [{\citenamefont {Rozsnyai}(2014)}]{Rozsnyai2014}%
  \BibitemOpen
  \bibfield  {author} {\bibinfo {author} {\bibfnamefont {B.~F.}\ \bibnamefont
  {Rozsnyai}},\ }\bibfield  {title} {\bibinfo {title} {Equation of state
  calculations based on the self-consistent ion-sphere and ion-correlation
  average atom models},\ }\href@noop {} {\bibfield  {journal} {\bibinfo
  {journal} {High Energy Density Physics}\ }\textbf {\bibinfo {volume} {10}},\
  \bibinfo {pages} {16} (\bibinfo {year} {2014})}\BibitemShut {NoStop}%
\bibitem [{\citenamefont {Starrett}\ and\ \citenamefont
  {Saumon}(2016)}]{Starrett2016Si}%
  \BibitemOpen
  \bibfield  {author} {\bibinfo {author} {\bibfnamefont {C.}~\bibnamefont
  {Starrett}}\ and\ \bibinfo {author} {\bibfnamefont {D.}~\bibnamefont
  {Saumon}},\ }\bibfield  {title} {\bibinfo {title} {Equation of state of dense
  plasmas with pseudoatom molecular dynamics},\ }\href@noop {} {\bibfield
  {journal} {\bibinfo  {journal} {Physical Review E}\ }\textbf {\bibinfo
  {volume} {93}},\ \bibinfo {pages} {063206} (\bibinfo {year}
  {2016})}\BibitemShut {NoStop}%
\bibitem [{\citenamefont {Pain}(2007)}]{Pain2007}%
  \BibitemOpen
  \bibfield  {author} {\bibinfo {author} {\bibfnamefont {J.}~\bibnamefont
  {Pain}},\ }\bibfield  {title} {\bibinfo {title} {Equation-of-state model for
  shock compression of hot dense matter},\ }\href
  {https://doi.org/https://doi.org/10.1016/j.physleta.2006.10.013} {\bibfield
  {journal} {\bibinfo  {journal} {Physics Letters A}\ }\textbf {\bibinfo
  {volume} {362}},\ \bibinfo {pages} {120 } (\bibinfo {year}
  {2007})}\BibitemShut {NoStop}%
\bibitem [{\citenamefont {Danel}\ \emph {et~al.}(2012)\citenamefont {Danel},
  \citenamefont {Kazandjian},\ and\ \citenamefont {Zérah}}]{Danel2012}%
  \BibitemOpen
  \bibfield  {author} {\bibinfo {author} {\bibfnamefont {J.-F.}\ \bibnamefont
  {Danel}}, \bibinfo {author} {\bibfnamefont {L.}~\bibnamefont {Kazandjian}},\
  and\ \bibinfo {author} {\bibfnamefont {G.}~\bibnamefont {Zérah}},\
  }\bibfield  {title} {\bibinfo {title} {Equation of state of dense plasmas by
  ab initio simulations: Bridging the gap between quantum molecular dynamics
  and orbital-free molecular dynamics at high temperature},\ }\href
  {https://doi.org/10.1063/1.4773191} {\bibfield  {journal} {\bibinfo
  {journal} {Physics of Plasmas}\ }\textbf {\bibinfo {volume} {19}},\ \bibinfo
  {pages} {122712} (\bibinfo {year} {2012})}\BibitemShut {NoStop}%
\bibitem [{\citenamefont {Danel}\ \emph {et~al.}(2014)\citenamefont {Danel},
  \citenamefont {Blottiau}, \citenamefont {Kazandjian}, \citenamefont {Piron},\
  and\ \citenamefont {Torrent}}]{Danel2014}%
  \BibitemOpen
  \bibfield  {author} {\bibinfo {author} {\bibfnamefont {J.-F.}\ \bibnamefont
  {Danel}}, \bibinfo {author} {\bibfnamefont {P.}~\bibnamefont {Blottiau}},
  \bibinfo {author} {\bibfnamefont {L.}~\bibnamefont {Kazandjian}}, \bibinfo
  {author} {\bibfnamefont {R.}~\bibnamefont {Piron}},\ and\ \bibinfo {author}
  {\bibfnamefont {M.}~\bibnamefont {Torrent}},\ }\bibfield  {title} {\bibinfo
  {title} {Equation of state of dense plasmas: Orbital-free molecular dynamics
  as the limit of quantum molecular dynamics for high-z elements},\ }\href
  {https://doi.org/10.1063/1.4897190} {\bibfield  {journal} {\bibinfo
  {journal} {Physics of Plasmas}\ }\textbf {\bibinfo {volume} {21}},\ \bibinfo
  {pages} {102701} (\bibinfo {year} {2014})}\BibitemShut {NoStop}%
\bibitem [{\citenamefont {More}\ \emph {et~al.}(1988)\citenamefont {More},
  \citenamefont {Warren}, \citenamefont {Young},\ and\ \citenamefont
  {Zimmerman}}]{QEOS}%
  \BibitemOpen
  \bibfield  {author} {\bibinfo {author} {\bibfnamefont {R.~M.}\ \bibnamefont
  {More}}, \bibinfo {author} {\bibfnamefont {K.~H.}\ \bibnamefont {Warren}},
  \bibinfo {author} {\bibfnamefont {D.~A.}\ \bibnamefont {Young}},\ and\
  \bibinfo {author} {\bibfnamefont {G.~B.}\ \bibnamefont {Zimmerman}},\
  }\href@noop {} {\bibfield  {journal} {\bibinfo  {journal} {Phys. Fluids}\
  }\textbf {\bibinfo {volume} {31}},\ \bibinfo {pages} {3059} (\bibinfo {year}
  {1988})}\BibitemShut {NoStop}%
\bibitem [{\citenamefont {Kerley}(1972)}]{Sesame}%
  \BibitemOpen
  \bibfield  {author} {\bibinfo {author} {\bibfnamefont {G.~I.}\ \bibnamefont
  {Kerley}},\ }\href@noop {} {\bibfield  {journal} {\bibinfo  {journal} {Phys.
  Earth Planet. Inter.}\ }\textbf {\bibinfo {volume} {6}},\ \bibinfo {pages}
  {78} (\bibinfo {year} {1972})}\BibitemShut {NoStop}%
\bibitem [{\citenamefont {Militzer}\ \emph {et~al.}(2020)\citenamefont
  {Militzer}, \citenamefont {Gonz{\'{a}}lez-Cataldo}, \citenamefont {Zhang},
  \citenamefont {Whitley}, \citenamefont {Swift},\ and\ \citenamefont
  {Millot}}]{Militzer2020}%
  \BibitemOpen
  \bibfield  {author} {\bibinfo {author} {\bibfnamefont {B.}~\bibnamefont
  {Militzer}}, \bibinfo {author} {\bibfnamefont {F.}~\bibnamefont
  {Gonz{\'{a}}lez-Cataldo}}, \bibinfo {author} {\bibfnamefont {S.}~\bibnamefont
  {Zhang}}, \bibinfo {author} {\bibfnamefont {H.~D.}\ \bibnamefont {Whitley}},
  \bibinfo {author} {\bibfnamefont {D.~C.}\ \bibnamefont {Swift}},\ and\
  \bibinfo {author} {\bibfnamefont {M.}~\bibnamefont {Millot}},\ }\bibfield
  {title} {\bibinfo {title} {{Nonideal mixing effects in warm dense matter
  studied with first-principles computer simulations}},\ }\href
  {https://doi.org/10.1063/5.0023232} {\bibfield  {journal} {\bibinfo
  {journal} {The Journal of Chemical Physics}\ }\textbf {\bibinfo {volume}
  {153}},\ \bibinfo {pages} {184101} (\bibinfo {year} {2020})}\BibitemShut
  {NoStop}%
\bibitem [{Sup()}]{SupplementalMaterial}%
  \BibitemOpen
  \href@noop {} {\bibinfo {title} {{Our EOS tables in ASCII text format, C++
  interpolation code, and Python graphics scripts will be published as
  supplemental material of our article in Physical Review E. We will also make
  them available at http://militzer.berkeley.edu/FPEOS and update them as
  needed.}}}\BibitemShut {Stop}%
\bibitem [{\citenamefont {Militzer}\ and\ \citenamefont
  {Ceperley}(2000)}]{MC00}%
  \BibitemOpen
  \bibfield  {author} {\bibinfo {author} {\bibfnamefont {B.}~\bibnamefont
  {Militzer}}\ and\ \bibinfo {author} {\bibfnamefont {D.~M.}\ \bibnamefont
  {Ceperley}},\ }\bibfield  {title} {\bibinfo {title} {{Path Integral Monte
  Carlo Calculation of the Deuterium Hugoniot}},\ }\href
  {https://link.aps.org/doi/10.1103/PhysRevLett.85.1890} {\bibfield  {journal}
  {\bibinfo  {journal} {Phys. Rev. Lett.}\ }\textbf {\bibinfo {volume} {85}},\
  \bibinfo {pages} {1890} (\bibinfo {year} {2000})}\BibitemShut {NoStop}%
\bibitem [{\citenamefont {Militzer}\ and\ \citenamefont
  {Ceperley}(2001)}]{MC01}%
  \BibitemOpen
  \bibfield  {author} {\bibinfo {author} {\bibfnamefont {B.}~\bibnamefont
  {Militzer}}\ and\ \bibinfo {author} {\bibfnamefont {D.~M.}\ \bibnamefont
  {Ceperley}},\ }\bibfield  {title} {\bibinfo {title} {{Path integral Monte
  Carlo simulation of the low-density hydrogen plasma}},\ }\href
  {https://link.aps.org/doi/10.1103/PhysRevE.63.066404} {\bibfield  {journal}
  {\bibinfo  {journal} {Phys. Rev. E}\ }\textbf {\bibinfo {volume} {63}},\
  \bibinfo {pages} {066404} (\bibinfo {year} {2001})}\BibitemShut {NoStop}%
\bibitem [{\citenamefont {Hu}\ \emph {et~al.}(2010)\citenamefont {Hu},
  \citenamefont {Militzer}, \citenamefont {Goncharov},\ and\ \citenamefont
  {Skupsky}}]{Hu2010}%
  \BibitemOpen
  \bibfield  {author} {\bibinfo {author} {\bibfnamefont {S.~X.}\ \bibnamefont
  {Hu}}, \bibinfo {author} {\bibfnamefont {B.}~\bibnamefont {Militzer}},
  \bibinfo {author} {\bibfnamefont {V.~N.}\ \bibnamefont {Goncharov}},\ and\
  \bibinfo {author} {\bibfnamefont {S.}~\bibnamefont {Skupsky}},\ }\bibfield
  {title} {\bibinfo {title} {Strong coupling and degeneracy effects in inertial
  confinement fusion implosions},\ }\href
  {https://doi.org/10.1103/PhysRevLett.104.235003} {\bibfield  {journal}
  {\bibinfo  {journal} {Phys. Rev. Lett.}\ }\textbf {\bibinfo {volume} {104}},\
  \bibinfo {pages} {235003} (\bibinfo {year} {2010})}\BibitemShut {NoStop}%
\bibitem [{\citenamefont {Hu}\ \emph {et~al.}(2011)\citenamefont {Hu},
  \citenamefont {Militzer}, \citenamefont {Goncharov},\ and\ \citenamefont
  {Skupsky}}]{Hu2011}%
  \BibitemOpen
  \bibfield  {author} {\bibinfo {author} {\bibfnamefont {S.~X.}\ \bibnamefont
  {Hu}}, \bibinfo {author} {\bibfnamefont {B.}~\bibnamefont {Militzer}},
  \bibinfo {author} {\bibfnamefont {V.~N.}\ \bibnamefont {Goncharov}},\ and\
  \bibinfo {author} {\bibfnamefont {S.}~\bibnamefont {Skupsky}},\ }\bibfield
  {title} {\bibinfo {title} {First-principles equation-of-state table of
  deuterium for inertial confinement fusion applications},\ }\href
  {https://doi.org/10.1103/PhysRevB.84.224109} {\bibfield  {journal} {\bibinfo
  {journal} {Phys. Rev. B}\ }\textbf {\bibinfo {volume} {84}},\ \bibinfo
  {pages} {224109} (\bibinfo {year} {2011})}\BibitemShut {NoStop}%
\bibitem [{\citenamefont {Militzer}\ \emph {et~al.}(2001)\citenamefont
  {Militzer}, \citenamefont {Ceperley}, \citenamefont {Kress}, \citenamefont
  {Johnson}, \citenamefont {Collins},\ and\ \citenamefont {Mazevet}}]{Mi01}%
  \BibitemOpen
  \bibfield  {author} {\bibinfo {author} {\bibfnamefont {B.}~\bibnamefont
  {Militzer}}, \bibinfo {author} {\bibfnamefont {D.~M.}\ \bibnamefont
  {Ceperley}}, \bibinfo {author} {\bibfnamefont {J.~D.}\ \bibnamefont {Kress}},
  \bibinfo {author} {\bibfnamefont {J.~D.}\ \bibnamefont {Johnson}}, \bibinfo
  {author} {\bibfnamefont {L.~A.}\ \bibnamefont {Collins}},\ and\ \bibinfo
  {author} {\bibfnamefont {S.}~\bibnamefont {Mazevet}},\ }\bibfield  {title}
  {\bibinfo {title} {Calculation of a deuterium double shock hugoniot from ab
  initio simulations},\ }\href
  {https://link.aps.org/doi/10.1103/PhysRevLett.87.275502} {\bibfield
  {journal} {\bibinfo  {journal} {Phys. Rev. Lett.}\ }\textbf {\bibinfo
  {volume} {87}},\ \bibinfo {pages} {275502} (\bibinfo {year}
  {2001})}\BibitemShut {NoStop}%
\bibitem [{\citenamefont {Militzer}\ \emph {et~al.}(1999)\citenamefont
  {Militzer}, \citenamefont {Magro},\ and\ \citenamefont {Ceperley}}]{Mi99}%
  \BibitemOpen
  \bibfield  {author} {\bibinfo {author} {\bibfnamefont {B.}~\bibnamefont
  {Militzer}}, \bibinfo {author} {\bibfnamefont {W.}~\bibnamefont {Magro}},\
  and\ \bibinfo {author} {\bibfnamefont {D.}~\bibnamefont {Ceperley}},\
  }\bibfield  {title} {\bibinfo {title} {Characterization of the state of
  hydrogen at high temperature and density},\ }\href
  {https://doi.org/10.1002/ctpp.2150390137} {\bibfield  {journal} {\bibinfo
  {journal} {Contributions to Plasma Physics}\ }\textbf {\bibinfo {volume}
  {39}},\ \bibinfo {pages} {151} (\bibinfo {year} {1999})}\BibitemShut
  {NoStop}%
\bibitem [{\citenamefont {Militzer}(2006)}]{Mi06}%
  \BibitemOpen
  \bibfield  {author} {\bibinfo {author} {\bibfnamefont {B.}~\bibnamefont
  {Militzer}},\ }\bibfield  {title} {\bibinfo {title} {First principles
  calculations of shock compressed fluid helium},\ }\href
  {https://link.aps.org/doi/10.1103/PhysRevLett.97.175501} {\bibfield
  {journal} {\bibinfo  {journal} {Phys. Rev. Lett.}\ }\textbf {\bibinfo
  {volume} {97}},\ \bibinfo {pages} {175501} (\bibinfo {year}
  {2006})}\BibitemShut {NoStop}%
\bibitem [{\citenamefont {Militzer}(2009)}]{Militzer2009}%
  \BibitemOpen
  \bibfield  {author} {\bibinfo {author} {\bibfnamefont {B.}~\bibnamefont
  {Militzer}},\ }\bibfield  {title} {\bibinfo {title} {{Path integral Monte
  Carlo and density functional molecular dynamics simulations of hot, dense
  helium}},\ }\href {https://doi.org/10.1103/PhysRevB.79.155105} {\bibfield
  {journal} {\bibinfo  {journal} {Phys. Rev. B}\ }\textbf {\bibinfo {volume}
  {79}},\ \bibinfo {pages} {155105} (\bibinfo {year} {2009})}\BibitemShut
  {NoStop}%
\bibitem [{\citenamefont {Zhang}\ \emph
  {et~al.}(2018{\natexlab{a}})\citenamefont {Zhang}, \citenamefont {Militzer},
  \citenamefont {Gregor}, \citenamefont {Caspersen}, \citenamefont {Yang},
  \citenamefont {Gaffney}, \citenamefont {Ogitsu}, \citenamefont {Swift},
  \citenamefont {Lazicki}, \citenamefont {Erskine}, \citenamefont {London},
  \citenamefont {Celliers}, \citenamefont {Nilsen}, \citenamefont {Sterne},\
  and\ \citenamefont {Whitley}}]{Zhang2018}%
  \BibitemOpen
  \bibfield  {author} {\bibinfo {author} {\bibfnamefont {S.}~\bibnamefont
  {Zhang}}, \bibinfo {author} {\bibfnamefont {B.}~\bibnamefont {Militzer}},
  \bibinfo {author} {\bibfnamefont {M.~C.}\ \bibnamefont {Gregor}}, \bibinfo
  {author} {\bibfnamefont {K.}~\bibnamefont {Caspersen}}, \bibinfo {author}
  {\bibfnamefont {L.~H.}\ \bibnamefont {Yang}}, \bibinfo {author}
  {\bibfnamefont {J.}~\bibnamefont {Gaffney}}, \bibinfo {author} {\bibfnamefont
  {T.}~\bibnamefont {Ogitsu}}, \bibinfo {author} {\bibfnamefont
  {D.}~\bibnamefont {Swift}}, \bibinfo {author} {\bibfnamefont
  {A.}~\bibnamefont {Lazicki}}, \bibinfo {author} {\bibfnamefont
  {D.}~\bibnamefont {Erskine}}, \bibinfo {author} {\bibfnamefont {R.~A.}\
  \bibnamefont {London}}, \bibinfo {author} {\bibfnamefont {P.~M.}\
  \bibnamefont {Celliers}}, \bibinfo {author} {\bibfnamefont {J.}~\bibnamefont
  {Nilsen}}, \bibinfo {author} {\bibfnamefont {P.~A.}\ \bibnamefont {Sterne}},\
  and\ \bibinfo {author} {\bibfnamefont {H.~D.}\ \bibnamefont {Whitley}},\
  }\bibfield  {title} {\bibinfo {title} {{Theoretical and experimental
  investigation of the equation of state of boron plasmas}},\ }\href
  {https://link.aps.org/doi/10.1103/PhysRevE.98.023205} {\bibfield  {journal}
  {\bibinfo  {journal} {Phys. Rev. E}\ }\textbf {\bibinfo {volume} {98}},\
  \bibinfo {pages} {023205} (\bibinfo {year} {2018}{\natexlab{a}})}\BibitemShut
  {NoStop}%
\bibitem [{\citenamefont {Driver}\ and\ \citenamefont
  {Militzer}(2012)}]{Driver2012}%
  \BibitemOpen
  \bibfield  {author} {\bibinfo {author} {\bibfnamefont {K.~P.}\ \bibnamefont
  {Driver}}\ and\ \bibinfo {author} {\bibfnamefont {B.}~\bibnamefont
  {Militzer}},\ }\bibfield  {title} {\bibinfo {title} {{All-Electron Path
  Integral Monte Carlo Simulations of Warm Dense Matter: Application to Water
  and Carbon Plasmas}},\ }\href
  {https://link.aps.org/doi/10.1103/PhysRevLett.108.115502} {\bibfield
  {journal} {\bibinfo  {journal} {Phys. Rev. Lett.}\ }\textbf {\bibinfo
  {volume} {108}},\ \bibinfo {pages} {115502} (\bibinfo {year}
  {2012})}\BibitemShut {NoStop}%
\bibitem [{\citenamefont {Benedict}\ \emph {et~al.}(2014)\citenamefont
  {Benedict}, \citenamefont {Driver}, \citenamefont {Hamel}, \citenamefont
  {Militzer}, \citenamefont {Qi}, \citenamefont {Correa}, \citenamefont
  {Saul},\ and\ \citenamefont {Schwegler}}]{Benedict2014}%
  \BibitemOpen
  \bibfield  {author} {\bibinfo {author} {\bibfnamefont {L.~X.}\ \bibnamefont
  {Benedict}}, \bibinfo {author} {\bibfnamefont {K.~P.}\ \bibnamefont
  {Driver}}, \bibinfo {author} {\bibfnamefont {S.}~\bibnamefont {Hamel}},
  \bibinfo {author} {\bibfnamefont {B.}~\bibnamefont {Militzer}}, \bibinfo
  {author} {\bibfnamefont {T.}~\bibnamefont {Qi}}, \bibinfo {author}
  {\bibfnamefont {A.~A.}\ \bibnamefont {Correa}}, \bibinfo {author}
  {\bibfnamefont {A.}~\bibnamefont {Saul}},\ and\ \bibinfo {author}
  {\bibfnamefont {E.}~\bibnamefont {Schwegler}},\ }\bibfield  {title} {\bibinfo
  {title} {A multiphase equation of state for carbon addressing high pressures
  and temperatures},\ }\href
  {https://link.aps.org/doi/10.1103/PhysRevB.89.224109} {\bibfield  {journal}
  {\bibinfo  {journal} {Phys. Rev. B}\ }\textbf {\bibinfo {volume} {89}},\
  \bibinfo {pages} {224109} (\bibinfo {year} {2014})}\BibitemShut {NoStop}%
\bibitem [{\citenamefont {Driver}\ and\ \citenamefont
  {Militzer}(2016)}]{DriverNitrogen2016}%
  \BibitemOpen
  \bibfield  {author} {\bibinfo {author} {\bibfnamefont {K.~P.}\ \bibnamefont
  {Driver}}\ and\ \bibinfo {author} {\bibfnamefont {B.}~\bibnamefont
  {Militzer}},\ }\bibfield  {title} {\bibinfo {title} {First-principles
  equation of state calculations of warm dense nitrogen},\ }\href
  {https://doi.org/10.1103/PhysRevB.93.064101} {\bibfield  {journal} {\bibinfo
  {journal} {Phys. Rev. B}\ }\textbf {\bibinfo {volume} {93}},\ \bibinfo
  {pages} {064101} (\bibinfo {year} {2016})}\BibitemShut {NoStop}%
\bibitem [{\citenamefont {Driver}\ \emph {et~al.}(2015)\citenamefont {Driver},
  \citenamefont {Soubiran}, \citenamefont {Zhang},\ and\ \citenamefont
  {Militzer}}]{Driver2015b}%
  \BibitemOpen
  \bibfield  {author} {\bibinfo {author} {\bibfnamefont {K.~P.}\ \bibnamefont
  {Driver}}, \bibinfo {author} {\bibfnamefont {F.}~\bibnamefont {Soubiran}},
  \bibinfo {author} {\bibfnamefont {S.}~\bibnamefont {Zhang}},\ and\ \bibinfo
  {author} {\bibfnamefont {B.}~\bibnamefont {Militzer}},\ }\bibfield  {title}
  {\bibinfo {title} {{First-principles equation of state and electronic
  properties of warm dense oxygen}},\ }\href
  {https://doi.org/10.1063/1.4934348} {\bibfield  {journal} {\bibinfo
  {journal} {J. Chem. Phys.}\ }\textbf {\bibinfo {volume} {143}},\ \bibinfo
  {pages} {164507} (\bibinfo {year} {2015})}\BibitemShut {NoStop}%
\bibitem [{\citenamefont {Driver}\ and\ \citenamefont
  {Militzer}(2015)}]{Driver2015}%
  \BibitemOpen
  \bibfield  {author} {\bibinfo {author} {\bibfnamefont {K.~P.}\ \bibnamefont
  {Driver}}\ and\ \bibinfo {author} {\bibfnamefont {B.}~\bibnamefont
  {Militzer}},\ }\bibfield  {title} {\bibinfo {title} {{First-principles
  simulations and shock Hugoniot calculations of warm dense neon}},\ }\href
  {https://link.aps.org/doi/10.1103/PhysRevB.91.045103} {\bibfield  {journal}
  {\bibinfo  {journal} {Phys. Rev. B}\ }\textbf {\bibinfo {volume} {91}},\
  \bibinfo {pages} {045103} (\bibinfo {year} {2015})}\BibitemShut {NoStop}%
\bibitem [{\citenamefont {Zhang}\ \emph
  {et~al.}(2017{\natexlab{a}})\citenamefont {Zhang}, \citenamefont {Driver},
  \citenamefont {Soubiran},\ and\ \citenamefont {Militzer}}]{ZhangSodium2017}%
  \BibitemOpen
  \bibfield  {author} {\bibinfo {author} {\bibfnamefont {S.}~\bibnamefont
  {Zhang}}, \bibinfo {author} {\bibfnamefont {K.~P.}\ \bibnamefont {Driver}},
  \bibinfo {author} {\bibfnamefont {F.}~\bibnamefont {Soubiran}},\ and\
  \bibinfo {author} {\bibfnamefont {B.}~\bibnamefont {Militzer}},\ }\bibfield
  {title} {\bibinfo {title} {{Equation of state and shock compression of warm
  dense sodium—A first-principles study}},\ }\href
  {https://doi.org/10.1063/1.4976559} {\bibfield  {journal} {\bibinfo
  {journal} {J. Chem. Phys.}\ }\textbf {\bibinfo {volume} {146}},\ \bibinfo
  {pages} {074505} (\bibinfo {year} {2017}{\natexlab{a}})}\BibitemShut
  {NoStop}%
\bibitem [{\citenamefont {Zhang}\ \emph {et~al.}(2016)\citenamefont {Zhang},
  \citenamefont {Driver}, \citenamefont {Soubiran},\ and\ \citenamefont
  {Militzer}}]{Zhang2016b}%
  \BibitemOpen
  \bibfield  {author} {\bibinfo {author} {\bibfnamefont {S.}~\bibnamefont
  {Zhang}}, \bibinfo {author} {\bibfnamefont {K.~P.}\ \bibnamefont {Driver}},
  \bibinfo {author} {\bibfnamefont {F.}~\bibnamefont {Soubiran}},\ and\
  \bibinfo {author} {\bibfnamefont {B.}~\bibnamefont {Militzer}},\ }\bibfield
  {title} {\bibinfo {title} {{Path integral Monte Carlo simulations of warm
  dense sodium}},\ }\href {https://doi.org/10.1016/j.hedp.2016.09.004}
  {\bibfield  {journal} {\bibinfo  {journal} {High Energ. Dens. Phys.}\
  }\textbf {\bibinfo {volume} {21}},\ \bibinfo {pages} {16} (\bibinfo {year}
  {2016})}\BibitemShut {NoStop}%
\bibitem [{\citenamefont {Gonz{\'{a}}lez-Cataldo}\ \emph
  {et~al.}(2020{\natexlab{a}})\citenamefont {Gonz{\'{a}}lez-Cataldo},
  \citenamefont {Soubiran},\ and\ \citenamefont
  {Militzer}}]{Gonzalez-Cataldo2020}%
  \BibitemOpen
  \bibfield  {author} {\bibinfo {author} {\bibfnamefont {F.}~\bibnamefont
  {Gonz{\'{a}}lez-Cataldo}}, \bibinfo {author} {\bibfnamefont {F.}~\bibnamefont
  {Soubiran}},\ and\ \bibinfo {author} {\bibfnamefont {B.}~\bibnamefont
  {Militzer}},\ }\bibfield  {title} {\bibinfo {title} {{Equation of state of
  hot, dense magnesium derived with first-principles computer simulations}},\
  }\href {https://doi.org/10.1063/5.0017555} {\bibfield  {journal} {\bibinfo
  {journal} {Physics of Plasmas}\ }\textbf {\bibinfo {volume} {27}},\ \bibinfo
  {pages} {092706} (\bibinfo {year} {2020}{\natexlab{a}})},\ \Eprint
  {https://arxiv.org/abs/2008.08459} {arXiv:2008.08459} \BibitemShut {NoStop}%
\bibitem [{\citenamefont {Driver}\ \emph {et~al.}(2018)\citenamefont {Driver},
  \citenamefont {Soubiran},\ and\ \citenamefont {Militzer}}]{Driver2018}%
  \BibitemOpen
  \bibfield  {author} {\bibinfo {author} {\bibfnamefont {K.~P.}\ \bibnamefont
  {Driver}}, \bibinfo {author} {\bibfnamefont {F.}~\bibnamefont {Soubiran}},\
  and\ \bibinfo {author} {\bibfnamefont {B.}~\bibnamefont {Militzer}},\
  }\bibfield  {title} {\bibinfo {title} {{Path integral Monte Carlo simulations
  of warm dense aluminum}},\ }\href
  {https://link.aps.org/doi/10.1103/PhysRevE.97.063207} {\bibfield  {journal}
  {\bibinfo  {journal} {Phys. Rev. E}\ }\textbf {\bibinfo {volume} {97}},\
  \bibinfo {pages} {063207} (\bibinfo {year} {2018})}\BibitemShut {NoStop}%
\bibitem [{\citenamefont {Militzer}\ and\ \citenamefont
  {Driver}(2015)}]{MilitzerDriver2015}%
  \BibitemOpen
  \bibfield  {author} {\bibinfo {author} {\bibfnamefont {B.}~\bibnamefont
  {Militzer}}\ and\ \bibinfo {author} {\bibfnamefont {K.~P.}\ \bibnamefont
  {Driver}},\ }\bibfield  {title} {\bibinfo {title} {{Development of Path
  Integral Monte Carlo Simulations with Localized Nodal Surfaces for Second-Row
  Elements}},\ }\href {https://link.aps.org/doi/10.1103/PhysRevLett.115.176403}
  {\bibfield  {journal} {\bibinfo  {journal} {Phys. Rev. Lett.}\ }\textbf
  {\bibinfo {volume} {115}},\ \bibinfo {pages} {176403} (\bibinfo {year}
  {2015})}\BibitemShut {NoStop}%
\bibitem [{\citenamefont {Hu}\ \emph {et~al.}(2016)\citenamefont {Hu},
  \citenamefont {Militzer}, \citenamefont {Collins}, \citenamefont {Driver},\
  and\ \citenamefont {Kress}}]{Hu2016}%
  \BibitemOpen
  \bibfield  {author} {\bibinfo {author} {\bibfnamefont {S.~X.}\ \bibnamefont
  {Hu}}, \bibinfo {author} {\bibfnamefont {B.}~\bibnamefont {Militzer}},
  \bibinfo {author} {\bibfnamefont {L.~A.}\ \bibnamefont {Collins}}, \bibinfo
  {author} {\bibfnamefont {K.~P.}\ \bibnamefont {Driver}},\ and\ \bibinfo
  {author} {\bibfnamefont {J.~D.}\ \bibnamefont {Kress}},\ }\bibfield  {title}
  {\bibinfo {title} {First-principles prediction of the softening of the
  silicon shock hugoniot curve},\ }\href
  {https://link.aps.org/doi/10.1103/PhysRevB.94.094109} {\bibfield  {journal}
  {\bibinfo  {journal} {Phys. Rev. B}\ }\textbf {\bibinfo {volume} {94}},\
  \bibinfo {pages} {094109} (\bibinfo {year} {2016})}\BibitemShut {NoStop}%
\bibitem [{\citenamefont {Driver}\ and\ \citenamefont
  {Militzer}(2017)}]{Driver2017}%
  \BibitemOpen
  \bibfield  {author} {\bibinfo {author} {\bibfnamefont {K.~P.}\ \bibnamefont
  {Driver}}\ and\ \bibinfo {author} {\bibfnamefont {B.}~\bibnamefont
  {Militzer}},\ }\bibfield  {title} {\bibinfo {title} {{First-principles
  simulations of warm dense lithium fluoride}},\ }\href
  {https://link.aps.org/doi/10.1103/PhysRevE.95.043205} {\bibfield  {journal}
  {\bibinfo  {journal} {Phys. Rev. E}\ }\textbf {\bibinfo {volume} {95}},\
  \bibinfo {pages} {043205} (\bibinfo {year} {2017})}\BibitemShut {NoStop}%
\bibitem [{\citenamefont {Zhang}\ \emph {et~al.}(2020)\citenamefont {Zhang},
  \citenamefont {Marshall}, \citenamefont {Yang}, \citenamefont {Sterne},
  \citenamefont {Militzer}, \citenamefont {D{\"{a}}ne}, \citenamefont
  {Gaffney}, \citenamefont {Shamp}, \citenamefont {Ogitsu}, \citenamefont
  {Caspersen}, \citenamefont {Lazicki}, \citenamefont {Erskine}, \citenamefont
  {London}, \citenamefont {Celliers}, \citenamefont {Nilsen},\ and\
  \citenamefont {Whitley}}]{Zhang_B4C_2020}%
  \BibitemOpen
  \bibfield  {author} {\bibinfo {author} {\bibfnamefont {S.}~\bibnamefont
  {Zhang}}, \bibinfo {author} {\bibfnamefont {M.~C.}\ \bibnamefont {Marshall}},
  \bibinfo {author} {\bibfnamefont {L.~H.}\ \bibnamefont {Yang}}, \bibinfo
  {author} {\bibfnamefont {P.~A.}\ \bibnamefont {Sterne}}, \bibinfo {author}
  {\bibfnamefont {B.}~\bibnamefont {Militzer}}, \bibinfo {author}
  {\bibfnamefont {M.}~\bibnamefont {D{\"{a}}ne}}, \bibinfo {author}
  {\bibfnamefont {J.~A.}\ \bibnamefont {Gaffney}}, \bibinfo {author}
  {\bibfnamefont {A.}~\bibnamefont {Shamp}}, \bibinfo {author} {\bibfnamefont
  {T.}~\bibnamefont {Ogitsu}}, \bibinfo {author} {\bibfnamefont
  {K.}~\bibnamefont {Caspersen}}, \bibinfo {author} {\bibfnamefont {A.~E.}\
  \bibnamefont {Lazicki}}, \bibinfo {author} {\bibfnamefont {D.}~\bibnamefont
  {Erskine}}, \bibinfo {author} {\bibfnamefont {R.~A.}\ \bibnamefont {London}},
  \bibinfo {author} {\bibfnamefont {P.~M.}\ \bibnamefont {Celliers}}, \bibinfo
  {author} {\bibfnamefont {J.}~\bibnamefont {Nilsen}},\ and\ \bibinfo {author}
  {\bibfnamefont {H.~D.}\ \bibnamefont {Whitley}},\ }\bibfield  {title}
  {\bibinfo {title} {{Benchmarking boron carbide equation of state using
  computation and experiment}},\ }\href
  {https://doi.org/10.1103/PhysRevE.102.053203} {\bibfield  {journal} {\bibinfo
   {journal} {Physical Review E}\ }\textbf {\bibinfo {volume} {102}},\ \bibinfo
  {pages} {053203} (\bibinfo {year} {2020})}\BibitemShut {NoStop}%
\bibitem [{\citenamefont {Zhang}\ \emph {et~al.}(2019)\citenamefont {Zhang},
  \citenamefont {Lazicki}, \citenamefont {Militzer}, \citenamefont {Yang},
  \citenamefont {Caspersen}, \citenamefont {Gaffney}, \citenamefont {D\"ane},
  \citenamefont {Pask}, \citenamefont {Johnson}, \citenamefont {Sharma},
  \citenamefont {Suryanarayana}, \citenamefont {Johnson}, \citenamefont
  {Smirnov}, \citenamefont {Sterne}, \citenamefont {Erskine}, \citenamefont
  {London}, \citenamefont {Coppari}, \citenamefont {Swift}, \citenamefont
  {Nilsen}, \citenamefont {Nelson},\ and\ \citenamefont
  {Whitley}}]{ZhangBN2019}%
  \BibitemOpen
  \bibfield  {author} {\bibinfo {author} {\bibfnamefont {S.}~\bibnamefont
  {Zhang}}, \bibinfo {author} {\bibfnamefont {A.}~\bibnamefont {Lazicki}},
  \bibinfo {author} {\bibfnamefont {B.}~\bibnamefont {Militzer}}, \bibinfo
  {author} {\bibfnamefont {L.~H.}\ \bibnamefont {Yang}}, \bibinfo {author}
  {\bibfnamefont {K.}~\bibnamefont {Caspersen}}, \bibinfo {author}
  {\bibfnamefont {J.~A.}\ \bibnamefont {Gaffney}}, \bibinfo {author}
  {\bibfnamefont {M.~W.}\ \bibnamefont {D\"ane}}, \bibinfo {author}
  {\bibfnamefont {J.~E.}\ \bibnamefont {Pask}}, \bibinfo {author}
  {\bibfnamefont {W.~R.}\ \bibnamefont {Johnson}}, \bibinfo {author}
  {\bibfnamefont {A.}~\bibnamefont {Sharma}}, \bibinfo {author} {\bibfnamefont
  {P.}~\bibnamefont {Suryanarayana}}, \bibinfo {author} {\bibfnamefont {D.~D.}\
  \bibnamefont {Johnson}}, \bibinfo {author} {\bibfnamefont {A.~V.}\
  \bibnamefont {Smirnov}}, \bibinfo {author} {\bibfnamefont {P.~A.}\
  \bibnamefont {Sterne}}, \bibinfo {author} {\bibfnamefont {D.}~\bibnamefont
  {Erskine}}, \bibinfo {author} {\bibfnamefont {R.~A.}\ \bibnamefont {London}},
  \bibinfo {author} {\bibfnamefont {F.}~\bibnamefont {Coppari}}, \bibinfo
  {author} {\bibfnamefont {D.}~\bibnamefont {Swift}}, \bibinfo {author}
  {\bibfnamefont {J.}~\bibnamefont {Nilsen}}, \bibinfo {author} {\bibfnamefont
  {A.~J.}\ \bibnamefont {Nelson}},\ and\ \bibinfo {author} {\bibfnamefont
  {H.~D.}\ \bibnamefont {Whitley}},\ }\bibfield  {title} {\bibinfo {title}
  {Equation of state of boron nitride combining computation, modeling, and
  experiment},\ }\href {https://doi.org/10.1103/PhysRevB.99.165103} {\bibfield
  {journal} {\bibinfo  {journal} {Phys. Rev. B}\ }\textbf {\bibinfo {volume}
  {99}},\ \bibinfo {pages} {165103} (\bibinfo {year} {2019})}\BibitemShut
  {NoStop}%
\bibitem [{\citenamefont {Zhang}\ \emph
  {et~al.}(2017{\natexlab{b}})\citenamefont {Zhang}, \citenamefont {Driver},
  \citenamefont {Soubiran},\ and\ \citenamefont {Militzer}}]{ZhangCH2017}%
  \BibitemOpen
  \bibfield  {author} {\bibinfo {author} {\bibfnamefont {S.}~\bibnamefont
  {Zhang}}, \bibinfo {author} {\bibfnamefont {K.~P.}\ \bibnamefont {Driver}},
  \bibinfo {author} {\bibfnamefont {F.}~\bibnamefont {Soubiran}},\ and\
  \bibinfo {author} {\bibfnamefont {B.}~\bibnamefont {Militzer}},\ }\bibfield
  {title} {\bibinfo {title} {{First-principles equation of state and shock
  compression predictions of warm dense hydrocarbons}},\ }\href
  {https://link.aps.org/doi/10.1103/PhysRevE.96.013204} {\bibfield  {journal}
  {\bibinfo  {journal} {Phys. Rev. E}\ }\textbf {\bibinfo {volume} {96}},\
  \bibinfo {pages} {013204} (\bibinfo {year} {2017}{\natexlab{b}})}\BibitemShut
  {NoStop}%
\bibitem [{\citenamefont {Zhang}\ \emph
  {et~al.}(2018{\natexlab{b}})\citenamefont {Zhang}, \citenamefont {Militzer},
  \citenamefont {Benedict}, \citenamefont {Soubiran}, \citenamefont {Sterne},\
  and\ \citenamefont {Driver}}]{ZhangCH2018}%
  \BibitemOpen
  \bibfield  {author} {\bibinfo {author} {\bibfnamefont {S.}~\bibnamefont
  {Zhang}}, \bibinfo {author} {\bibfnamefont {B.}~\bibnamefont {Militzer}},
  \bibinfo {author} {\bibfnamefont {L.~X.}\ \bibnamefont {Benedict}}, \bibinfo
  {author} {\bibfnamefont {F.}~\bibnamefont {Soubiran}}, \bibinfo {author}
  {\bibfnamefont {P.~A.}\ \bibnamefont {Sterne}},\ and\ \bibinfo {author}
  {\bibfnamefont {K.~P.}\ \bibnamefont {Driver}},\ }\bibfield  {title}
  {\bibinfo {title} {{Path integral Monte Carlo simulations of dense
  carbon-hydrogen plasmas}},\ }\href {https://doi.org/10.1063/1.5001208}
  {\bibfield  {journal} {\bibinfo  {journal} {J. Chem. Phys.}\ }\textbf
  {\bibinfo {volume} {148}},\ \bibinfo {pages} {102318} (\bibinfo {year}
  {2018}{\natexlab{b}})}\BibitemShut {NoStop}%
\bibitem [{\citenamefont {Soubiran}\ \emph {et~al.}(2019)\citenamefont
  {Soubiran}, \citenamefont {Gonz{\'{a}}lez-Cataldo}, \citenamefont {Driver},
  \citenamefont {Zhang},\ and\ \citenamefont {Militzer}}]{Soubiran2019}%
  \BibitemOpen
  \bibfield  {author} {\bibinfo {author} {\bibfnamefont {F.}~\bibnamefont
  {Soubiran}}, \bibinfo {author} {\bibfnamefont {F.}~\bibnamefont
  {Gonz{\'{a}}lez-Cataldo}}, \bibinfo {author} {\bibfnamefont {K.~P.}\
  \bibnamefont {Driver}}, \bibinfo {author} {\bibfnamefont {S.}~\bibnamefont
  {Zhang}},\ and\ \bibinfo {author} {\bibfnamefont {B.}~\bibnamefont
  {Militzer}},\ }\bibfield  {title} {\bibinfo {title} {{Magnesium oxide at
  extreme temperatures and pressures studied with first-principles
  simulations}},\ }\href {https://doi.org/10.1063/1.5126624} {\bibfield
  {journal} {\bibinfo  {journal} {The Journal of Chemical Physics}\ }\textbf
  {\bibinfo {volume} {151}},\ \bibinfo {pages} {214104} (\bibinfo {year}
  {2019})}\BibitemShut {NoStop}%
\bibitem [{\citenamefont {Gonz{\'{a}}lez-Cataldo}\ \emph
  {et~al.}(2020{\natexlab{b}})\citenamefont {Gonz{\'{a}}lez-Cataldo},
  \citenamefont {Soubiran}, \citenamefont {Peterson},\ and\ \citenamefont
  {Militzer}}]{Gonzalez2020}%
  \BibitemOpen
  \bibfield  {author} {\bibinfo {author} {\bibfnamefont {F.}~\bibnamefont
  {Gonz{\'{a}}lez-Cataldo}}, \bibinfo {author} {\bibfnamefont {F.}~\bibnamefont
  {Soubiran}}, \bibinfo {author} {\bibfnamefont {H.}~\bibnamefont {Peterson}},\
  and\ \bibinfo {author} {\bibfnamefont {B.}~\bibnamefont {Militzer}},\
  }\bibfield  {title} {\bibinfo {title} {{Path integral Monte Carlo and density
  functional molecular dynamics simulations of warm dense MgSiO$_3$}},\ }\href
  {https://doi.org/10.1103/PhysRevB.101.024107} {\bibfield  {journal} {\bibinfo
   {journal} {Physical Review B}\ }\textbf {\bibinfo {volume} {101}},\ \bibinfo
  {pages} {024107} (\bibinfo {year} {2020}{\natexlab{b}})}\BibitemShut
  {NoStop}%
\bibitem [{\citenamefont {Gonz{\'{a}}lez-Cataldo}\ and\ \citenamefont
  {Militzer}(2020)}]{GonzalezMilitzer2020}%
  \BibitemOpen
  \bibfield  {author} {\bibinfo {author} {\bibfnamefont {F.}~\bibnamefont
  {Gonz{\'{a}}lez-Cataldo}}\ and\ \bibinfo {author} {\bibfnamefont
  {B.}~\bibnamefont {Militzer}},\ }\bibfield  {title} {\bibinfo {title}
  {{Thermal and pressure ionization in warm, dense MgSiO$_3$ studied with
  first-principles computer simulations}},\ }in\ \href
  {https://doi.org/10.1063/12.0000793} {\emph {\bibinfo {booktitle} {AIP
  Conference Proceedings}}},\ Vol.\ \bibinfo {volume} {2272}\ (\bibinfo {year}
  {2020})\ p.\ \bibinfo {pages} {090001},\ \Eprint
  {https://arxiv.org/abs/2002.12163} {arXiv:2002.12163} \BibitemShut {NoStop}%
\bibitem [{\citenamefont {Pollock}\ and\ \citenamefont
  {Ceperley}(1984)}]{PC84}%
  \BibitemOpen
  \bibfield  {author} {\bibinfo {author} {\bibfnamefont {E.~L.}\ \bibnamefont
  {Pollock}}\ and\ \bibinfo {author} {\bibfnamefont {D.~M.}\ \bibnamefont
  {Ceperley}},\ }\bibfield  {title} {\bibinfo {title} {Simulation of quantum
  many-body systems by path-integral methods},\ }\href
  {https://doi.org/10.1103/PhysRevB.30.2555} {\bibfield  {journal} {\bibinfo
  {journal} {Phys. Rev. B}\ }\textbf {\bibinfo {volume} {30}},\ \bibinfo
  {pages} {2555} (\bibinfo {year} {1984})}\BibitemShut {NoStop}%
\bibitem [{\citenamefont {Pollock}\ and\ \citenamefont
  {Ceperley}(1987)}]{PC87}%
  \BibitemOpen
  \bibfield  {author} {\bibinfo {author} {\bibfnamefont {E.}~\bibnamefont
  {Pollock}}\ and\ \bibinfo {author} {\bibfnamefont {D.~M.}\ \bibnamefont
  {Ceperley}},\ }\href@noop {} {\bibfield  {journal} {\bibinfo  {journal}
  {Phys. Rev. B}\ }\textbf {\bibinfo {volume} {{36}}},\ \bibinfo {pages} {8343}
  (\bibinfo {year} {1987})}\BibitemShut {NoStop}%
\bibitem [{\citenamefont {Ceperley}(1995)}]{Ce95}%
  \BibitemOpen
  \bibfield  {author} {\bibinfo {author} {\bibfnamefont {D.~M.}\ \bibnamefont
  {Ceperley}},\ }\bibfield  {title} {\bibinfo {title} {Path integrals in the
  theory of condensed helium},\ }\href
  {https://doi.org/10.1103/RevModPhys.67.279} {\bibfield  {journal} {\bibinfo
  {journal} {Rev. Mod. Phys.}\ }\textbf {\bibinfo {volume} {67}},\ \bibinfo
  {pages} {279} (\bibinfo {year} {1995})}\BibitemShut {NoStop}%
\bibitem [{\citenamefont {Ceperley}(1991)}]{Ce91}%
  \BibitemOpen
  \bibfield  {author} {\bibinfo {author} {\bibfnamefont {D.~M.}\ \bibnamefont
  {Ceperley}},\ }\bibfield  {title} {\bibinfo {title} {{Fermion nodes}},\
  }\href@noop {} {\bibfield  {journal} {\bibinfo  {journal} {Journal of
  Statistical Physics}\ }\textbf {\bibinfo {volume} {63}},\ \bibinfo {pages}
  {1237} (\bibinfo {year} {1991})}\BibitemShut {NoStop}%
\bibitem [{\citenamefont {Ceperley}(1992)}]{Ce92}%
  \BibitemOpen
  \bibfield  {author} {\bibinfo {author} {\bibfnamefont {D.~M.}\ \bibnamefont
  {Ceperley}},\ }\bibfield  {title} {\bibinfo {title} {Path-integral
  calculations of normal liquid $^{3}\mathrm{He}$},\ }\href
  {https://doi.org/10.1103/PhysRevLett.69.331} {\bibfield  {journal} {\bibinfo
  {journal} {Phys. Rev. Lett.}\ }\textbf {\bibinfo {volume} {69}},\ \bibinfo
  {pages} {331} (\bibinfo {year} {1992})}\BibitemShut {NoStop}%
\bibitem [{\citenamefont {Ceperley}(1996)}]{Ce96}%
  \BibitemOpen
  \bibfield  {author} {\bibinfo {author} {\bibfnamefont {D.}~\bibnamefont
  {Ceperley}},\ }\bibfield  {title} {\bibinfo {title} {Monte carlo and
  molecular dynamics of condensed matter systems}\ }(\bibinfo  {publisher}
  {Editrice Compositori, Bologna, Italy},\ \bibinfo {year} {1996})\ p.\
  \bibinfo {pages} {443}\BibitemShut {NoStop}%
\bibitem [{\citenamefont {Pierleoni}\ \emph {et~al.}(1994)\citenamefont
  {Pierleoni}, \citenamefont {Ceperley}, \citenamefont {Bernu},\ and\
  \citenamefont {Magro}}]{PC94}%
  \BibitemOpen
  \bibfield  {author} {\bibinfo {author} {\bibfnamefont {C.}~\bibnamefont
  {Pierleoni}}, \bibinfo {author} {\bibfnamefont {D.~M.}\ \bibnamefont
  {Ceperley}}, \bibinfo {author} {\bibfnamefont {B.}~\bibnamefont {Bernu}},\
  and\ \bibinfo {author} {\bibfnamefont {W.~R.}\ \bibnamefont {Magro}},\
  }\bibfield  {title} {\bibinfo {title} {Equation of state of the hydrogen
  plasma by path integral monte carlo simulation},\ }\href@noop {} {\bibfield
  {journal} {\bibinfo  {journal} {Phys. Rev. Lett.}\ }\textbf {\bibinfo
  {volume} {73}},\ \bibinfo {pages} {2145} (\bibinfo {year}
  {1994})}\BibitemShut {NoStop}%
\bibitem [{\citenamefont {Magro}\ \emph {et~al.}(1996)\citenamefont {Magro},
  \citenamefont {Ceperley}, \citenamefont {Pierleoni},\ and\ \citenamefont
  {Bernu}}]{Ma96}%
  \BibitemOpen
  \bibfield  {author} {\bibinfo {author} {\bibfnamefont {W.~R.}\ \bibnamefont
  {Magro}}, \bibinfo {author} {\bibfnamefont {D.~M.}\ \bibnamefont {Ceperley}},
  \bibinfo {author} {\bibfnamefont {C.}~\bibnamefont {Pierleoni}},\ and\
  \bibinfo {author} {\bibfnamefont {B.}~\bibnamefont {Bernu}},\ }\bibfield
  {title} {\bibinfo {title} {Molecular dissociation in hot, dense hydrogen},\
  }\href {https://doi.org/10.1103/PhysRevLett.76.1240} {\bibfield  {journal}
  {\bibinfo  {journal} {Phys. Rev. Lett.}\ }\textbf {\bibinfo {volume} {76}},\
  \bibinfo {pages} {1240} (\bibinfo {year} {1996})}\BibitemShut {NoStop}%
\bibitem [{\citenamefont {Jones}\ and\ \citenamefont {Ceperley}(1996)}]{JC96}%
  \BibitemOpen
  \bibfield  {author} {\bibinfo {author} {\bibfnamefont {M.~D.}\ \bibnamefont
  {Jones}}\ and\ \bibinfo {author} {\bibfnamefont {D.~M.}\ \bibnamefont
  {Ceperley}},\ }\bibfield  {title} {\bibinfo {title} {Crystallization of the
  one-component plasma at finite temperature},\ }\href@noop {} {\bibfield
  {journal} {\bibinfo  {journal} {Phys. Rev. Lett.}\ }\textbf {\bibinfo
  {volume} {76}},\ \bibinfo {pages} {4572} (\bibinfo {year}
  {1996})}\BibitemShut {NoStop}%
\bibitem [{\citenamefont {Pollock}\ and\ \citenamefont
  {Militzer}(2004)}]{MP04}%
  \BibitemOpen
  \bibfield  {author} {\bibinfo {author} {\bibfnamefont {E.~L.}\ \bibnamefont
  {Pollock}}\ and\ \bibinfo {author} {\bibfnamefont {B.}~\bibnamefont
  {Militzer}},\ }\bibfield  {title} {\bibinfo {title} {Dense plasma effects on
  nuclear reaction rates},\ }\href@noop {} {\bibfield  {journal} {\bibinfo
  {journal} {Phys. Rev. Lett.}\ }\textbf {\bibinfo {volume} {92}},\ \bibinfo
  {pages} {021101} (\bibinfo {year} {2004})}\BibitemShut {NoStop}%
\bibitem [{\citenamefont {Militzer}\ and\ \citenamefont
  {Pollock}(2005)}]{MP05}%
  \BibitemOpen
  \bibfield  {author} {\bibinfo {author} {\bibfnamefont {B.}~\bibnamefont
  {Militzer}}\ and\ \bibinfo {author} {\bibfnamefont {E.~L.}\ \bibnamefont
  {Pollock}},\ }\bibfield  {title} {\bibinfo {title} {Equilibrium contact
  probabilities in dense plasmas},\ }\href@noop {} {\bibfield  {journal}
  {\bibinfo  {journal} {Phys. Rev. B}\ }\textbf {\bibinfo {volume} {71}},\
  \bibinfo {pages} {134303} (\bibinfo {year} {2005})}\BibitemShut {NoStop}%
\bibitem [{\citenamefont {Militzer}(2005)}]{Mi05}%
  \BibitemOpen
  \bibfield  {author} {\bibinfo {author} {\bibfnamefont {B.}~\bibnamefont
  {Militzer}},\ }\bibfield  {title} {\bibinfo {title} {Hydrogen--helium
  mixtures at high pressure},\ }\href@noop {} {\bibfield  {journal} {\bibinfo
  {journal} {Journal of Low Temperature Physics}\ }\textbf {\bibinfo {volume}
  {139}},\ \bibinfo {pages} {739} (\bibinfo {year} {2005})}\BibitemShut
  {NoStop}%
\bibitem [{\citenamefont {Natoli}\ and\ \citenamefont {Ceperley}(1995)}]{Na95}%
  \BibitemOpen
  \bibfield  {author} {\bibinfo {author} {\bibfnamefont {V.}~\bibnamefont
  {Natoli}}\ and\ \bibinfo {author} {\bibfnamefont {D.~M.}\ \bibnamefont
  {Ceperley}},\ }\bibfield  {title} {\bibinfo {title} {An optimized method for
  treating long-range potentials},\ }\href@noop {} {\bibfield  {journal}
  {\bibinfo  {journal} {Journal of Computational Physics}\ }\textbf {\bibinfo
  {volume} {117}},\ \bibinfo {pages} {171} (\bibinfo {year}
  {1995})}\BibitemShut {NoStop}%
\bibitem [{\citenamefont {Militzer}(2016)}]{BM2016}%
  \BibitemOpen
  \bibfield  {author} {\bibinfo {author} {\bibfnamefont {B.}~\bibnamefont
  {Militzer}},\ }\bibfield  {title} {\bibinfo {title} {Computation of the high
  temperature coulomb density matrix in periodic boundary conditions},\
  }\href@noop {} {\bibfield  {journal} {\bibinfo  {journal} {Comp. Phys.
  Comm.}\ }\textbf {\bibinfo {volume} {204}},\ \bibinfo {pages} {88} (\bibinfo
  {year} {2016})}\BibitemShut {NoStop}%
\bibitem [{\citenamefont {Militzer}\ \emph
  {et~al.}(2019{\natexlab{b}})\citenamefont {Militzer}, \citenamefont
  {Pollock},\ and\ \citenamefont {Ceperley}}]{Militzer2019}%
  \BibitemOpen
  \bibfield  {author} {\bibinfo {author} {\bibfnamefont {B.}~\bibnamefont
  {Militzer}}, \bibinfo {author} {\bibfnamefont {E.}~\bibnamefont {Pollock}},\
  and\ \bibinfo {author} {\bibfnamefont {D.}~\bibnamefont {Ceperley}},\
  }\bibfield  {title} {\bibinfo {title} {{Path integral Monte Carlo calculation
  of the momentum distribution of the homogeneous electron gas at finite
  temperature}},\ }\href {https://doi.org/10.1016/j.hedp.2018.12.004
  https://linkinghub.elsevier.com/retrieve/pii/S1574181818300995} {\bibfield
  {journal} {\bibinfo  {journal} {High Energy Density Physics}\ }\textbf
  {\bibinfo {volume} {30}},\ \bibinfo {pages} {13} (\bibinfo {year}
  {2019}{\natexlab{b}})}\BibitemShut {NoStop}%
\bibitem [{\citenamefont {Militzer}(2000)}]{MilitzerThesis}%
  \BibitemOpen
  \bibfield  {author} {\bibinfo {author} {\bibfnamefont {B.}~\bibnamefont
  {Militzer}},\ }\emph {\bibinfo {title} {Path Integral Monte Carlo Simulations
  of Hot Dense Hydrogen}},\ \href@noop {} {Ph.D. thesis},\ \bibinfo  {school}
  {University of Illinois at Urbana-Champaign} (\bibinfo {year}
  {2000})\BibitemShut {NoStop}%
\bibitem [{\citenamefont {Kresse}\ and\ \citenamefont
  {Joubert}(1999)}]{Kresse1999}%
  \BibitemOpen
  \bibfield  {author} {\bibinfo {author} {\bibfnamefont {G.}~\bibnamefont
  {Kresse}}\ and\ \bibinfo {author} {\bibfnamefont {D.}~\bibnamefont
  {Joubert}},\ }\bibfield  {title} {\bibinfo {title} {{From ultrasoft
  pseudopotentials to the projector augmented-wave method}},\ }\href@noop {}
  {\bibfield  {journal} {\bibinfo  {journal} {Phys. Rev. B}\ }\textbf {\bibinfo
  {volume} {59}},\ \bibinfo {pages} {1758} (\bibinfo {year}
  {1999})}\BibitemShut {NoStop}%
\bibitem [{\citenamefont {Bl{\"{o}}chl}(1994)}]{Blochl1994}%
  \BibitemOpen
  \bibfield  {author} {\bibinfo {author} {\bibfnamefont {P.~E.}\ \bibnamefont
  {Bl{\"{o}}chl}},\ }\bibfield  {title} {\bibinfo {title} {{Projector
  augmented-wave method}},\ }\href {https://doi.org/10.1103/PhysRevB.50.17953}
  {\bibfield  {journal} {\bibinfo  {journal} {Phys. Rev. B}\ }\textbf {\bibinfo
  {volume} {50}},\ \bibinfo {pages} {17953} (\bibinfo {year}
  {1994})}\BibitemShut {NoStop}%
\bibitem [{\citenamefont {Perdew}\ \emph {et~al.}(1996)\citenamefont {Perdew},
  \citenamefont {Burke},\ and\ \citenamefont {Ernzerhof}}]{PBE}%
  \BibitemOpen
  \bibfield  {author} {\bibinfo {author} {\bibfnamefont {J.~P.}\ \bibnamefont
  {Perdew}}, \bibinfo {author} {\bibfnamefont {K.}~\bibnamefont {Burke}},\ and\
  \bibinfo {author} {\bibfnamefont {M.}~\bibnamefont {Ernzerhof}},\ }\bibfield
  {title} {\bibinfo {title} {{Generalized Gradient Approximation Made
  Simple}},\ }\href@noop {} {\bibfield  {journal} {\bibinfo  {journal} {Phys.
  Rev. Lett.}\ }\textbf {\bibinfo {volume} {77}},\ \bibinfo {pages} {3865}
  (\bibinfo {year} {1996})}\BibitemShut {NoStop}%
\bibitem [{\citenamefont {Ceperley}\ and\ \citenamefont
  {Alder}(1980)}]{Ceperley1980}%
  \BibitemOpen
  \bibfield  {author} {\bibinfo {author} {\bibfnamefont {D.~M.}\ \bibnamefont
  {Ceperley}}\ and\ \bibinfo {author} {\bibfnamefont {B.~J.}\ \bibnamefont
  {Alder}},\ }\bibfield  {title} {\bibinfo {title} {Ground state of the
  electron gas by a stochastic method},\ }\href@noop {} {\bibfield  {journal}
  {\bibinfo  {journal} {Phys. Rev. Lett.}\ }\textbf {\bibinfo {volume} {45}},\
  \bibinfo {pages} {566} (\bibinfo {year} {1980})}\BibitemShut {NoStop}%
\bibitem [{\citenamefont {Perdew}\ and\ \citenamefont
  {Zunger}(1981)}]{Perdew81}%
  \BibitemOpen
  \bibfield  {author} {\bibinfo {author} {\bibfnamefont {J.~P.}\ \bibnamefont
  {Perdew}}\ and\ \bibinfo {author} {\bibfnamefont {A.}~\bibnamefont
  {Zunger}},\ }\bibfield  {title} {\bibinfo {title} {{Self-interaction
  correction to density-functional approximations for many-electron systems}},\
  }\href@noop {} {\bibfield  {journal} {\bibinfo  {journal} {Phys. Rev. B}\
  }\textbf {\bibinfo {volume} {23}},\ \bibinfo {pages} {5048} (\bibinfo {year}
  {1981})}\BibitemShut {NoStop}%
\bibitem [{\citenamefont {Mermin}(1965)}]{Mermin1965}%
  \BibitemOpen
  \bibfield  {author} {\bibinfo {author} {\bibfnamefont {N.~D.}\ \bibnamefont
  {Mermin}},\ }\bibfield  {title} {\bibinfo {title} {Thermal properties of the
  inhomogeneous electron gas},\ }\href@noop {} {\bibfield  {journal} {\bibinfo
  {journal} {Phys. Rev.}\ }\textbf {\bibinfo {volume} {137}},\ \bibinfo {pages}
  {A1441} (\bibinfo {year} {1965})}\BibitemShut {NoStop}%
\bibitem [{\citenamefont {Suryanarayana}\ \emph {et~al.}(2018)\citenamefont
  {Suryanarayana}, \citenamefont {Pratapa}, \citenamefont {Sharma},\ and\
  \citenamefont {Pask}}]{surprapa2018}%
  \BibitemOpen
  \bibfield  {author} {\bibinfo {author} {\bibfnamefont {P.}~\bibnamefont
  {Suryanarayana}}, \bibinfo {author} {\bibfnamefont {P.~P.}\ \bibnamefont
  {Pratapa}}, \bibinfo {author} {\bibfnamefont {A.}~\bibnamefont {Sharma}},\
  and\ \bibinfo {author} {\bibfnamefont {J.~E.}\ \bibnamefont {Pask}},\
  }\bibfield  {title} {\bibinfo {title} {Sqdft: Spectral quadrature method for
  large-scale parallel o(n) kohn–sham calculations at high temperature},\
  }\href {https://doi.org/https://doi.org/10.1016/j.cpc.2017.12.003} {\bibfield
   {journal} {\bibinfo  {journal} {Computer Physics Communications}\ }\textbf
  {\bibinfo {volume} {224}},\ \bibinfo {pages} {288 } (\bibinfo {year}
  {2018})}\BibitemShut {NoStop}%
\bibitem [{\citenamefont {Shulenburger}\ and\ \citenamefont
  {Mattson}(2013)}]{Shulenburger2013}%
  \BibitemOpen
  \bibfield  {author} {\bibinfo {author} {\bibfnamefont {T.}~\bibnamefont
  {Shulenburger}}\ and\ \bibinfo {author} {\bibfnamefont {T.~R.}\ \bibnamefont
  {Mattson}},\ }\href@noop {} {\bibfield  {journal} {\bibinfo  {journal} {Phys.
  Rev. B}\ }\textbf {\bibinfo {volume} {88}},\ \bibinfo {pages} {245117}
  (\bibinfo {year} {2013})}\BibitemShut {NoStop}%
\bibitem [{\citenamefont {Wahl}\ \emph
  {et~al.}(2017{\natexlab{b}})\citenamefont {Wahl}, \citenamefont {Hubbard},
  \citenamefont {Militzer}, \citenamefont {Guillot}, \citenamefont {Miguel},
  \citenamefont {Movshovitz}, \citenamefont {Kaspi}, \citenamefont {Helled},
  \citenamefont {Reese}, \citenamefont {Galanti}, \citenamefont {Levin},
  \citenamefont {Connerney},\ and\ \citenamefont {Bolton}}]{Wahl2017a}%
  \BibitemOpen
  \bibfield  {author} {\bibinfo {author} {\bibfnamefont {S.~M.}\ \bibnamefont
  {Wahl}}, \bibinfo {author} {\bibfnamefont {W.~B.}\ \bibnamefont {Hubbard}},
  \bibinfo {author} {\bibfnamefont {B.}~\bibnamefont {Militzer}}, \bibinfo
  {author} {\bibfnamefont {T.}~\bibnamefont {Guillot}}, \bibinfo {author}
  {\bibfnamefont {Y.}~\bibnamefont {Miguel}}, \bibinfo {author} {\bibfnamefont
  {N.}~\bibnamefont {Movshovitz}}, \bibinfo {author} {\bibfnamefont
  {Y.}~\bibnamefont {Kaspi}}, \bibinfo {author} {\bibfnamefont
  {R.}~\bibnamefont {Helled}}, \bibinfo {author} {\bibfnamefont
  {D.}~\bibnamefont {Reese}}, \bibinfo {author} {\bibfnamefont
  {E.}~\bibnamefont {Galanti}}, \bibinfo {author} {\bibfnamefont
  {S.}~\bibnamefont {Levin}}, \bibinfo {author} {\bibfnamefont {J.~E.}\
  \bibnamefont {Connerney}},\ and\ \bibinfo {author} {\bibfnamefont {S.~J.}\
  \bibnamefont {Bolton}},\ }\bibfield  {title} {\bibinfo {title} {{Comparing
  Jupiter interior structure models to Juno gravity measurements and the role
  of a dilute core}},\ }\href {https://doi.org/10.1002/2017GL073160} {\bibfield
   {journal} {\bibinfo  {journal} {Geophys. Res. Lett.}\ }\textbf {\bibinfo
  {volume} {44}},\ \bibinfo {pages} {4649} (\bibinfo {year}
  {2017}{\natexlab{b}})},\ \Eprint {https://arxiv.org/abs/1707.01997}
  {arXiv:1707.01997} \BibitemShut {NoStop}%
\bibitem [{\citenamefont {Kritcher}(2020)}]{Kritcher2020}%
  \BibitemOpen
  \bibfield  {author} {\bibinfo {author} {\bibfnamefont {A.~L. e.~a.}\
  \bibnamefont {Kritcher}},\ }\bibfield  {title} {\bibinfo {title} {A
  measurement of the equation of state of carbon envelopes of white dwarfs},\
  }\href {https://doi.org/10.1038/s41586-020-2535-y} {\bibfield  {journal}
  {\bibinfo  {journal} {Nature}\ }\textbf {\bibinfo {volume} {584}},\ \bibinfo
  {pages} {51} (\bibinfo {year} {2020})}\BibitemShut {NoStop}%
\bibitem [{\citenamefont {McCoy}\ \emph {et~al.}(2019)\citenamefont {McCoy},
  \citenamefont {Marshall}, \citenamefont {Polsin}, \citenamefont
  {Fratanduono}, \citenamefont {Celliers}, \citenamefont {Meyerhofer},\ and\
  \citenamefont {Boehly}}]{McCoy2019}%
  \BibitemOpen
  \bibfield  {author} {\bibinfo {author} {\bibfnamefont {C.~A.}\ \bibnamefont
  {McCoy}}, \bibinfo {author} {\bibfnamefont {M.~C.}\ \bibnamefont {Marshall}},
  \bibinfo {author} {\bibfnamefont {D.~N.}\ \bibnamefont {Polsin}}, \bibinfo
  {author} {\bibfnamefont {D.~E.}\ \bibnamefont {Fratanduono}}, \bibinfo
  {author} {\bibfnamefont {P.~M.}\ \bibnamefont {Celliers}}, \bibinfo {author}
  {\bibfnamefont {D.~D.}\ \bibnamefont {Meyerhofer}},\ and\ \bibinfo {author}
  {\bibfnamefont {T.~R.}\ \bibnamefont {Boehly}},\ }\bibfield  {title}
  {\bibinfo {title} {{Hugoniot, sound velocity, and shock temperature of MgO to
  2300 GPa}},\ }\href@noop {} {\bibfield  {journal} {\bibinfo  {journal} {Phys.
  Rev. B}\ }\textbf {\bibinfo {volume} {100}},\ \bibinfo {pages} {014106}
  (\bibinfo {year} {2019})}\BibitemShut {NoStop}%
\bibitem [{\citenamefont {Millot}\ \emph {et~al.}(2020)\citenamefont {Millot},
  \citenamefont {Zhang}, \citenamefont {Fratanduono}, \citenamefont {Coppari},
  \citenamefont {Hamel}, \citenamefont {Militzer}, \citenamefont {Simonova},
  \citenamefont {Shcheka}, \citenamefont {Dubrovinskaia}, \citenamefont
  {Dubrovinsky},\ and\ \citenamefont {Eggert}}]{Millot2020}%
  \BibitemOpen
  \bibfield  {author} {\bibinfo {author} {\bibfnamefont {M.}~\bibnamefont
  {Millot}}, \bibinfo {author} {\bibfnamefont {S.}~\bibnamefont {Zhang}},
  \bibinfo {author} {\bibfnamefont {D.~E.}\ \bibnamefont {Fratanduono}},
  \bibinfo {author} {\bibfnamefont {F.}~\bibnamefont {Coppari}}, \bibinfo
  {author} {\bibfnamefont {S.}~\bibnamefont {Hamel}}, \bibinfo {author}
  {\bibfnamefont {B.}~\bibnamefont {Militzer}}, \bibinfo {author}
  {\bibfnamefont {D.}~\bibnamefont {Simonova}}, \bibinfo {author}
  {\bibfnamefont {S.}~\bibnamefont {Shcheka}}, \bibinfo {author} {\bibfnamefont
  {N.}~\bibnamefont {Dubrovinskaia}}, \bibinfo {author} {\bibfnamefont
  {L.}~\bibnamefont {Dubrovinsky}},\ and\ \bibinfo {author} {\bibfnamefont
  {J.~H.}\ \bibnamefont {Eggert}},\ }\bibfield  {title} {\bibinfo {title}
  {{Recreating Giants Impacts in the Laboratory: Shock Compression of
  Bridgmanite to 14 Mbar}},\ }\href {https://doi.org/10.1029/2019GL085476}
  {\bibfield  {journal} {\bibinfo  {journal} {Geophysical Research Letters}\
  }\textbf {\bibinfo {volume} {47}},\ \bibinfo {pages} {1} (\bibinfo {year}
  {2020})}\BibitemShut {NoStop}%
\bibitem [{\citenamefont {Crandall}\ \emph {et~al.}(2020)\citenamefont
  {Crandall}, \citenamefont {Rygg}, \citenamefont {Spaulding}, \citenamefont
  {Boehly}, \citenamefont {Brygoo}, \citenamefont {Celliers}, \citenamefont
  {Eggert}, \citenamefont {Fratanduono}, \citenamefont {Henderson},
  \citenamefont {Huff}, \citenamefont {Jeanloz}, \citenamefont {Lazicki},
  \citenamefont {Marshall}, \citenamefont {Polsin}, \citenamefont {Zaghoo},
  \citenamefont {Millot},\ and\ \citenamefont {Collins}}]{Crandall2020}%
  \BibitemOpen
  \bibfield  {author} {\bibinfo {author} {\bibfnamefont {L.~E.}\ \bibnamefont
  {Crandall}}, \bibinfo {author} {\bibfnamefont {J.~R.}\ \bibnamefont {Rygg}},
  \bibinfo {author} {\bibfnamefont {D.~K.}\ \bibnamefont {Spaulding}}, \bibinfo
  {author} {\bibfnamefont {T.~R.}\ \bibnamefont {Boehly}}, \bibinfo {author}
  {\bibfnamefont {S.}~\bibnamefont {Brygoo}}, \bibinfo {author} {\bibfnamefont
  {P.~M.}\ \bibnamefont {Celliers}}, \bibinfo {author} {\bibfnamefont {J.~H.}\
  \bibnamefont {Eggert}}, \bibinfo {author} {\bibfnamefont {D.~E.}\
  \bibnamefont {Fratanduono}}, \bibinfo {author} {\bibfnamefont {B.~J.}\
  \bibnamefont {Henderson}}, \bibinfo {author} {\bibfnamefont {M.~F.}\
  \bibnamefont {Huff}}, \bibinfo {author} {\bibfnamefont {R.}~\bibnamefont
  {Jeanloz}}, \bibinfo {author} {\bibfnamefont {A.}~\bibnamefont {Lazicki}},
  \bibinfo {author} {\bibfnamefont {M.~C.}\ \bibnamefont {Marshall}}, \bibinfo
  {author} {\bibfnamefont {D.~N.}\ \bibnamefont {Polsin}}, \bibinfo {author}
  {\bibfnamefont {M.}~\bibnamefont {Zaghoo}}, \bibinfo {author} {\bibfnamefont
  {M.}~\bibnamefont {Millot}},\ and\ \bibinfo {author} {\bibfnamefont {G.~W.}\
  \bibnamefont {Collins}},\ }\bibfield  {title} {\bibinfo {title} {{Equation of
  State of ${\mathrm{CO}}_{2}$ Shock Compressed to 1 TPa}},\ }\href
  {https://doi.org/10.1103/PhysRevLett.125.165701} {\bibfield  {journal}
  {\bibinfo  {journal} {Phys. Rev. Lett.}\ }\textbf {\bibinfo {volume} {125}},\
  \bibinfo {pages} {165701} (\bibinfo {year} {2020})}\BibitemShut {NoStop}%
\bibitem [{\citenamefont {Hugoniot}(1887)}]{Hugoniot1887}%
  \BibitemOpen
  \bibfield  {author} {\bibinfo {author} {\bibfnamefont {H.}~\bibnamefont
  {Hugoniot}},\ }\bibfield  {title} {\bibinfo {title} {Memoir on the
  propagation of movements in bodies, especially perfect gases (first part)},\
  }\href@noop {} {\bibfield  {journal} {\bibinfo  {journal} {J. de l’Ecole
  Polytechnique}\ }\textbf {\bibinfo {volume} {57}},\ \bibinfo {pages} {3}
  (\bibinfo {year} {1887})}\BibitemShut {NoStop}%
\bibitem [{\citenamefont {Hugoniot}(1889)}]{Hugoniot1889}%
  \BibitemOpen
  \bibfield  {author} {\bibinfo {author} {\bibfnamefont {H.}~\bibnamefont
  {Hugoniot}},\ }\bibfield  {title} {\bibinfo {title} {Memoir on the
  propagation of movements in bodies, especially perfect gases (second part)},\
  }\href@noop {} {\bibfield  {journal} {\bibinfo  {journal} {J. de l’Ecole
  Polytechnique}\ }\textbf {\bibinfo {volume} {58}},\ \bibinfo {pages} {1}
  (\bibinfo {year} {1889})}\BibitemShut {NoStop}%
\bibitem [{\citenamefont {D\"oppner}\ \emph {et~al.}(2018)\citenamefont
  {D\"oppner}, \citenamefont {Swift}, \citenamefont {Kritcher}, \citenamefont
  {Bachmann}, \citenamefont {Collins}, \citenamefont {Chapman}, \citenamefont
  {Hawreliak}, \citenamefont {Kraus}, \citenamefont {Nilsen}, \citenamefont
  {Rothman}, \citenamefont {Benedict}, \citenamefont {Dewald}, \citenamefont
  {Fratanduono}, \citenamefont {Gaffney}, \citenamefont {Glenzer},
  \citenamefont {Hamel}, \citenamefont {Landen}, \citenamefont {Lee},
  \citenamefont {LePape}, \citenamefont {Ma}, \citenamefont {MacDonald},
  \citenamefont {MacPhee}, \citenamefont {Milathianaki}, \citenamefont
  {Millot}, \citenamefont {Neumayer}, \citenamefont {Sterne}, \citenamefont
  {Tommasini},\ and\ \citenamefont {Falcone}}]{Doeppner2018}%
  \BibitemOpen
  \bibfield  {author} {\bibinfo {author} {\bibfnamefont {T.}~\bibnamefont
  {D\"oppner}}, \bibinfo {author} {\bibfnamefont {D.~C.}\ \bibnamefont
  {Swift}}, \bibinfo {author} {\bibfnamefont {A.~L.}\ \bibnamefont {Kritcher}},
  \bibinfo {author} {\bibfnamefont {B.}~\bibnamefont {Bachmann}}, \bibinfo
  {author} {\bibfnamefont {G.~W.}\ \bibnamefont {Collins}}, \bibinfo {author}
  {\bibfnamefont {D.~A.}\ \bibnamefont {Chapman}}, \bibinfo {author}
  {\bibfnamefont {J.}~\bibnamefont {Hawreliak}}, \bibinfo {author}
  {\bibfnamefont {D.}~\bibnamefont {Kraus}}, \bibinfo {author} {\bibfnamefont
  {J.}~\bibnamefont {Nilsen}}, \bibinfo {author} {\bibfnamefont
  {S.}~\bibnamefont {Rothman}}, \bibinfo {author} {\bibfnamefont {L.~X.}\
  \bibnamefont {Benedict}}, \bibinfo {author} {\bibfnamefont {E.}~\bibnamefont
  {Dewald}}, \bibinfo {author} {\bibfnamefont {D.~E.}\ \bibnamefont
  {Fratanduono}}, \bibinfo {author} {\bibfnamefont {J.~A.}\ \bibnamefont
  {Gaffney}}, \bibinfo {author} {\bibfnamefont {S.~H.}\ \bibnamefont
  {Glenzer}}, \bibinfo {author} {\bibfnamefont {S.}~\bibnamefont {Hamel}},
  \bibinfo {author} {\bibfnamefont {O.~L.}\ \bibnamefont {Landen}}, \bibinfo
  {author} {\bibfnamefont {H.~J.}\ \bibnamefont {Lee}}, \bibinfo {author}
  {\bibfnamefont {S.}~\bibnamefont {LePape}}, \bibinfo {author} {\bibfnamefont
  {T.}~\bibnamefont {Ma}}, \bibinfo {author} {\bibfnamefont {M.~J.}\
  \bibnamefont {MacDonald}}, \bibinfo {author} {\bibfnamefont {A.~G.}\
  \bibnamefont {MacPhee}}, \bibinfo {author} {\bibfnamefont {D.}~\bibnamefont
  {Milathianaki}}, \bibinfo {author} {\bibfnamefont {M.}~\bibnamefont
  {Millot}}, \bibinfo {author} {\bibfnamefont {P.}~\bibnamefont {Neumayer}},
  \bibinfo {author} {\bibfnamefont {P.~A.}\ \bibnamefont {Sterne}}, \bibinfo
  {author} {\bibfnamefont {R.}~\bibnamefont {Tommasini}},\ and\ \bibinfo
  {author} {\bibfnamefont {R.~W.}\ \bibnamefont {Falcone}},\ }\bibfield
  {title} {\bibinfo {title} {Absolute equation-of-state measurement for
  polystyrene from 25 to 60 mbar using a spherically converging shock wave},\
  }\href {https://doi.org/10.1103/PhysRevLett.121.025001} {\bibfield  {journal}
  {\bibinfo  {journal} {Phys. Rev. Lett.}\ }\textbf {\bibinfo {volume} {121}},\
  \bibinfo {pages} {025001} (\bibinfo {year} {2018})}\BibitemShut {NoStop}%
\bibitem [{\citenamefont {Khairallah}\ and\ \citenamefont
  {Militzer}(2008)}]{KM08}%
  \BibitemOpen
  \bibfield  {author} {\bibinfo {author} {\bibfnamefont {S.~A.}\ \bibnamefont
  {Khairallah}}\ and\ \bibinfo {author} {\bibfnamefont {B.}~\bibnamefont
  {Militzer}},\ }\href@noop {} {\bibfield  {journal} {\bibinfo  {journal}
  {Phys. Rev. Lett.}\ }\textbf {\bibinfo {volume} {101}},\ \bibinfo {pages}
  {106407} (\bibinfo {year} {2008})}\BibitemShut {NoStop}%
\bibitem [{\citenamefont {French}\ \emph {et~al.}(2009)\citenamefont {French},
  \citenamefont {Mattsson}, \citenamefont {Nettelmann},\ and\ \citenamefont
  {Redmer}}]{French2009}%
  \BibitemOpen
  \bibfield  {author} {\bibinfo {author} {\bibfnamefont {M.}~\bibnamefont
  {French}}, \bibinfo {author} {\bibfnamefont {T.~R.}\ \bibnamefont
  {Mattsson}}, \bibinfo {author} {\bibfnamefont {N.}~\bibnamefont
  {Nettelmann}},\ and\ \bibinfo {author} {\bibfnamefont {R.}~\bibnamefont
  {Redmer}},\ }\bibfield  {title} {\bibinfo {title} {{Equation of state and
  phase diagram of water at ultrahigh pressures as in planetary interiors}},\
  }\href@noop {} {\bibfield  {journal} {\bibinfo  {journal} {Phys. Rev. B}\
  }\textbf {\bibinfo {volume} {79}},\ \bibinfo {pages} {5} (\bibinfo {year}
  {2009})}\BibitemShut {NoStop}%
\bibitem [{\citenamefont {Knudson}\ \emph
  {et~al.}(2012{\natexlab{b}})\citenamefont {Knudson}, \citenamefont
  {Desjarlais}, \citenamefont {Lemke}, \citenamefont {Mattsson}, \citenamefont
  {French}, \citenamefont {Nettelmann},\ and\ \citenamefont
  {Redmer}}]{Knudson12}%
  \BibitemOpen
  \bibfield  {author} {\bibinfo {author} {\bibfnamefont {M.~D.}\ \bibnamefont
  {Knudson}}, \bibinfo {author} {\bibfnamefont {M.~P.}\ \bibnamefont
  {Desjarlais}}, \bibinfo {author} {\bibfnamefont {R.~W.}\ \bibnamefont
  {Lemke}}, \bibinfo {author} {\bibfnamefont {T.~R.}\ \bibnamefont {Mattsson}},
  \bibinfo {author} {\bibfnamefont {M.}~\bibnamefont {French}}, \bibinfo
  {author} {\bibfnamefont {N.}~\bibnamefont {Nettelmann}},\ and\ \bibinfo
  {author} {\bibfnamefont {R.}~\bibnamefont {Redmer}},\ }\href@noop {}
  {\bibfield  {journal} {\bibinfo  {journal} {Phys. Rev. Lett.}\ }\textbf
  {\bibinfo {volume} {108}},\ \bibinfo {pages} {091102} (\bibinfo {year}
  {2012}{\natexlab{b}})}\BibitemShut {NoStop}%
\bibitem [{\citenamefont {Podurets}\ \emph {et~al.}(1972)\citenamefont
  {Podurets}, \citenamefont {Simakov}, \citenamefont {Trunin}, \citenamefont
  {Popov},\ and\ \citenamefont {Moiseev}}]{Podurets1972}%
  \BibitemOpen
  \bibfield  {author} {\bibinfo {author} {\bibfnamefont {M.~A.}\ \bibnamefont
  {Podurets}}, \bibinfo {author} {\bibfnamefont {G.~V.}\ \bibnamefont
  {Simakov}}, \bibinfo {author} {\bibfnamefont {R.}~\bibnamefont {Trunin}},
  \bibinfo {author} {\bibfnamefont {L.~V.}\ \bibnamefont {Popov}},\ and\
  \bibinfo {author} {\bibfnamefont {B.}~\bibnamefont {Moiseev}},\ }\href@noop
  {} {\bibfield  {journal} {\bibinfo  {journal} {Sov. Phys. JETP}\ }\textbf
  {\bibinfo {volume} {35}},\ \bibinfo {pages} {375} (\bibinfo {year}
  {1972})}\BibitemShut {NoStop}%
\bibitem [{\citenamefont {Wilson}\ \emph {et~al.}(2013)\citenamefont {Wilson},
  \citenamefont {Wong},\ and\ \citenamefont {Militzer}}]{Wilson2013}%
  \BibitemOpen
  \bibfield  {author} {\bibinfo {author} {\bibfnamefont {H.~F.}\ \bibnamefont
  {Wilson}}, \bibinfo {author} {\bibfnamefont {M.~L.}\ \bibnamefont {Wong}},\
  and\ \bibinfo {author} {\bibfnamefont {B.}~\bibnamefont {Militzer}},\
  }\bibfield  {title} {\bibinfo {title} {{Superionic to Superionic Phase Change
  in Water: Consequences for the Interiors of Uranus and Neptune}},\
  }\href@noop {} {\bibfield  {journal} {\bibinfo  {journal} {Phys. Rev. Lett.}\
  }\textbf {\bibinfo {volume} {110}},\ \bibinfo {pages} {151102} (\bibinfo
  {year} {2013})}\BibitemShut {NoStop}%
\bibitem [{\citenamefont {Johnson}(1997)}]{JDJohnson1997}%
  \BibitemOpen
  \bibfield  {author} {\bibinfo {author} {\bibfnamefont {J.~D.}\ \bibnamefont
  {Johnson}},\ }\href@noop {} {\bibfield  {journal} {\bibinfo  {journal} {Phys.
  Rev. E}\ }\textbf {\bibinfo {volume} {59}},\ \bibinfo {pages} {3727}
  (\bibinfo {year} {1997})}\BibitemShut {NoStop}%
\bibitem [{\citenamefont {Ozaki}\ \emph {et~al.}(2016)\citenamefont {Ozaki},
  \citenamefont {Nellis}, \citenamefont {Mashimo}, \citenamefont {Ramzan},
  \citenamefont {Ahuja}, \citenamefont {Kaewmaraya}, \citenamefont {Kimura},
  \citenamefont {Knudson}, \citenamefont {Miyanishi}, \citenamefont {Sakawa}
  \emph {et~al.}}]{Ozaki2016}%
  \BibitemOpen
  \bibfield  {author} {\bibinfo {author} {\bibfnamefont {N.}~\bibnamefont
  {Ozaki}}, \bibinfo {author} {\bibfnamefont {W.}~\bibnamefont {Nellis}},
  \bibinfo {author} {\bibfnamefont {T.}~\bibnamefont {Mashimo}}, \bibinfo
  {author} {\bibfnamefont {M.}~\bibnamefont {Ramzan}}, \bibinfo {author}
  {\bibfnamefont {R.}~\bibnamefont {Ahuja}}, \bibinfo {author} {\bibfnamefont
  {T.}~\bibnamefont {Kaewmaraya}}, \bibinfo {author} {\bibfnamefont
  {T.}~\bibnamefont {Kimura}}, \bibinfo {author} {\bibfnamefont
  {M.}~\bibnamefont {Knudson}}, \bibinfo {author} {\bibfnamefont
  {K.}~\bibnamefont {Miyanishi}}, \bibinfo {author} {\bibfnamefont
  {Y.}~\bibnamefont {Sakawa}}, \emph {et~al.},\ }\bibfield  {title} {\bibinfo
  {title} {Dynamic compression of dense oxide (gd 3 ga 5 o 12) from 0.4 to 2.6
  tpa: Universal hugoniot of fluid metals},\ }\href@noop {} {\bibfield
  {journal} {\bibinfo  {journal} {Scientific reports}\ }\textbf {\bibinfo
  {volume} {6}},\ \bibinfo {pages} {26000} (\bibinfo {year}
  {2016})}\BibitemShut {NoStop}%
\end{thebibliography}

\providecommand{\noopsort}[1]{}\providecommand{\singleletter}[1]{#1}%

\end{document}